\documentclass{aa}
\usepackage{graphicx}
\usepackage{amsmath,amssymb}
\usepackage{verbatim}
\usepackage{chngcntr}
\usepackage{float}
\usepackage{placeins}
\usepackage{color}
\usepackage{hyperref}
\usepackage[varg]{txfonts}

\bibpunct{(}{)}{;}{a}{}{,} 

\def\lapprox{\lower.4ex\hbox{$\;\buildrel <\over{\scriptstyle\sim}\;$}}
\def\gapprox{\lower.4ex\hbox{$\;\buildrel >\over{\scriptstyle\sim}\;$}}

\DeclareRobustCommand{\ion}[2]{%
\relax\ifmmode
\ifx\testbx\f@series
{\mathbf{#1\,\mathsc{#2}}}\else
{\mathrm{#1\,\mathsc{#2}}}\fi
\else\textup{#1\,{\mdseries\textsc{#2}}}%
\fi}

\begin{document}

\title{FUV line emission, gas kinematics, and discovery of [Fe XXI]  \\ $\lambda$1354.1 in the sightline toward a filament in M87}

\author{Michael E. Anderson \inst{1} 
\and
Rashid Sunyaev\inst{1,2,3} 
}

\institute{Max-Planck Institute for Astrophysics, Garching bei Muenchen, Germany\\ \email{michevan@mpa-garching.mpg.de}
\and
Space Research Institute (IKI),  Russian Academy of Sciences, Profsoyuznaya 84/32, Moscow 117997, Russia
\and
Institute for Advanced Study, 1 Einstein Drive, Princeton, NJ, USA
}

\titlerunning{FUV line emission toward M87}
\authorrunning{Anderson \& Sunyaev}

\abstract{

We present new Hubble Space Telescope - Cosmic Origins Spectrograph (HST-COS) G130M spectroscopy which we have obtained for a sightline toward a filament projected 1.9 kpc from the nucleus of M87, near the edge of the inner radio lobe to the east of the nucleus. The combination of the sensitivity of COS and the proximity of M87 allows us to study the structure of this filament in unparalleled detail. We propose that the filament is composed of many cold clumps, each surrounded by an FUV-emitting boundary layer, with the filament having a radius $r_c \sim 10$ pc and the clumps filling the cylinder with a low volume filling factor. The observed velocity dispersion in emission lines from the filament results from the random motions of these clumps within the filament. 

We measure fluxes and kinematics for emission lines of Ly$\alpha$, \ion{C}{ii} $\lambda$1335, and \ion{N}{v} $\lambda1238$, finding $v_r = 147\pm2$ km s$^{-1}$, $138\pm18$ km s$^{-1}$, and $148^{+14}_{-16}$ km s$^{-1}$ relative to M87, and line broadenings $\sigma_r = 171\pm2$ km s$^{-1}$, $189^{+12}_{-11}$ km s$^{-1}$, and $128^{+23}_{-17}$ km s$^{-1}$ respectively.  We associate these three lines, as well as archival measurements of H$\alpha$, \ion{C}{iv} $\lambda$1549, and \ion{He}{ii} $\lambda$1640, with a multitemperature boundary layer around clumps which are moving with supersonic random motions in the filament. This boundary layer is a significant coolant of the hot gas. We show that the [\ion{C}{ii}] $\lambda$158$\mu$m flux observed by Herschel-PACS from this region implies the existence of a massive  cold ($T \sim 10^3$ K) component in the filament which contains significantly more mass $(M \sim 8000 M_{\odot}$ within our $r \approx 100$ pc sightline) than the FUV-emitting boundary layer. It has about the same bulk velocity and velocity dispersion as the boundary layer.

We also detect [\ion{Fe}{xxi}] $\lambda$1354 in emission at $4-5\sigma$. This line is emitted from 1 keV ($T \approx 10^7$ K) plasma, and we use it to measure the bulk radial velocity ($v_r = -92^{+34}_{-22}$ km s$^{-1}$) and velocity dispersion ($\sigma_r = 69^{+79}_{-27}$ km s$^{-1}$) of the plasma at this temperature. In contrast to the intermediate-temperature FUV lines, [\ion{Fe}{xxi}] is blueshifted relative to M87 and matches the bulk velocity of a nearby filament to the south. We hypothesize that this line arises from the approaching face of the radio bubble expanding through this sightline, while the filament lies on the receding side of the bubble. A byproduct of our observations is the detection of absorption from interstellar gas in our Galaxy, observed in \ion{C}{ii} $\lambda$1335 and Ly$\alpha$.}

\maketitle

\section{Introduction}

In this paper we discuss a filament projected 1.9 kpc from the nucleus of M87, for which we have obtained a new far-ultraviolet (FUV) spectrum with HST-COS. Using this spectrum, we measure with 10-15 km s$^{-1}$ relative precision the broadening and bulk velocities of \ion{C}{ii} $\lambda$1335, \ion{N}{v} $\lambda$1238, and Ly$\alpha$, which originate from plasma associated with the filament. We also update kinematic measurements of \ion{C}{iv} $\lambda$1549 and \ion{He}{ii} $\lambda$1640 from archival low-resolution spectra of the same sightline. These FUV lines provide us with unique information about the properties of the transition zone between hot gas and cold core of the filament, as these lines are bright in collisional excitation at a broad range of intermediate temperatures. We cannot prove that the boundary layer with which we associate these lines is powered only via electron thermal conductivity, but nevertheless this is a rare case when we see in detail and are able to measure random motions and outflow velocities, as well as line intensities and emission measures in many emission lines corresponding to different temperatures of collisionally excited plasma. Until now we had a similar situation only in the cases of the solar chromosphere and corona during solar flares, and in the upper atmospheres of some nearby stars.

We also report the detection of [\ion{Fe}{xxi}] $\lambda$1354 in this sightline, and show that it exhibits bulk motion in the opposite direction from the other FUV lines. The permitted FUV lines exhibit velocities from random motions $\sigma_r \approx 130$ km s$^{-1}$ and bulk velocities $v_r \approx 140$ km s$^{-1}$, consistent with previous measurements in other wavebands, and Ly$\alpha$ appears slightly broader due to resonant scattering.  [\ion{Fe}{xxi}] probably exhibits a lower velocity dispersion than these other lines, although it has significantly larger uncertainties due to a lower SN. We propose a model wherein the [\ion{Fe}{xxi}] originates from the approaching face of the radio bubble which is expanding through out line of sight, while the filamentary emission originates from the receding edge of this radio lobe.

These observations also represent a first chance to combine \ion{C}{ii} $\lambda$1335 observations and far-infrared (FIR) [\ion{C}{ii}] $\lambda$158$\mu$m observations for the same filament. The latter come from Herschel-PACS data downloaded from the archive which have been previously discussed by \citet{Werner2013}. As we will show, we gain insight into the structure of both the cold ($T < 10^4$ K) and warm $(10^4$ K$ \lapprox T \lapprox 10^5$ K) phases of the filament through the simultaneous study of permitted and forbidden lines for the same ion, with different excitation temperatures and critical densities.  We will show that the FIR and FUV \ion{C}{ii} lines exhibit similar radial velocities and velocity dispersions. The general agreement in kinematics for sightline (between the permitted FUV lines in this paper, archival observations of H$\alpha$, and the forbidden [\ion{C}{ii}] line) is remarkable. These lines have characteristic temperatures spanning more than three orders of magnitude and the observed $\sigma_r$ slightly exceeds the sound speed for the warm gas but is highly supersonic for the cold gas. The observed random motions in this sightline must be driven by external forcing -- probably the expansion of the radio lobe into the M87 ISM -- and not internal turbulent motions in the multiphase gas itself.

We will also conclude that the cold phase contains orders of magnitude more mass than the FUV-emitting boundary layer. We propose a geometry for the filaments where this cold phase is located in numerous clumps with a low volume filling factor, and each clump is surrounded by a very thin FUV-emitting boundary layer. The random motions of these clumps produce the observed velocity dispersion in the permitted FUV lines. Our measurements also demonstrate the effectiveness of the UV lines (especially Ly$\alpha$ and \ion{He}{ii} $\lambda$304, and to a lesser degree [\ion{C}{ii}]  $\lambda$158$\mu$m), as coolants for the interstellar medium of M87. \ion{He}{ii} $\lambda$304 can also contribute to the photoionization of H and other elements. A by-product of our observations is the detection of absorption from interstellar gas in our Galaxy, observed in \ion{C}{ii} $\lambda$1335 and Ly$\alpha$.

The phenomenon of filaments in the centers of galaxy clusters remains somewhat mysterious. Initially discovered through their bright H$\alpha$ emission \citep{Arp1967}, they are generally thought to be connected with cooling instabilities originating in the intracluster medium (ICM), as the cooling times at typical ICM temperatures near the center of the cluster (several keV) and densities (0.01 - 0.1 cm$^{-3}$) are significantly shorter than the timescales for galaxy evolution. Empirically, this connection has support as well, in the observed correlation between the presence of a cool core in the ICM (corresponding to a low central entropy and a short cooling time) and the existence of H$\alpha$ filaments \citep{Cavagnolo2008}.

At the same time, we know that the brightest central galaxies of galaxy clusters contain supermassive black holes (SMBHs). The energy released during accretion onto these SMBHs is powerful enough to prevent the formation of cooling flows (e.g., \citealt{Tucker1997}, \citealt{Churazov2001}, \citealt{Churazov2002}). Nevertheless, networks of filaments persist in such systems. 

M87, the central galaxy of the Virgo Cluster, has an especially well-studied SMBH (\citealt{Sargent1978}, \citealt{Young1978}, \citealt{Macchetto1997}, \citealt{Gebhardt2009}, \citealt{Walsh2013}) due to its mass and proximity. The SMBH is relatively dormant while still approximately balancing cooling from the Virgo ICM (\citealt{Churazov2001}, \citealt{Forman2005}), but M87 also has an enormous system of low-temperature filaments extending from the nucleus several kpc to the east. In contrast to many other galaxy cluster filament systems, the filaments in M87 are not forming stars (\citealt{Salome2008}, \citealt{Sparks2012}), and represent a prototypical example of this class of "AGN-dominated" or "shock-dominated" filaments.

The bulk of the previous work on these filaments has been connected with attempts to image the filament region in the optical and X-ray and to find evidence for spectral lines corresponding to temperatures in the unstable range between $10^4$ K and $10^7$ K. (\citealt{Arp1967}, \citealt{Ford1979}, \citealt{Sparks1993}, \citealt{Ford1994},  \citealt{Harms1994}, \citealt{Macchetto1997}, \citealt{Simionescu2008}, \citealt{Sparks2009}, \citealt{Sparks2012}, \citealt{Werner2013}). In this paper, we use data we obtained from HST-COS, which trace this temperature range directly but also permit us to measure velocity structure. These velocity measurements are an essential ingredient for theory and for numerical simulations, as they in principle can prove whether the cooling material is falling onto the SMBH or is uplifted in an outflow. 

The [\ion{Fe}{xxi}] line merits some additional discussion. As we showed in \citet{Anderson2016}, there are several forbidden lines in the FUV from various species of highly ionized Iron, but in emission [\ion{Fe}{xxi}] is the brightest, and in collisional excitation equilibrium (CIE) its emissivity peaks at log T = 7.05 K. This line is well-studied in solar flare spectra, and its ionization potential is 1.7 keV so photoionization is not expected to be important outside of AGN environments. While $T \sim 10^7$ K plasma generally radiates primarily in X-rays, the ability to study its emission in the FUV allows us to measure kinematics of the hot plasma, which is generally impossible without microcalorimeters or mm wavelength observations of the kinematic Sunyaev-Zel'dovich effect from bulk motions and turbulence inside clusters of galaxies \citep{Inogamov2003}.

In \citet{Anderson2016} we used archival HST-COS data to provide tentative evidence for [\ion{Fe}{xxi}] emission in several sightlines toward galaxy clusters with known reservoirs of $10^7$ K plasma, including the nucleus of M87 and the filament region studied here. \citet{Danforth2016} were able to confirm our suggestion of [\ion{Fe}{xxi}] in the M87 nucleus, and here we will confirm its presence in the filaments. In this paper we present a detection of [\ion{Fe}{xxi}] in the ISM of M87 at $4-5\sigma$ significance, which represents the second direct measurement to date of motions in the hot ICM of nearby galaxy clusters, after the measurement in Perseus with Hitomi \citep{Hitomi2016}. The turbulent velocity we measure for [\ion{Fe}{xxi}] in the core of Virgo is also similar to the turbulent velocity measured by Hitomi in the core of Perseus. [\ion{Fe}{xxi}] is a useful complement to the exciting science possible with microcalorimeters, since it probes a slightly cooler phase ($T \approx 1$ keV) of the ICM using independent instrumentation. Additionally, until the launch of another mission with a microcalorimeter, these FUV lines are the only currently accessible direct probe of turbulence in the hot ICM.

One question which has remained open is the excitation mechanism of the line emission from these filaments. In the nucleus of M87, there is a rotating cloud with $r\sim35$ pc which is thought to be photoionized by emission from the central source (\citealt{Sabra2003}, Anderson and Sunyaev in preparation), but the filamentary system studied here is projected 1.9 kpc from the nucleus, and it is not obvious that photoionization from the nucleus can ionize filaments at this distance.  A similar issue of insufficient ionizing radiation is seen in the cores of many other LINER-type systems as well (\citealt{Binette1993}, \citealt{Ho2008}, \citealt{Eracleous2010}), but the proximity of M87 offers a unique chance to find a resolution.

One possibility is strong variability in the nuclear emission, so that photoionization is still the primary excitation mechanism in the filaments. In this case, an AGN outburst several millennia ago would have provided the necessary photons to ionize the filaments at large radii. This is consistent with many observations that show high variability with a short duty cycle in AGN \citep{Shankar2009}, including in M87 (\citealt{Harris1997}, \citealt{Perlman2011}). In filaments which are actively forming stars, radiation from the young stars can also power the photoionization (e.g., \citealt{Odea2004}, \citealt{McDonald2012}), although as we mentioned above, young stars are not observed in M87.

The other possibility is a transition from photoionization in the nucleus to collisional excitation at larger radii. Collisional excitation is commonly proposed as a mechanism for heating filaments, either via thermal conduction (\citealt{Boehringer1989}, \citealt{Nipoti2004}, \citealt{Inogamov2010}) or via collisions with suprathermal electrons from the hot ICM \citep{Fabian2011}. \citet{Sparks2012} argue for thermal conduction in the filaments of M87 as well. As we will show, there are likely shocks present in the filament as well, due to collisions of cold cloudlets moving with supersonic random velocities relative to one another in the core of the filament, which are another plausible mechanism for collisional excitation.

We use our spectroscopic results to distinguish between these cases, and they point to a mixture of collisional excitation and recombination in  Ly$\alpha$ and H$\alpha$. However, with the exceptions of H$\alpha$ and \ion{He}{ii} $\lambda$1640 (which is the Balmer-$\alpha$ transition in He$^{+}$), a simple collisional model is able to reproduce the intensities and line ratios of the other FUV lines in this filament. 
 
The structure of this paper is as follows. In section 2 we describe and present our COS G130M observations of this system, and in section 3 we discuss the FUV continuum and present fits to permitted emission lines  such as \ion{C}{ii}, \ion{N}{v}, and Ly$\alpha$. In Section 4 we present the forbidden FUV emission line [\ion{Fe}{xxi}], and in section 5 we summarize the FUV line luminosities of this filament region and discuss their contribution to the total cooling budget of the ICM. In section 6 we analyze these lines, discussing the excitation mechanisms for this line emission, the differential emission measure for the sightline, and the volume filling factor of the FUV-emitting plasma. In Section 7 we discuss [\ion{Fe}{xxi}] and we propose a possible physical model to explain the different bulk velocity observed in [\ion{Fe}{xxi}] compared to the other FUV lines. Finally, in section 8 we discuss the cold phase of the filaments as traced by [\ion{C}{ii}] $\lambda$158 $\mu$m, and estimate the mass, dimensions, and filling factor of the filamentary gas in this phase. Throughout this paper we use a velocity relative to M87 of 1283 km s$^{-1}$ \citep{Makarov2014} and a distance $d = 16.7$ Mpc \citep{Blakeslee2009} for M87. Errorbars are quoted at the 1$\sigma$ level unless otherwise stated. 

 \section{HST-COS observations}

This paper focuses on one particular H$\alpha$-emitting filament located in M87. As mentioned above, the filament is part of a network of filaments projected about 1.9 kpc from the nucleus of the galaxy, and has been observed in H$\alpha$+[\ion{N}{ii}] narrowband imaging, FUV imaging \citep{Sparks2009} and FUV spectroscopy \citep{Sparks2012} with HST. Additionally, spectrophometry with Herschel-PACS \citep{Poglitsch2010} has mapped out the filament in [\ion{C}{ii}] 158$\mu$m emission as well \citep{Werner2013}. Deep X-ray studies of the region have also been performed (\citealt{Young2002}, \citealt{Werner2013}), showing that the ICM around these filaments is multiphase, and that the X-ray emission can be described with a model containing at least two or three separate components. 
 
Figure 1 shows the region of interest. Our sightline indicated with the red circle targets one of the filaments in the northern section of the system. The filament is clearly visible in H$\alpha$+[\ion{N}{ii}] emission, and also lies in an area which shows evidence of having a multiphase ICM, including a strong 1 keV component and even a tentative 0.5 keV component \citep{Werner2013}. The especially large emission measure of 1 keV plasma around this filament motivated our selection of this sightline. We proposed to observe this sightline with HST-COS in order to detect [\ion{Fe}{xxi}]$\lambda$1354.1 emission and to measure with medium-resolution spectroscopy the other FUV lines expected from the intermediate-temperature gas associated with the filament.

\begin{figure}
\begin{center}
{\includegraphics[width=8.5cm]{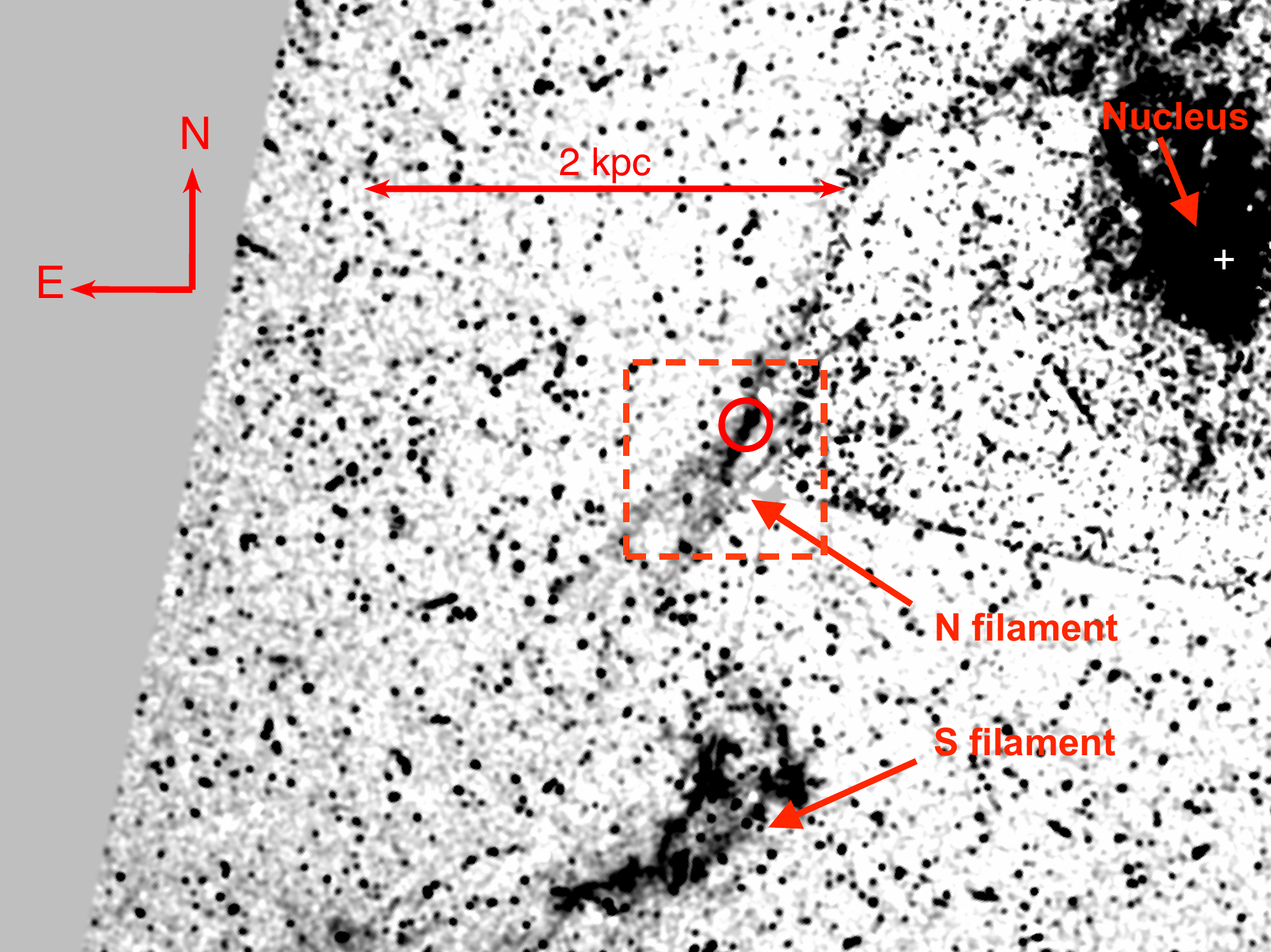}}
\end{center}
\caption{ H$\alpha$+[\ion{N}{ii}] image of the multiphase filamentary structures east of the nucleus of M87, constructed from archival HST-WFPC2 images as described in sect. 6.2. The red circle in the center of the image indicates the location of our COS aperture, which targets the northern of two large filaments projected east of the nucleus. This filament is clearly separated from the other filament to the south, as there is no H$\alpha$ bridge connecting the two filaments, and as we will show they have different velocities and likely originate on opposite sides of an expanding radio lobe. The dashed red box indicates the approximate size of the Herschel-PACS region from which we measure a [\ion{C}{ii}] 158$\mu$m spectrum (see, e.g., sect. 2.1). Comparison of this image to figure 2 of Werner et al. (2013) shows that the hot ICM in this sightline is multiphase and rich in 1 keV plasma, although the majority of the hot gas is in the slightly hotter 2 keV phase.}
\end{figure}

We successfully obtained 14 orbits with HST-COS in cycle 24 (proposal id 14623) to observe this filament using the G130M grating. Our observation was split across five different visits, each using a central wavelength of 1318\AA. We combined the \verb"x1dsum" files from these five visits directly in counts space and then binned the results. We use 0.70\AA{} bins in Figure 2 for display purposes, but for analysis and subsequent figures we use 0.21\AA{} bins, which corresponds to approximately three resels. For Ly$\alpha$, which is much brighter than the rest of the spectrum, we use 0.07\AA{} bins.  We estimate errors using Poisson statistics on the gross counts within each bin. For plotting purposes we show 1$\sigma$ errors estimated using the \citet{Gehrels1986} approximation, in similar manner to \citet{Tumlinson2013} and to the Hubble Spectroscopic Legacy Archive \citep{Peeples2017}, but for model fitting we compute the likelihoods of our models using Poisson statistics without the Gehrels approximation. 

Since the continuum emission may be spatially extended (see sect. 2.2 and 3.1) we also checked the effect of the pipeline background subtraction by constructing alternate spectra reduced using the \verb"BOXCAR" algorithm, but the results were unchanged within the uncertainties, and no qualitative differences were seen. We use the pipeline products downloaded from the HST archive for our scientific analysis here, which use the \verb"TWOZONE" algorithm. 

The resulting spectrum is shown in Figure 2. A faint continuum with a few bright emission lines can be seen, and the \ion{N}{i} $\lambda$1200, \ion{H}{i} $\lambda$1216, and \ion{O}{i} $\lambda$1302,1305 airglow lines are also clearly visible. In the next section, we discuss this spectrum in more detail.  

\begin{figure*}
\begin{center}
{\includegraphics[width=17.4cm]{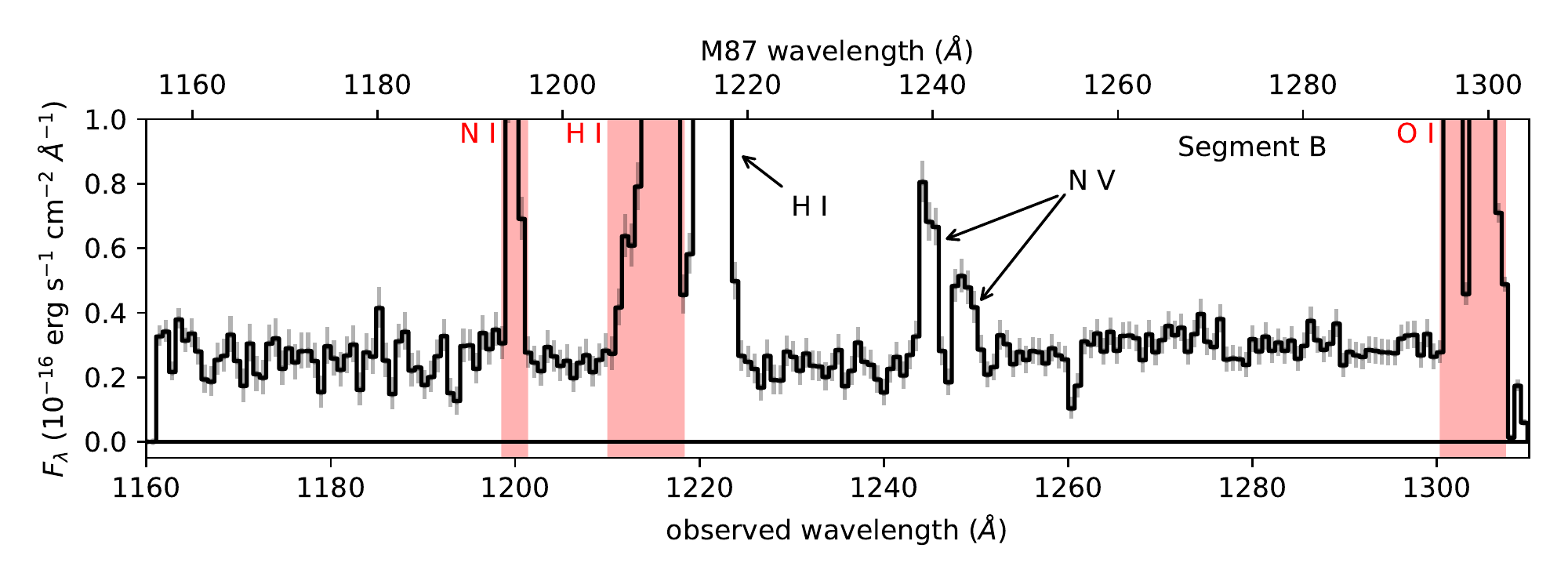}}
{\includegraphics[width=17.4cm]{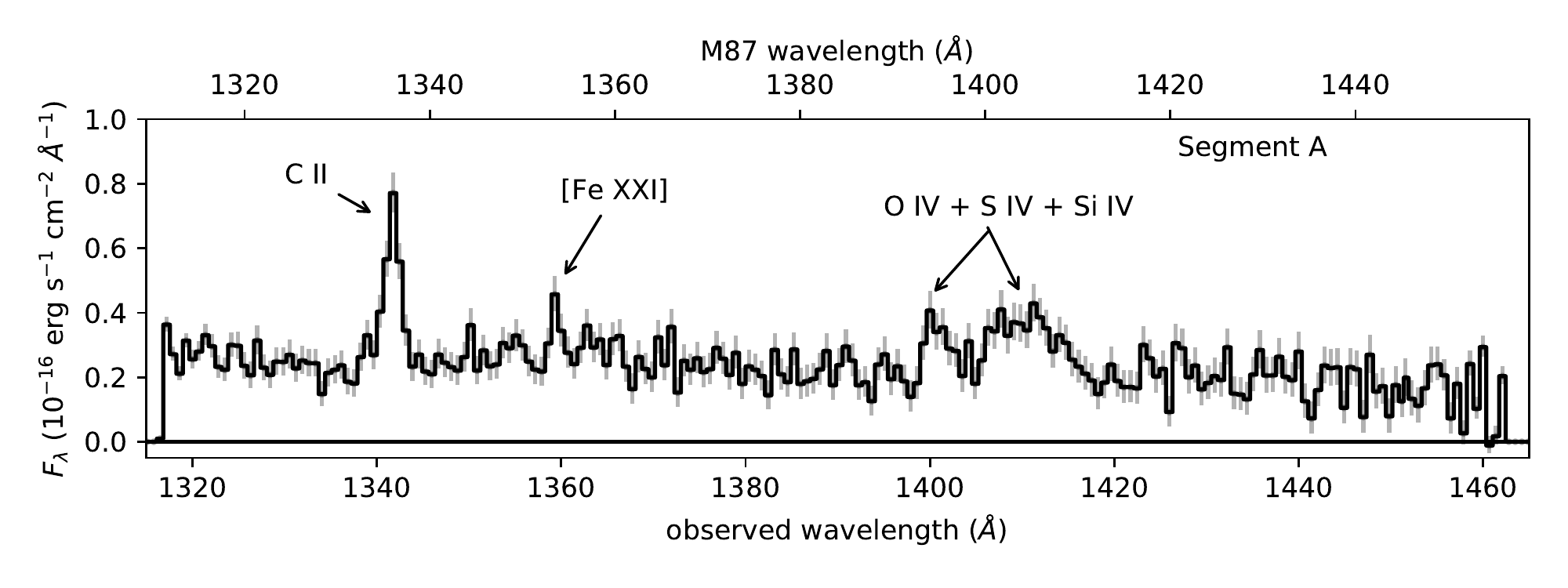}}
\end{center}
\caption{Our COS G130M spectrum of the M87 filament, divided into segments B (top) and A (bottom) for visual clarity. We have binned the spectrum in this plot into 0.70\AA{} bins, and display 1$\sigma$ errors on each bin as described in the text (sect. 2). Prominent emission lines are identified and discussed in sects. 3 and 4, and airglow lines are indicated with the shaded red regions and identified with red text.}
\end{figure*}

This spectrum also suffers from reddening caused by Galactic dust. In this paper we display and perform modeling on reddened spectra, but in sects. 5 and 6 we also cite de-reddened integrated fluxes. We compute the reddening correction using the \citet{Cardelli1989} extinction curves with $R_V$ = 3.05 and $E(B-V) \approx 0.02$, with the latter value based on a number of estimates for the extinction toward M87 but excluding the nucleus (e.g., \citealt{Dopita1997}, \citealt{Danforth2016}).  While some of the filaments in M87 are dusty, there is little evidence of dust in M87 around the location of this filament, with an upper limit to the intrinsic absorption of $A_V < 0.02$ \citep{Sparks1993}. 

\subsection{Discussion of aperture sizes}
The COS aperture has a circular footprint on the sky with a radius of 1.25", and this corresponds to a physical radius of 101 pc at the distance of M87. Our sightline is projected 1.9 kpc from the nucleus of M87, and the filament which passes through our sightline has a projected length of a little over 1 kpc (of which just 200 pc falls within our field of view). Thus, the luminosities we measure in our spectrum only correspond to a portion of the total line luminosity of the filament, but in general we cite the measured aperture fluxes and luminosities in this paper. 

There is also an issue of the absolute flux calibration for COS, which is highly accurate for point sources but can be more uncertain for extended emission. As we will discuss in the next subsection, the FUV emission is indeed somewhat extended along the cross-dispersion axis, which can affect the flux calibration. We checked in sect. 2 that changing the pipeline background subtraction algorithm does not affect the measured fluxes within the uncertainties, so this does not appear to be a major issue for our analysis, probably because the source is fairly faint already, so the vignetted wings of the extended emission in the cross-dispersion direction are not very significant. 

Additionally, we will also refer to archival Herschel-PACS data in this work. We use the level 2 rebinned data products, which have $9.4$"$\times9.4$" spaxels. These cover an area 18 times larger than the COS aperture (see Figure 1), so care must be taken in comparing our measurements from Herschel-PACS with those from HST-COS. In some cases we will also quote the measurements of \citet{Werner2013} for comparison, who also analyzed the Herschel observations but reprojected the data into 6"$\times$6" regions for their analysis. 

Another issue with the Herschel data is that the [\ion{C}{ii}]$\lambda$158$\mu$m emission which we detect with Herschel-PACS is not likely to be spatially uniform, since we associate it with the filaments, as we discuss in detail in sect. 8. This complicates the conversion of the measured flux within the $9.4$"$\times9.4$" or 6"$\times$6" box into the estimated flux within the COS aperture, so that the conversion is probably not accurate to better than a factor of a few. For the one instance in this work we attempt such a conversion (sect. 8), our conclusions are therefore uncertain by a similar amount. We note this caveat here and emphasize it in that section, but our results are robust against uncertainties at that level, so it it does not qualitatively affect our conclusions.

\subsection{Discussion of extended emission and interpretation of velocity dispersion}

We will also be measuring the broadening of emission lines in this paper, and in general we will be interpreting this broadening as the result of physical motions (i.e., velocity dispersion) within the FUV-emitting gas. There are other effects which can cause these emission lines to appear broadened, however, and we briefly address them here. 

One effect is thermal broadening. Using a typical expression for the line of sight thermal velocity dispersion $\sigma_r = \sqrt{2 k T / M}$, for the temperatures at which these species are at the peak of their emissivity, we expect thermal broadening for Ly$\alpha$, \ion{C}{ii}, \ion{C}{iv}, and \ion{N}{v} of 17 km s$^{-1}$, 8 km s$^{-1}$, 12 km s$^{-1}$, and 12 km s$^{-1}$ respectively. These are all negligible compared to the $\sigma_r \gapprox 120$ km s$^{-1}$ which we will measure for these lines. [\ion{Fe}{xxi}] is emitted at a much higher temperature, but Fe ions are also significantly heavier, so that the thermal broadening expected for [\ion{Fe}{xxi}] is 58 km s$^{-1}$. In this case the thermal velocity is not negligible, and we discuss the physical interpretation of this line's broadening more in sect. 7.3. 

The other effect results from the filament being extended on the sky. The COS aperture is circular instead of having a narrow slit, and so extended emission along the dispersion axis can artificially broaden emission lines. Fortunately, the relatively good spectral resolution of G130M makes this much less of an issue than G140L (see \citealt{Anderson2016}), but it is not necessarily insignificant. We first consider the worst case of uniform aperture-filling emission, and measure the profile of the Ly$\alpha$ airglow line in our spectra. We find $\sigma \approx 76$ km s$^{-1}$ for this line, and we can use this as an upper limit on the effect of artificial spectral broadening due to extended emission in this work.

We now estimate the degree to which the emission is extended along the dispersion axis in our observations.  The five G130M observations of this filament targeted identical sky coordinates and have position angles of -7$^{\circ}$, -10$^{\circ}$, -10$^{\circ}$, -10$^{\circ}$, and -25$^{\circ}$, so the filament subtends roughly the same portion of the COS field of view in each observation. These position angles place the filament roughly along the cross-dispersion axis, so we can expect the emission from the filament to be much more extended along the cross-dispersion axis than along the dispersion axis. 

We show in Appendix A that the FUV emission from the filament is indeed somewhat extended along the cross-dispersion axis, but this does not therefore imply that there is a significant effect from extended emission from the filament along the dispersion axis. We claim in sect. 8.1 that the diameter of the filament is about 0.24", for which if the emission is uniformly distributed across the filament would correspond to an artificial velocity dispersion of about $\sigma = 7$ km s$^{-1}$. This would be added in quadrature to the physical broadening from the filament itself to obtain the measured velocity dispersions $\sigma_r \gapprox  120$ km s$^{-1}$. Thus we expect that the effect of extended emission is negligible here. 

The one exception is [\ion{Fe}{xxi}], which we argue is not physically associated with the filament. In this case, it would be possible for the [\ion{Fe}{xxi}] emission to fill the aperture. However, since the measured  velocity dispersion for this line is rather uncertain ($\sigma_r = 69^{+79}_{-27}$ km s$^{-1}$), it is difficult to constrain the degree of artificial broadening for this line. If the line is aperture-filling, and we add this artificial broadening in quadrature to the  the thermal broadening of 58 km s$^{-1}$ estimated above, the total broadening would be 95 km s$^{-1}$, which can explain all of the observed broadening in the line, with no need for turbulence at all. We discuss this more in sect. 7.3.

\section{Observational results for FUV continuum and permitted lines}

\subsection{FUV continuum}

Now we turn to the reduced FUV spectrum and discuss the observational results. The first result is that there is a faint FUV continuum visible across the G130M bandpass, which is about 50 times fainter than the FUV continuum in the nucleus of M87. We attribute this continuum to the so-called ''FUV excess" in M87. The FUV excess in some early-type galaxies was first detected by the Orbiting Astronomical Observatory \citep{Code1969} and was measured in M87 by the International Ultraviolet Explorer  \citep{Bertola1980}. The observed COS G130M continuum is similar in shape to the IUE spectrum. The origin of the FUV excess is somewhat unclear, but FUV excesses are common in elliptical galaxies and seem to be tied to some late stage of stellar evolution (see reviews by \citealt{Oconnell1999} and \citealt{Brown1999} and more recent work by \citealt{Yi2011} and \citealt{Boissier2018}). Fitting a low-order polynomial to the observed FUV continuum, we estimate the observed FUV continuum luminosity (from 1160\AA{} to 1450\AA{}) in this aperture to be $2.4\times10^{38}$ erg s$^{-1}$. The dereddened value is $2.9\times10^{38}$ erg s$^{-1}$. This is consistent with the observed FUV continuum surface brightness measured by \citet{Ohl1998} in an overlapping waveband. We also remind the reader that there is no evidence for star formation in the M87 filaments (\citealt{Salome2008}, \citealt{Sparks2012}).

\subsection{C II}

\ion{C}{ii} $\lambda$1335 is a triplet for singly-ionized Carbon from the $^2$D to $^2$P$^0$ states, with one of the lines transitioning to the ground state. The lines are at 1334.53\AA, 1335.66\AA{}, and 1335.71\AA{}   in roughly a 1:0.2:2 ratio in emission in the laboratory \citep{Moore1970}. In absorption, the latter two lines are frequently not detected, since they are transitions from excited states while the former line is resonant. 

For our line fitting we employ Gaussian models with three free parameters: integrated flux $F_0$, velocity dispersion along the line of sight $\sigma_r$, and mean radial velocity $v_r$. For \ion{C}{ii} we specify three Gaussians, one for each component of the triplet. We give each component the same $\sigma_r$ and $v_r$ and freeze the fluxes to a 1:0.2:2 ratio, yielding a total of three free parameters for the fit. We generate a 3-dimensional grid of values for these model parameters, and compute the likelihood at each location in the grid for the data given the corresponding model. From the grid of likelihoods, we then compute medians and central 68\% credible intervals for each parameter, marginalizing over the others. We verify in Appendix B that it is safe to neglect the COS line spread function in our modeling, which is roughly a Gaussian with $\sigma \approx 8$ km s$^{-1}$ in the core and features non-Gaussian wings.

After fitting the \ion{C}{ii} emission from M87, we explore Galactic \ion{C}{ii} absorption, which is slightly displaced from the emission due to the redshift of M87. The 1334.5\AA{} line is resonant and produces very clear absorption (probably saturated) but the non-resonant lines are not robustly detected. We perform a double Gaussian fit to the absorption profile, using a single Gaussian for the saturated resonant line (the SN is not good enough to perform a more detailed fit) and allowing for the presence of a second absorption feature at the location of the strong 1335.7\AA{} non-resonant line. We fix the relative velocities of these two absorption features to the value from atomic physics (264 km s$^{-1}$) relative to the 1334.5\AA{} line, but allow them to have free normalizations and velocity dispersions (since the resonant line is saturated). There are therefore five free parameters in the absorption model. We perform a fit with these five parameters to the Galactic \ion{C}{ii} absorption profile in the same manner as above.

\begin{figure}
\begin{center}
{\includegraphics[width=8.5cm]{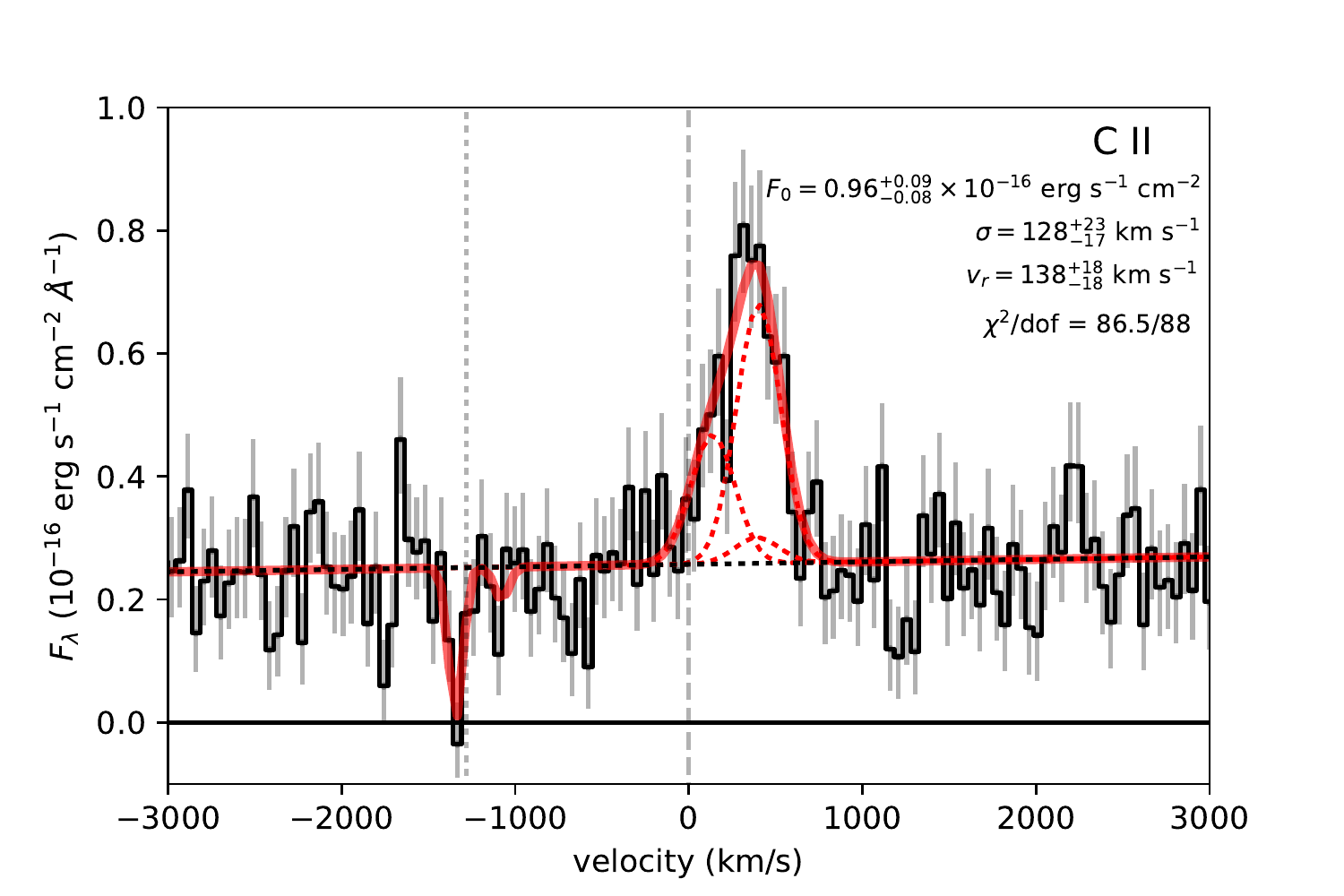}}
\end{center}
\caption{Portion of the spectrum around the \ion{C}{ii} $\lambda$1335 triplet, with velocities relative to the 1334.5\AA{} resonant line in the rest frame of M87. The three components of the triplet are indicated with dotted red lines, and a linear estimate of the local background is indicated with the black dotted line. The overall model is shown with the red solid line, which includes redshifted \ion{C}{ii} emission as well as two Gaussians for Galactic \ion{C}{ii} absorption. Saturated absorption near the LSR (indicated with the vertical dotted line) is visible, as well as a tentative sign of absorption from the non-resonant 1335.7\AA{} line. }
\end{figure}

The results can be seen in Figure 3. The simple Gaussian fit to the M87 \ion{C}{ii} emission profile is a good fit (using the $1\sigma$ errors from Figure 3 for a $\chi^2$ goodness-of-fit test, we find $\chi^2 = 86.5$ for 88 d.o.f.), and the 1:0.2:2 ratio for the three components of the \ion{C}{ii} triplet seems reasonable. The best-fit and $1\sigma$ uncertainties for the model parameters are: $F_0 = 0.96^{+0.09}_{-0.08} \times10^{-16}$ erg s$^{-1}$ cm$^{-2}$ (sum of all three components), $\sigma_r = 128^{+23}_{-17}$ km s$^{-1}$, and $v_r = 138 \pm 18$ km s$^{-1}$. 

For comparison, \citet{Werner2013} also observed this same field in [\ion{C}{ii}] $\lambda$158$\mu$m with Herschel-PACS, and for this region they found a velocity dispersion $\sigma_r \approx 130$ km s$^{-1}$ and $v_r \approx 140$ km s$^{-1}$ (no uncertainties given) within their 6"$\times$6" region. We analyzed the Herschel-PACS data ourselves using HIPE v. 15.0.1, and for the $9.4$"$\times9.4$" spaxel containing our COS sightline the line is visible above the detector noise but not the uncertainties are fairly large. We find $v_r = 108 \pm 24$ km s$^{-1}$ and $\sigma_r = 107 \pm 20$ km s$^{-1}$. This is reasonably consistent with both the \citet{Werner2013} results and with our FUV data, although it is clear that the FUV data have a higher SN and better precision. 

This match confirms the FUV \ion{C}{ii} and the FIR [\ion{C}{ii}] originate from the same system, and the Herschel-PACS imaging shows that this material is also approximately co-spatial with and has the same bulk velocity as the H$\alpha$ filaments measured by \citet{Sparks1993}, implying the \ion{C}{ii} is associated with the H$\alpha$-emitting filaments, although the Herschel-PACS resolution is too poor to localize [\ion{C}{ii}] to the exact location of the H$\alpha$ emission within the COS aperture.

Given the good results from our fitting above and the agreement in velocity dispersions between the FIR and FUV measurements, we expect that \ion{C}{ii} is not heavily affected by scattering or absorption in the resonant line, so that the velocity dispersion we measure can be fully attributed to motions in the filament.  In order to check this we fit an additional model which unties the width of the resonant line from the widths of the other two lines, in order to see if additional broadening in the resonant line is preferred by the data. A slightly broader line is indeed preferred, but the improvement is only $\Delta \chi^2 = 1.3$ for 1 dof, which is not significant. We conclude that there is no evidence for resonant scattering affecting our measurements of \ion{C}{ii}, and the measured velocity dispersions can best be attributed to intrinsic motions. 

We also compare our FUV line flux to the FIR flux in [\ion{C}{ii}] $\lambda$158$\mu$m. \citet{Werner2013} measure a line flux of $1.5\times10^{-15}$ erg s$^{-1}$ cm$^{-2}$ over the 6"$\times$6" region which covers our sightline, and we measure a line flux of $2.4\pm0.6\times10^{-15}$ erg s$^{-1}$ cm$^{-2}$ in the larger 9.4"$\times$9.4" spaxel. This scaling of flux with length of the box (instead of area) supports a picture where the  [\ion{C}{ii}] $\lambda$158$\mu$m emission is associated with a long filament, and in this case the expected flux within the $r=1.25$" COS aperture would be $6\times10^{-16}$ erg s$^{-1}$ cm$^{-2}$, or about a factor of five brighter than the FUV line flux. This corresponds to several thousand times as many 158$\mu$m photons as compared to 1335\AA{} photons.  We will discuss the implications of this result in much more detail in sect. 8.

For the absorption lines, we measure a velocity dispersion of $\sigma_r=41^{+35}_{-20}$ km s$^{-1}$ and a mean velocity of $v_r = -60^{+18}_{-13}$ km s$^{-1}$ relative to the LSR for the 1334.5\AA{} resonant line, although the line is saturated so the velocity dispersion is not physical. This corresponds to a lower limit on the Galactic C$^{+}$ column density $N_{C+} > 2\times10^{14}$ cm$^{-2}$, if the line were on the linear portion of the curve of growth. We weakly detect absorption at the expected location of the 1335.7\AA{} non-resonant line, but the significance is not very high ($\Delta \chi^2 = 1.5$ for two parameters) and the line parameters are not well-constrained by the data, so our model for this line is strongly affected by the implicitly uniform priors assumed for each of the line parameters.

\subsection{N V}

\ion{N}{v} is the $^2$P-$^2$S transition in Lithium-like Nitrogen, analogous to \ion{C}{iv} $\lambda$1549, and it produces a widely spaced doublet with a 2:1 flux ratio for lines at 1238.82\AA{} and 1242.81\AA{} respectively. We again employ a Gaussian model for \ion{N}{v} emission, with two components corresponding to the two lines in the doublet. Again we fix the relative separation between the lines and fix their velocity dispersions, and require the lines to have the fixed 2:1 flux ratio so that there are three free parameters in total. The spectrum and our fit are shown in Figure 4.

Our Gaussian model does not yield quite as good a fit for \ion{N}{v} as it does for \ion{C}{ii}, but it still gives an adequate fit ($\chi^2,$dof = 60.9,52). The best-fit and $1\sigma$ uncertainties for the model parameters are: $F_0 = 1.71^{+0.12}_{-0.10} \times10^{-16}$ erg s$^{-1}$ cm$^{-2}$ (sum of both components), $\sigma_r = 189^{+12}_{-11}$ km s$^{-1}$, and $v_r = 148^{+14}_{-16}$ km s$^{-1}$. The mean velocity approximately matches the velocity of \ion{C}{ii}, but the velocity dispersion is higher for \ion{N}{v} and the \ion{N}{v} profiles do not look precisely Gaussian to the eye, unlike the \ion{C}{ii} triplet. 

No Galactic \ion{N}{v} absorption is visible, and we can put an upper limit on the Galactic N$^{4+}$ column. Restricting fits to Gaussians with a velocity within 100 km s$^{-1}$ of the LSR and velocity dispersions $\sigma_r < 200$ km s$^{-1}$, we find a 2$\sigma$ upper limit of $N_{N^{4+}} < 2\times10^{14}$ for Galactic absorption. There are a few other features in this spectrum as well. The weak emission line at +2070 km s$^{-1}$ is probably \ion{C}{iii} $\lambda$1247.4 in M87. A Gaussian fit to this line yields $F_0 = 0.08\pm0.05 \times10^{-16}$ erg s$^{-1}$ cm$^{-2}$, $v_r = 40^{+100}_{-70}$ km s$^{-1}$, and $\sigma_r = 110\pm60$ km s$^{-1}$. We have not been able to identify the strong absorption line at -500 km s$^{-1}$, although it appears to be weakly statistically significant.

\begin{figure}
\begin{center}
{\includegraphics[width=8.5cm]{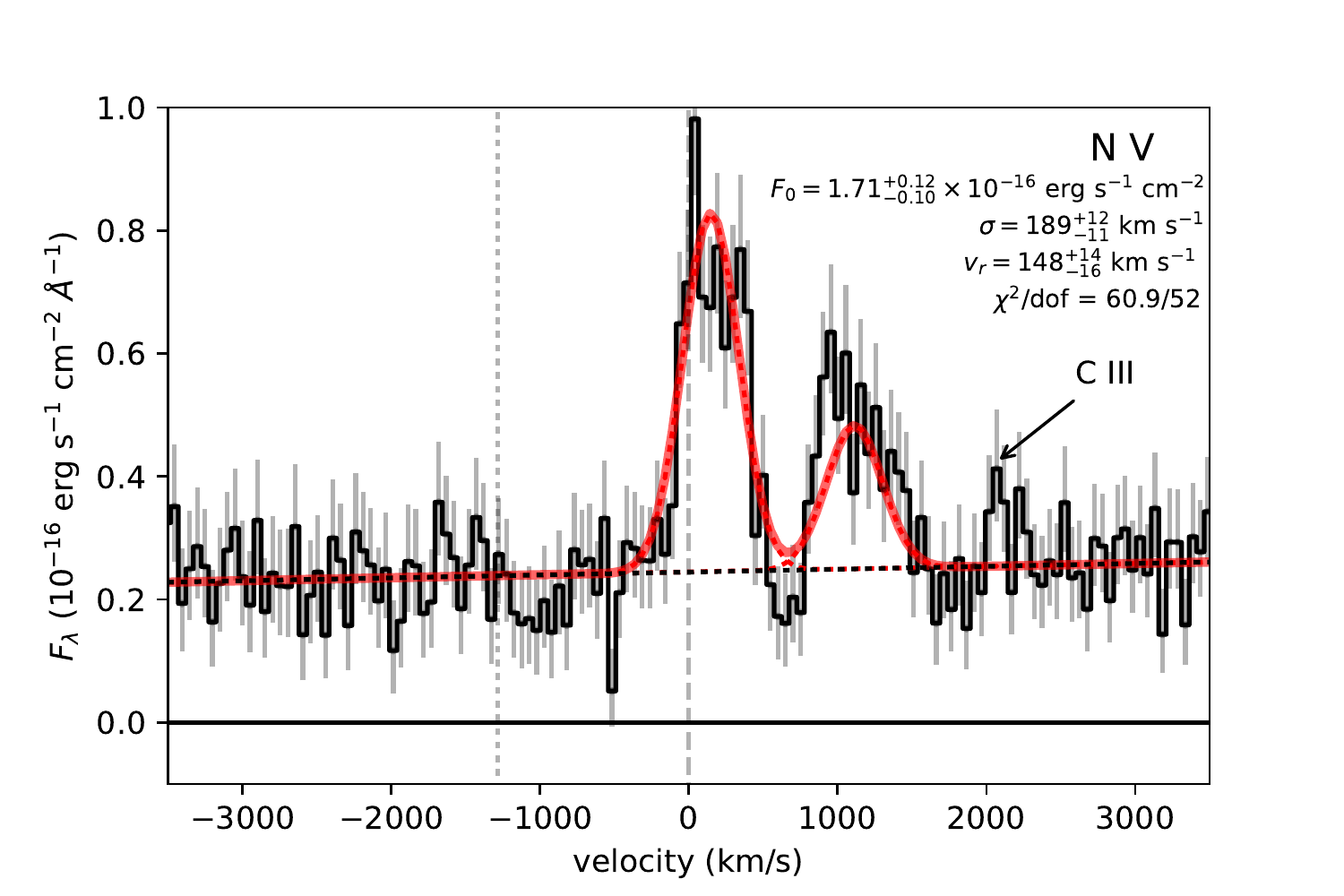}}
\end{center}
\caption{Portion of the spectrum around the \ion{N}{v} $\lambda$1239,1243 doublet, with velocities relative to the 1239\AA{} resonant line in the rest frame of M87. A linear estimate of the local background is indicated with the black dotted line, and the overall model is shown with the red solid line. We have also identified and labeled a faint C III $\lambda$1247.4 emission line in this spectrum. }
\end{figure}

\subsection{Ly$\alpha$}

The most prominent line in the spectrum is Ly$\alpha$, which we display in Figure 5. This line requires a more complex model, since it is optically thick and since it is affected by Galactic neutral Hydrogen absorption. We neglect detailed modeling of the optical depth of the line in favor of a phenomenological fit using two Gaussians, one with positive integrated flux for the bulk of the line and one with negative integrated flux  to produce the self-absorption profile at the top of the line. For the Galactic absorption, we include the damping wings of the Galactic Ly$\alpha$ profile in our model, multiplying both Gaussians by the Lorentzian wings of the Voigt profile. Following \citet{Danforth2016}, who studied the profile of Ly$\alpha$ from the nucleus of M87, we adopt a column density of $N_H = 1.4\times10^{20}$ cm$^{-2}$ for the Galactic absorption and place it at -1283 km s$^{-1}$, the velocity difference between the Milky Way and M87.

Our overall model therefore has six free parameters, corresponding to the integrated fluxes, velocity dispersions, and mean velocities of both the positive and negative Gaussians. Both Gaussians are multiplied by the the Lorentzian wings of the Voigt profile specified above, but the latter has no free parameters. The result is shown in Figure 5.

\begin{figure}
\begin{center}
{\includegraphics[width=8.5cm]{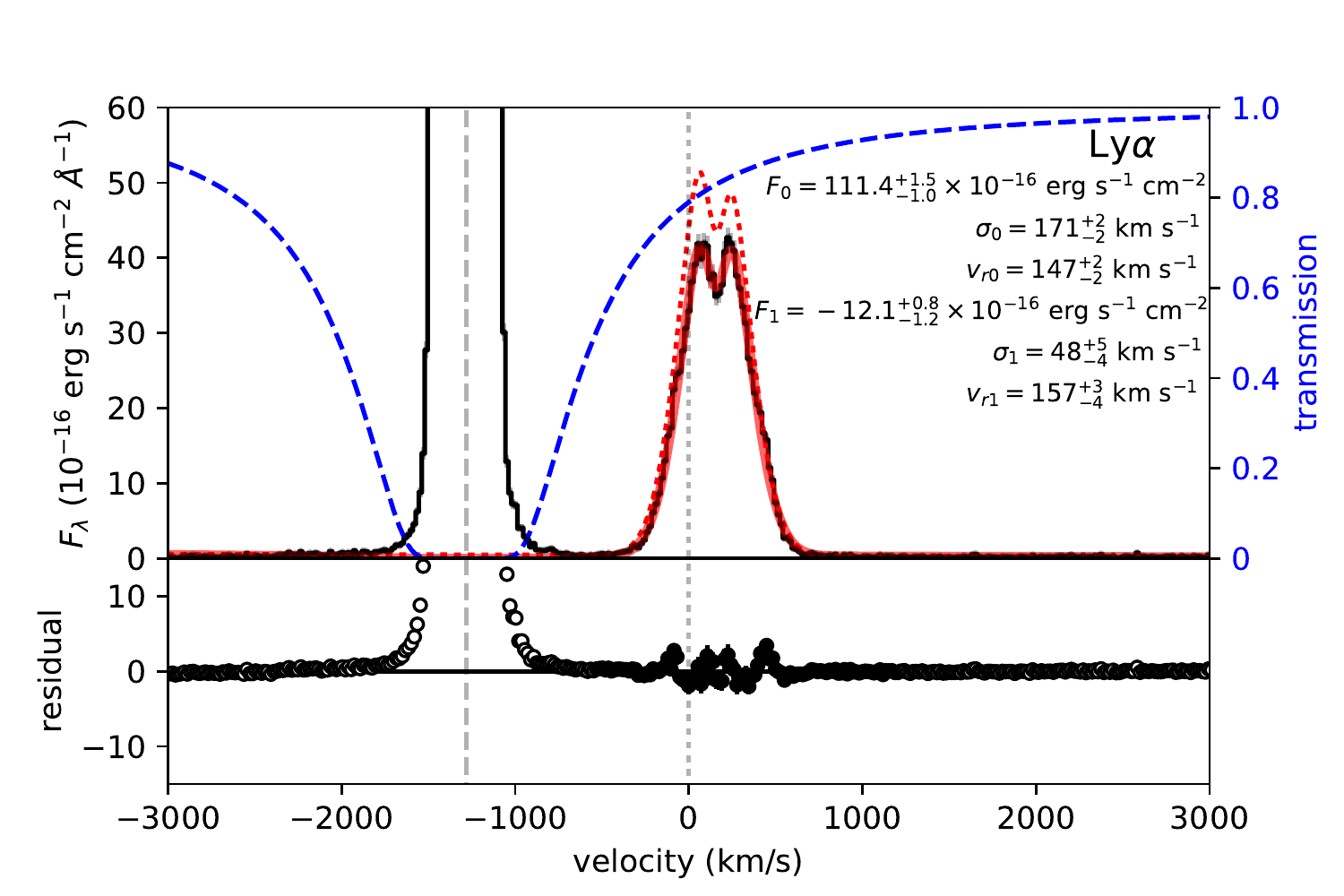}}
\end{center}
\caption{Portion of the spectrum around Ly$\alpha$. Geocoronal Ly$\alpha$ airglow contaminates the spectrum significantly at -1283 km s$^{-1}$ relative to M87 (indicated with the dashed vertical line), and Galactic Ly$\alpha$ absorption also affects the spectrum considerably. We model the transmission through the Galactic disk with the blue dashed line, assuming a neutral Hydrogen column toward M87 of $1.4\times10^{20}$ cm$^{-2}$ (Danforth et al. 2016). We fit the Ly$\alpha$ from M87 with two Gaussians - one positive containing most of the line flux and a small negative one which models the self-absorption - and multiply both Gaussians by the Galactic transmission profile. The red solid line is the overall model and the red dotted line shows the model without Galactic absorption. The lower panel shows the residuals between the model and the data. There are several small-scale features in the Ly$\alpha$ which are not captured with our simple model. In order to estimate uncertainties in model parameters, we include the rms of these residuals as a systematic uncertainty term in the error budget, which changes the $\chi^2$ from 182.8 to 39.1, while keeping the best-fit model parameters essentially unchanged. }
\end{figure}

The fit appears good by eye, although it is formally unacceptable ($\chi^2$ = 182.8 for 93 dof) due to small-scale features which are more easily visible in the residuals panel of Figure 5. Modeling these would require a much more extensive model for the filament, which is outside the scope of this paper. Instead, we compute the rms of these residuals (0.96 erg s$^{-1}$ cm$^{-2}$ \AA$^{-1}$) and incorporate it into our error budget as a systematic error. This changes the $\chi^2$ to 39.1 and increases the uncertainties on model parameters to more appropriate values, while keeping the best-fit values for these parameters essentially unchanged. Our final results for the two Gaussians are: $(F0, F1)$ = ($111.4^{+1.5}_{-1.0}$, $-12.1^{+0.8}_{-1.2}$)$\times10^{-16}$ erg s$^{-1}$ cm$^{-2}$,  ($\sigma_{r0}$,$\sigma_{r1}$) = ($171\pm2$, $48^{+5}_{-4}$) km s$^{-1}$, ($v_{r0}$,$v_{r1}$) = ($147\pm2$, $157^{+3}_{-4}$) km s$^{-1}$. Galactic absorption removes about 18\% of the flux, so that the unabsorbed line luminosity within the COS aperture is $3.3\times10^{38}$ erg s$^{-1}$ and the observed line luminosity is $2.7\times10^{38}$ erg s$^{-1}$. It is also interesting to note that Ly$\alpha$ has the same mean velocity as \ion{C}{ii} and \ion{N}{v}. This implies that all three lines (plus H$\alpha$) are being emitted from the same system.

 It is also notable that the profile is asymmetric after correcting for Galactic absorption; this is evidence that the Ly$\alpha$ photons within the optically thick filament are not being injected at the rest-frame frequency of Ly$\alpha$, in the frame of the filament. In particular, the blueward shift implies a blueshift in the filament frame. The characteristic equation for the double-peaked profile of optically thick Ly$\alpha$ is derived by \citet{Harrington1973}, \citet{Neufeld1990}, and \citet{Dijkstra2006} for optically thick slab and cloud geometries. It has three free parameters: temperature $T$, optical depth at line center $\tau$, and injection velocity $v_i$. For a temperature of $1.6\times10^4$ K and a line center optical depth of $10^4$ (based on the column densities inferred in sect. 6.4), we estimate that the asymmetry can be explained by an injection velocity of about 10 km s$^{-1}$. This is small compared to the velocity of random motions $\sigma_r \approx 130$ km s$^{-1}$ estimated for the filament from \ion{C}{ii}, H$\alpha$, and \ion{N}{v}, and we hypothesize that this injection velocity is due to an outflow of matter from the filament (see sect. 8.7). This injection velocity is also comparable to the offset between $v_{r0}$ and $v_{r1}$ in our modeling of the line.

We also consider the uncertainty in our adopted value $N_H$, which as we mentioned is fixed at the value measured by \citet{Danforth2016} for the nucleus of M87. Their sightline is projected about 20" from our sightline, which for a distance of $d$ kpc corresponds to a physical separation of $0.1d$ pc. Thus, since the absorption stems from neutral Hydrogen in the Galactic ISM $(d \lapprox 10)$, our value of $N_H$ could be different if the cold neutral medium varies on sub-pc scales, which is certainly possible \citep{Heiles2003}. We therefore perform a seven-parameter fit, allowing $N_H$ to vary, and estimate the allowable range for $N_H$. We find a 90\% credible interval spanning $1.0\times10^{20}$ cm$^{-2} - 2.0\times10^{20}$ cm$^{-2}$. 

Finally, we also considered modeling Ly$\alpha$ as the sum of two positive Gaussians, as \citet{Danforth2016} did for Ly$\alpha$ from the M87 nucleus. The number of degrees of freedom is the same as our fiducial model in Figure 5, but the $\chi^2$ is worse -- higher by 18.8 -- and the fit is qualitatively unable to capture the two peaks adequately. We therefore do not consider such a model further in this work.

\subsection{O V]}
 \ion{O}{v}] $\lambda$1218.34 is an intercombination line and corresponds to a higher effective temperature in CIE ($\sim2\times10^5$ K) than Ly$\alpha$. It is almost always blended with Ly$\alpha$ and usually neglected since it is much fainter. However, it can reach 10\% of the flux of Ly$\alpha$ in both CIE (\citealt{Doschek1976}, \citealt{Roussel1982}) and photoionization (\citealt{Kwan1981}, \citealt{Kwan1984}, \citealt{Netzer1987}) conditions, so it should not be dismissed out of hand when high-resolution spectroscopy is available. We searched for this line in our Ly$\alpha$ spectrum, but we see no evidence for it in our spectrum.
 
In a fit with Ly$\alpha$ as above, plus \ion{O}{v}] fixed at the same velocity as the positive Ly$\alpha$ component, with free normalization and free velocity dispersion (capped at 200 km s$^{-1}$), our 2$\sigma$ upper limit on the integrated \ion{O}{v}] flux is $2\times10^{-16}$ erg s$^{-1}$ cm$^{-2}$.

\subsection{Other lines}

\subsubsection{O IV, S IV, and Si IV}

Around 1400\AA{} there is a complex of blended emission lines from \ion{O}{iv}, \ion{S}{iv}, and \ion{Si}{iv}. The complex is clearly visible in Figure 2, but the SN is not high enough to separate these features from one another. For completeness, we fit two broad Gaussians to this complex in order to estimate the total flux from the these three species (which is $1.9\pm0.3\times10^{-16}$ erg s$^{-1}$ cm$^{-2}$), but we do not attempt to disentangle them into individual lines.

\ion{Si}{iv} has resonant absorption lines at $1393.8$\AA{} and $1402.8$\AA, and we detect this absorption from the Galaxy. The best-fit velocity is $v_{\text{LSR}} = -7^{+26}_{-14}$ km s$^{-1}$, the width of the line is $\sigma_r = 4\pm3$ km s$^{-1}$ (marginally consistent with the expected thermal velocity of 7 km s$^{-1}$, and the absorbing column is at least $N_{Si^{3+}} > 3\times10^{12}$ cm$^{-2}$. We again express the column as a 2$\sigma$ lower limit, although the absorption does not appear to be saturated, the SN of the continuum is not good enough to exclude this possibility definitively.

\subsubsection{C IV and He II}

\ion{C}{iv} $\lambda$1549 and \ion{He}{ii} $\lambda$1640 lie outside the range of G130M, but they are detected in the low-resolution COS G140L spectrum of the same sightline \citep{Sparks2012}. We measured fluxes for these lines in \citet{Anderson2016}, using data from propsal ids 12271 and 13439, and show the new fits in Figure 6 based on our new estimates of the error vectors. We use these new values in Table 1 in this paper as well. The G140L observations have the same coordinates as our G130M sightline to within the HST precision ($\sim 0.1"$) and similar position angles as well.

\begin{figure}
\begin{center}
{\includegraphics[width=8.5cm]{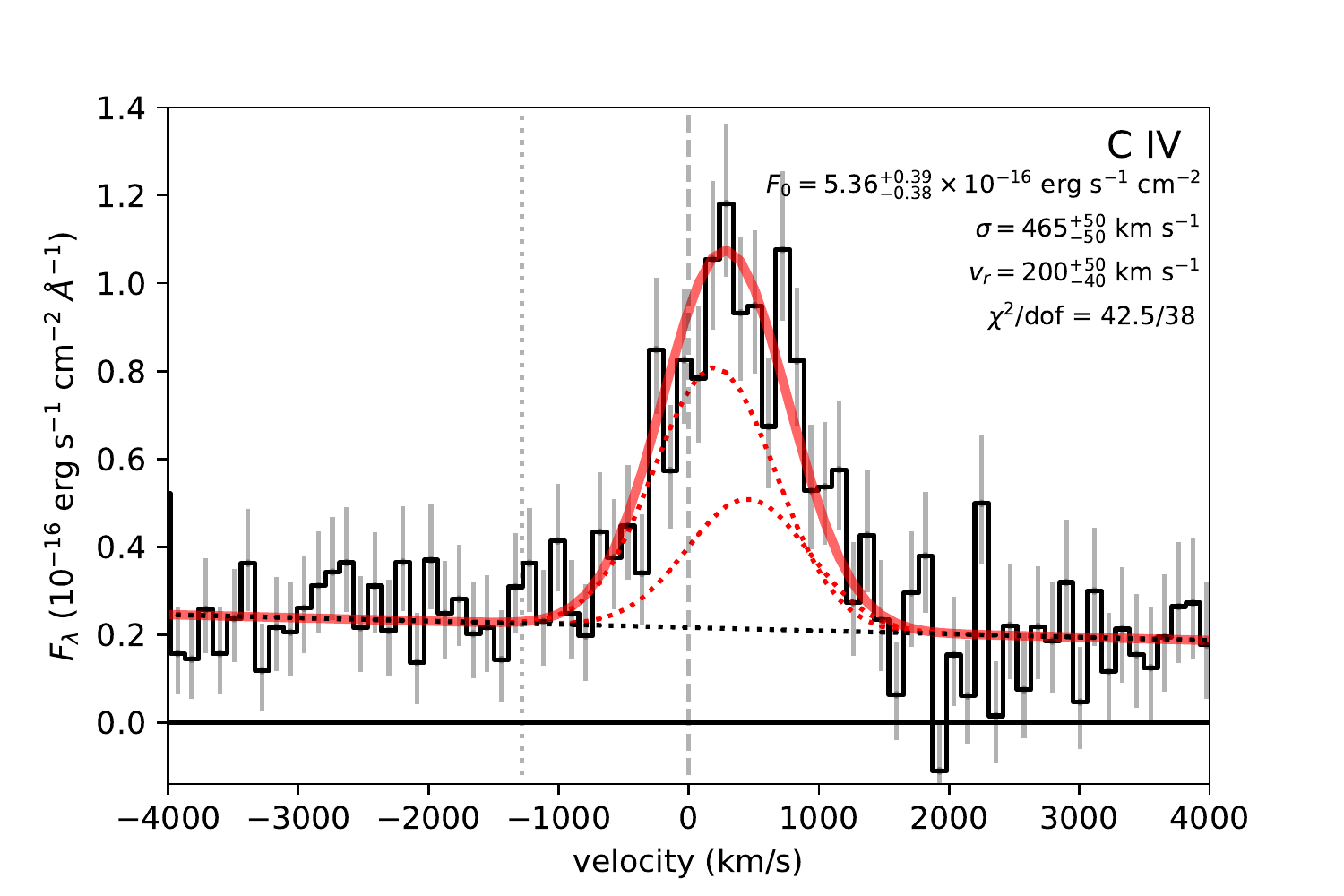}}
{\includegraphics[width=8.5cm]{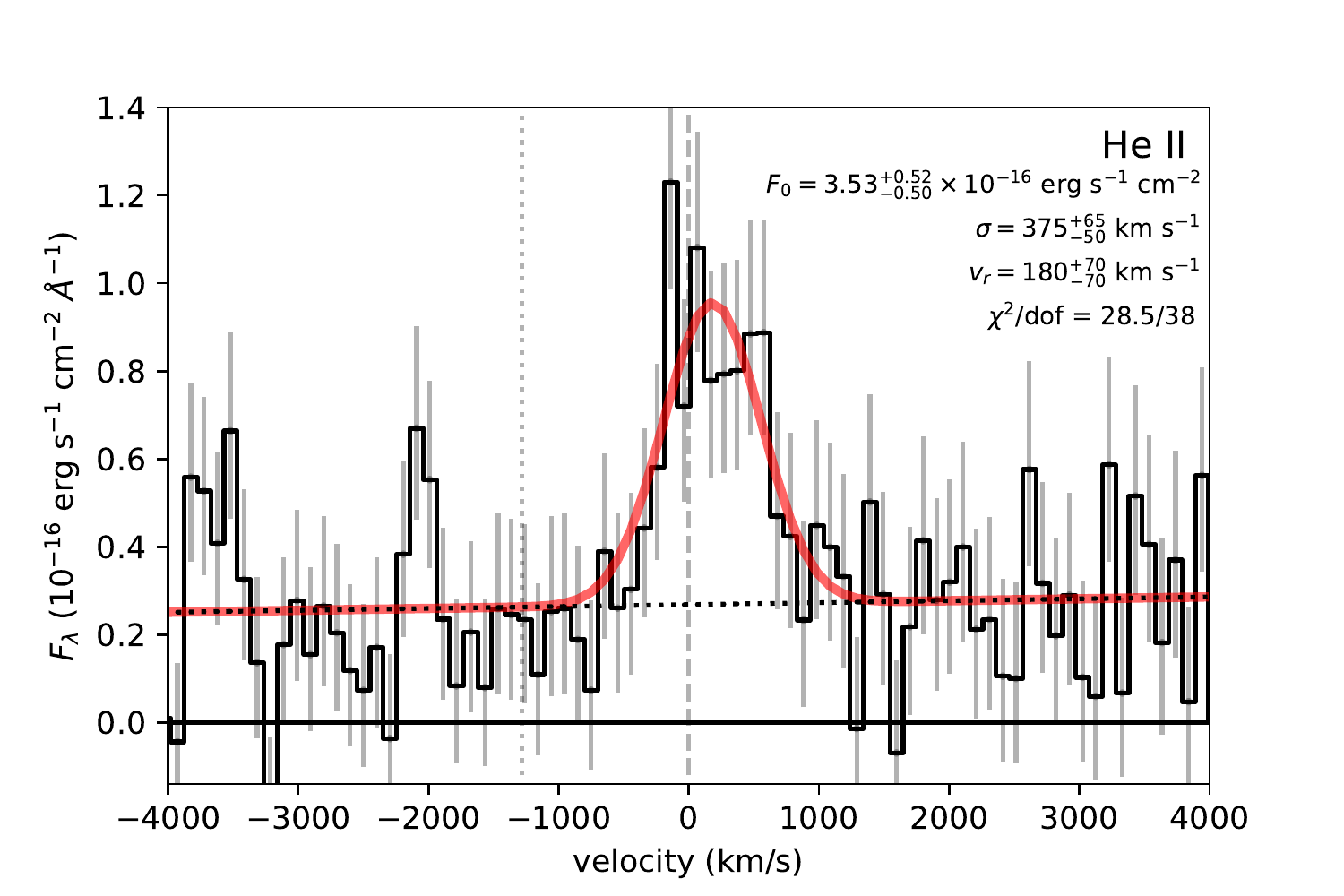}}
\end{center}
\caption{Portion of the archival low-resolution COS G140L spectrum around the \ion{C}{iv} $\lambda$1549 doublet (upper panel) and \ion{He}{ii} $\lambda$1640 multiplet (lower panel), with velocities in the upper panel relative to the 1548\AA{} strong line in the rest frame of M87. A linear estimate of the local background is indicated with the black dotted line, and the overall model is shown with the red solid line. These lines have similar $v_r$ to the other FUV lines in this paper, but larger $\sigma_r$ due to the extended nature of the emission (see sect. 8.1).}
\end{figure}

\section{Observations of [\texorpdfstring{F\MakeLowercase{e}}{Fe} XXI]}
In this section we consider one additional emission line -- the forbidden FUV line [\ion{Fe}{xxi}] $\lambda$1354.1. This is different from all the other FUV lines discussed in the previous section in that its characteristic temperature in CIE is $1.1\times10^7$ K, so it traces the hot (1 keV) intracluster medium instead of the intermediate-temperature gas in the filaments. We discussed this line in \citet{Anderson2016}, where we showed tentative ($\approx 2.2\sigma$) evidence of it in this same sightline, based on low-resolution COS G140L archival spectra. 

With the higher-resolution G130M data, we see stronger evidence for this line. Figure 7 shows the spectrum around [\ion{Fe}{xxi}] $\lambda$1354.1, and the line is clearly visible. Unlike the other lines we have discussed, however, this line is slightly blueshifted. Our best-fit model has $F_0 = 1.8^{+0.7}_{-0.5} \times10^{-17}$ erg s$^{-1}$ cm$^{-2}$, $\sigma_r = 69^{+79}_{-27}$ km s$^{-1}$, and $v_r = -92^{+34}_{-22}$ km s$^{-1}$.  As we discussed in \citet{Anderson2016}, geocoronal \ion{O}{i} has multiplets at 1304\AA{} and 1356\AA{}; the latter doublet falls within 1000 km s$^{-1}$ of redshifted [\ion{Fe}{xxi}] and with the low resolution and SN of the archival G140L data (which can have linewidths of more than 1000 km s$^{-1}$ for emission lines that fill the aperture) we pointed out the possibility of contamination by this doublet in the putative [\ion{Fe}{xxi}] feature. 

In the new G130M observations, geocoronal \ion{O}{i} $\lambda$1304 is several times weaker than in the G140L archival data, and no \ion{O}{i} $\lambda$1356 airglow is detected (with the superior spectral resolution of G130M, it would have been easily identifiable and separable if present). We also checked using the \verb"timefilter" routine that a G130M spectrum composed only of data taken during orbital night shows the same features as the full G130M spectrum, which further confirms that geocoronal \ion{O}{i} $\lambda$1356 is not important in the G130M data. We therefore attribute the higher flux, width, and blueshift of [\ion{Fe}{xxi}] in the archival G140L data to contamination from geocoronal \ion{O}{i} $\lambda$1356 in those data, and we favor the new values presented here, which have better spectral resolution and no geocoronal contamination. 

With our new data, the uncertainties are reduced enormously, and while the line is unfortunately at the faint end of the possible range of fluxes inferred from the G140L data, a blueshift persists in both observations. This demonstrates that the observed clump of $T \sim 10^7$ K plasma is moving in a direction with a substantially different radial velocity as compared to the filament in our sightline, although the difference in velocity is only significant at about 2.5$\sigma$. The overall significance of the G130M detection is 4-5$\sigma$, depending on the binning and on the value of the pseudo-random seed used during the \verb"RANDCORR" step in the COS calibration pipeline. It is also the most significant positive residual in the spectrum other than the lines which we fit in sect. 3, and this can also be seen visually in Figure 2. For comparison, the nearby positive residual at $+800$ km s$^{-1}$ is only significant at $2.7\sigma$ (or lower, depending again on the binning).

\begin{figure}
\begin{center}
{\includegraphics[width=8.5cm]{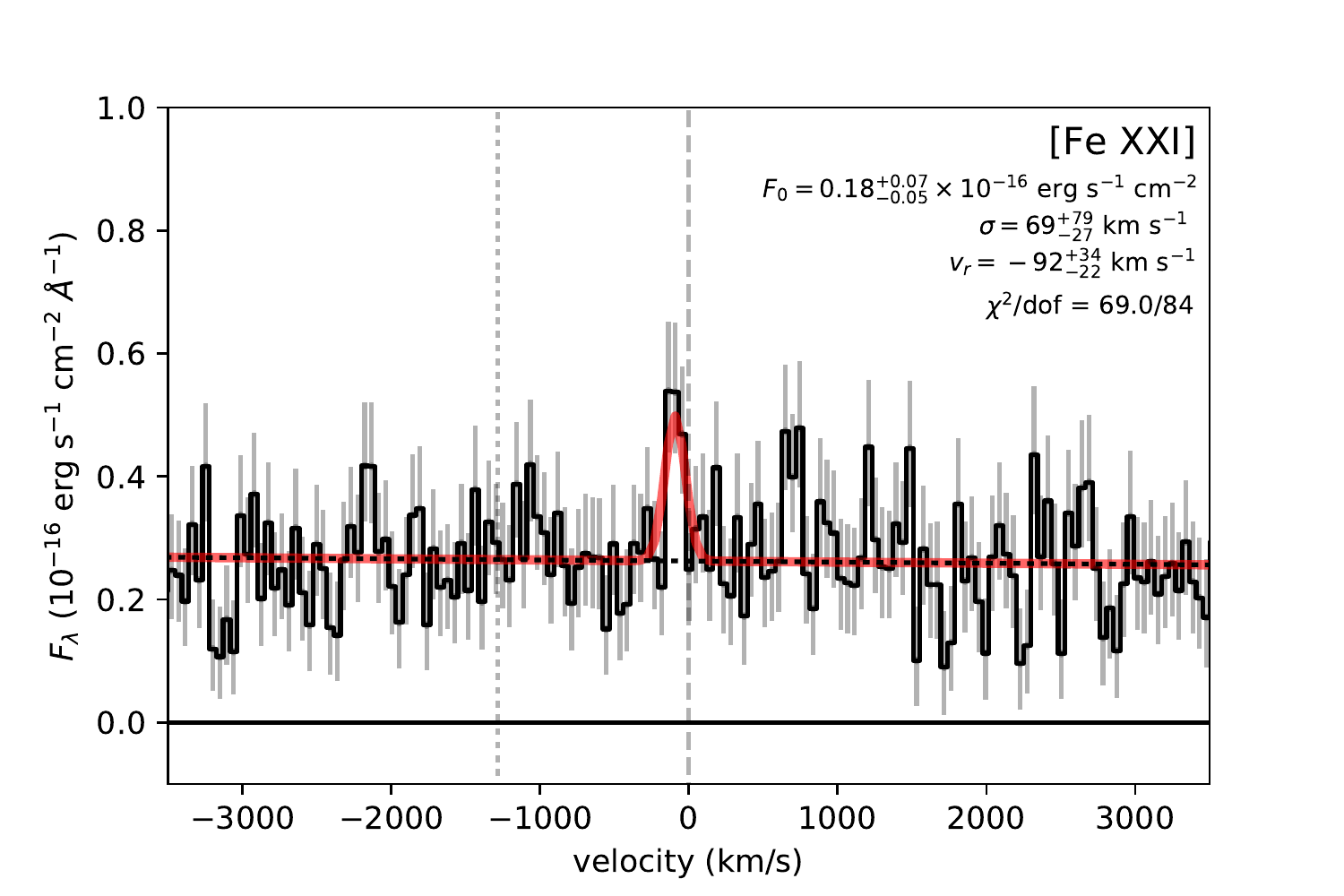}}
\end{center}
\caption{Portion of the spectrum around the [\ion{Fe}{xxi}] $\lambda$1354.1, with velocities relative to it in the rest frame of M87. A linear estimate of the local background is indicated with the black dotted line, and the overall model is shown with the red solid line. Our best-fit model for this line has $F_0 = 0.18^{+0.07}_{-0.05} \times10^{-16}$ erg s$^{-1}$ cm$^{-2}$, $\sigma_r = 69^{+79}_{-27}$ km s$^{-1}$, and $v_r = -92^{+34}_{-22}$ km s$^{-1}$, and the $2\sigma$ upper limit on the velocity dispersion is $\sigma_r < 229$ km s$^{-1}$. Unlike the other lines presented in this paper, [\ion{Fe}{xxi}] traces $T\sim10^7$ K plasma, and it appears to be slightly blueshifted while all the other FUV lines are redshifted. We discuss the implications of this blueshift, and potential geometries which can explain it, in sect. 7.1.} 
\end{figure}

\section{FUV line luminosities and cooling budget of M87}

In Table 1, we summarize the emission line modeling for our COS spectrum. We also compute the luminosities for each of the FUV lines, using the dereddened fluxes and assuming a distance of 16.7 Mpc. For comparison, the H$\alpha$ line luminosity within this aperture is $1.8-2.4\times10^{37}$ erg s$^{-1}$ (see sect. 6.2) and the [\ion{C}{ii}] $\lambda$158$\mu$m line luminosity within this aperture (assuming the emission is associated with the filament) is $2\times10^{37}$ erg s$^{-1}$ (see sect. 3.2). The 0.5-7.0 keV X-ray luminosity within this aperture is $3\times10^{38}$ erg s$^{-1}$ (see sect. 6.4). 

The [\ion{C}{ii}] $\lambda$158$\mu$m FIR line is therefore comparable in luminosity to the permitted FUV lines, and the strongest of the FUV lines -- Ly$\alpha$ -- is comparable in luminosity to the entire column of hot X-ray emitting ICM along the line of sight. 

We also consider the luminosities in some of these lines integrated over the whole nucleus of M87. \citet{Sparks1993} measure an H$\alpha$+[\ion{N}{ii}] line luminosity of $3\times10^{40}$ erg s$^{-1}$ for the entire M87 filamentary nebula, and for the H$\alpha$ : [\ion{N}{ii}] ratio discussed below (sect. 6.2) this corresponds to an H$\alpha$ luminosity of $9\times10^{39}$ erg s$^{-1}$. If the nebula has a spatially constant Ly$\alpha$ : H$\alpha$ ratio, as is suggested by their similarity in this filament (sect. 6.2) and in the central 100 pc of the nucleus (Anderson and Sunyaev, in preparation), then the Ly$\alpha$ luminosity over the entire filamentary nebula is $2\times10^{41}$ erg s$^{-1}$. Finally, from the Herschel-PACS data, the  [\ion{C}{ii}] $\lambda$158$\mu$m luminosity from the full 47" $\times$ 47" PACS field of view is $1.3\times10^{39}$ erg s$^{-1}$. It is difficult to know how much cold gas lies in the filaments closer to the nucleus of M87, but the total  [\ion{C}{ii}] $\lambda$158$\mu$m luminosity from M87 very likely exceeds $10^{40}$ erg s$^{-1}$. 
 
Based on the $\beta$-model fit of \citet{Churazov2008} and the profiles in \citet{Russell2015}, we estimate a broad-band (0.5-8.0 kev) X-ray luminosity of $\approx 1\times10^{42}$ erg s$^{-1}$ from the hot gas in the central 30" of M87. Thus, we predict that the Ly$\alpha$ emission from the filamentary nebula covering the nucleus of M87 is just a factor of $\approx 5$ less luminous than the volume-filling hot gas in the M87 interstellar medium.

Finally, we include for comparison the V-band luminosity of M87 in our sightline, taken from the surface brightness profile measured by \citet{Kormendy2009}. This optical emission is produced by the stars in M87 and completely dwarfs the ultraviolet luminosity, which is not surprising near the center of such a massive elliptical galaxy.

\begin{table*}
\begin{minipage}{180mm}
\caption{Fits to Emission Lines}
\begin{tabular}{llllll}
\hline
line & $F_0$ (reddened) & $F_0$ (dereddened) & $\sigma_r$ & $v_r$ & L (dereddened)  \\
 & ($10^{-16}$ erg s$^{-1}$ cm$^{-2}$) & ($10^{-16}$ erg $^{-1}$ cm$^{-2}$)& (km s$^{-1}$) & (km s$^{-1}$)  & ($10^{36}$ erg s$^{-1}$)\\
\hline 
{[}\ion{Fe}{xxi}] $\lambda$1354 & $0.18^{+0.07}_{-0.05}$ & $0.22^{+0.08}_{-0.06}$&$69^{+79}_{-27}$&$-92^{+34}_{-22}$ & $0.73^{+0.27}_{-0.20}$\\
Ly$\alpha$ & $99.3^{+1.5}_{-1.0}$& $120.9^{+1.8}_{-1.2}$ &$171\pm2$&$147\pm2$ & $403.5^{+6.0}_{-4.0}$ \\
\ion{O}{iv}+\ion{S}{iv}+\ion{Si}{iv} $\lambda$1400 & $1.9\pm0.3$ & $2.3\pm0.4$&-&- &$7.7\pm1.3$ \\
\ion{N}{v} $\lambda$1240 & $1.71^{+0.12}_{-0.10}$ & $2.02^{+0.14}_{-0.12}$& $189^{+12}_{-11}$&$148^{+14}_{-16}$ &  $6.74^{+0.47}_{-0.41}$\\
\ion{C}{ii} $\lambda$1335 & 0.96$^{+0.09}_{-0.08}$ & $1.17^{+0.11}_{-0.10}$ & $128^{+23}_{-17}$ & $138\pm18$ & $3.90^{+0.37}_{-0.33}$ \\
\hline
 \ion{C}{iv} $\lambda$1549 & $5.36^{+0.39}_{-0.38}$ & $6.20^{+0.45}_{-0.44}$ &$465\pm50$*& $200^{+50}_{-40}$ & $20.7\pm1.5$ \\
\ion{He}{ii} $\lambda$1640 & $3.53^{+0.52}_{-0.50}$ & $4.07^{+0.60}_{-0.58}$ &$375^{+65}_{-50}$*& $180\pm70$ & $13.6^{+2.0}_{-1.9}$ \\
\hline
H$\alpha$ & $6.0\pm0.9$ & $6.2\pm1.0$ & -&- & $21\pm3$ \\
\ion{He}{ii} $\lambda$304 & - & - & -&- & $360$ (predicted) \\
{[}\ion{C}{ii}]  $\lambda$158$\mu$m & $6\pm2$ & $6\pm2$ & -&- & $20\pm5$ \\
\hline
 X-ray continuum & & & & & 300\\
FUV continuum & & & & &290\\
V-band (stars) & & & $300\pm10$& $0\pm10$ & $4\times10^4$\\
\hline
\end{tabular}
\\
\small{Fits to emission lines in this paper. $F_0$ is the integrated flux (sum of all components for multiplet lines), and the second value is corrected for reddening assuming $E(B-V) = 0.02$ (see sect. 2). $\sigma_r$ is the velocity dispersion along the line of sight, and $v_r$ is the mean radial velocity of the line relative to M87. For convenience we also convert dereddened fluxes into luminosities assuming a distance of 16.7 Mpc. The top panel shows our fits to observed FUV lines. For Ly$\alpha$, both values of $F_0$ are corrected for Galactic absorption as well as discussed in sect. 3.4. The second panel shows new fits we have made for \ion{C}{iv} and \ion{He}{ii}, which are observed in the same sightline but based on COS G140L data as discussed in Anderson and Sunyaev (2016) since they fall outside the COS G130M bandpass (sect. 3.6.2). These two lines have artificially broadened measured velocity dispersions due to the extended emission covering multiple pixels of the low-resolution G140L spectrograph (sect. 8.1), so we mark these dispersions with asterisks. In the third panel we list estimates for selected lines in other portions of the electromagnetic spectrum. We obtain H$\alpha$ from HST-WFPC2 data (see sect. 6.2) and [\ion{C}{ii}] $\lambda$158$\mu$m from Herschel-PACS data (rescaling the latter to the HST-COS aperture, see sects. 2.1 and 3.2). \ion{He}{ii} $\lambda$304 is not observed, but we estimate its strength in sect. 6.3.2 and note that it may be up to six times larger if the effects of photoionization are included as well. The three lines in this panel have additional systematic uncertainties which are discussed in the text but not included in the measurement errors quoted in this table. Finally, the lower panel shows various continuum luminosities within our aperture for comparison. The X-ray luminosity is evaluated in the 0.5-7.0 keV band (sect. 6.4) and the FUV luminosity from 1160\AA{} to 1450\AA{} (sect. 3.1). We note that Ly$\alpha$ and \ion{He}{ii} $\lambda$304 are more luminous than these continua. The V-band luminosity in our aperture is calculated from the stellar surface brightness profile measured by \citet{Kormendy2009}, and kinematics for the stars are included on this line as well, as measured by \citet{Gebhardt2009} using absorption lines.} 
\end{minipage}
\end{table*}

In Table 2, we present measurements and/or upper limits for Galactic absorption obtained in the previous section. The lower limit on the C$^{+}$ column comes from the saturation of the line, and the upper limits on the N$^{3+}$ and O$^{3+}$ columns come from our non-detections of these lines in absorption at the rest-frame velocity of the Galaxy. The errorbars on the neutral Hydrogen column are based on the FUV absorption measurements (sect. 3.4). 

One interesting result which we can see from this table is that the ratio of C$^{+}$ to neutral Hydrogen in the Galactic ISM along this sightline is $> 1.4\times10^{-6}$. At solar abundance \citep{Grevesse1998}, the C:H ratio is $3.3\times10^{-4}$, so this implies that at least 0.4\% of the Carbon in the neutral ISM is ionized.

\begin{table}
\begin{minipage}{80mm}
\caption{HST-COS Constraints on Galactic Absorption Lines}
\begin{tabular}{ll}
\hline
line & $N_i$  \\
 & (cm$^{-2}$)\\
\hline 
\ion{C}{ii} & $>2\times10^{14}$ \\
\ion{N}{v} & $<2\times10^{14}$ \\
\ion{H}{i} & $1.4^{+0.6}_{-0.4} \times10^{20}$ \\
\ion{O}{iv} & $>3\times10^{12}$ \\
\hline
\end{tabular}
\\
\small{Upper limits, lower limits, and/or fits to Galactic absorption lines in this paper. Limits are quoted at $2\sigma$ and the errorbars for \ion{H}{i} in this table span the central 90\% credible interval, unlike the typical $1\sigma$ errorbars quoted in the rest of this paper.  }
\end{minipage}
\end{table}

\section{Analysis of FUV results, part 1: emission from the filament}

\subsection{Are the FUV lines and the FUV continuum connected?}

In this subsection, we show that the FUV lines are not produced by the stars which produce the FUV continuum, and that the FUV continuum is also not bright enough to produce the observed line emission either through photoexcitation or recombinations. 

For the first issue, as we review in \citet{Anderson2016}, the FUV excess continuum spectrum does not typically produce line emission, and indeed it would be very surprising to see FUV line emission from such an old stellar population. Moreover, in this paper we have showed that the typical line profiles of the FUV emission lines are narrow and slightly redshifted, whereas the stars in M87 near this filament have zero net velocity (\citealt{Emsellem2011}; \citealt{Krajnovic2011}) and velocity dispersion $\sigma_r \approx 300$ km s$^{-1}$ \citep{Gebhardt2009}. Thus the line emission is not associated with the stars in M87, but rather to the gaseous filament in our aperture.

For the second issue, we computed in sect. 3.1 the dereddened FUV continuum luminosity in our sightline, integrated from 1160\AA{} to 1450\AA{}, which is $2.9\times10^{38}$ erg s$^{-1}$. The dereddened Ly$\alpha$ luminosity from this sightline is already $3.0\times10^{38}$ erg s$^{-1}$, and the FUV excess spectrum gradually decreases toward shorter wavelengths, so it is clear this continuum does not provide the photons necessary to power these lines from the filament.

\subsection{Excitation mechanisms for Ly$\alpha$ and H$\alpha$}

The ratio of Ly$\alpha$ to H$\alpha$ is a useful diagnostic of the excitation conditions in a cloud of gas. For recombination radiation at $T_e \sim 10^4$ K, as would be expected from a photoionized plasma, this ratio is about 8.2. For collisional excitation of Hydrogen, the ratio can be 10 or 20 times higher (see tables 11.4 and 11.5 in \citealt{Osterbrock2006}).

We measured the Ly$\alpha$ luminosity (corrected for Galactic absorption) in sect. 3.4. Here we also measure the H$\alpha$ luminosity for this same sightline using archival data. We follow the procedure of \citet{Werner2013}, downloading from the archive an HST-WFPC2 image of this region using the F658N filter (proposal id 5122), which covers the redshifted H$\alpha$+[\ion{N}{ii}] lines from M87 as well as some optical continuum. We subtract the latter by scaling an F814W image of the same region (proposal id 5941) to match the narrowband image in regions without filament emission. The resulting H$\alpha$+[\ion{N}{ii}]  count rate within the 1.25" radius of our COS aperture is 0.34-0.46 ct s$^{-1}$. 

Next, we use the \verb"pysynphot" package v.0.9.8.7 \citep{Lim2015} in order to convert this count rate into a flux. As in \citet{Werner2013} we specify a ratio of 0.81:1:1.45 for [\ion{N}{ii}]$\lambda$6548 : H$\alpha$ : [\ion{N}{ii}]$\lambda$6583 (based on measurements of the southern filament by \citealt{Ford1979}) and we give each line a velocity dispersion $\sigma_r = 130$ km s$^{-1}$ at the redshift of M87. This yields an H$\alpha$+[\ion{N}{ii}] flux of $2.2-2.9\times10^{-15}$ erg s$^{-1}$ cm$^{-2}$ within our aperture, or an H$\alpha$ flux of $5.1-6.8\times10^{-16}$ erg s$^{-1}$ cm$^{-2}$. 

This is approximately consistent with the result from \citet{Sparks2009} that the ratio of \ion{C}{iv} to H$\alpha$+[\ion{N}{ii}] flux appears uniform across the filamentary nebula, with a value of 0.5. Using this ratio and the same H$\alpha$ : [\ion{N}{ii}] ratio as above we can convert our \ion{C}{iv} flux into an H$\alpha$ flux, which yields $7.2\times10^{-16}$ erg s$^{-1}$ cm$^{-2}$. 

Correcting our H$\alpha$ by a factor of 1.04 to account for extinction, the resulting Ly$\alpha$ : H$\alpha$ ratio is therefore around 17-23. This is clearly an intermediate value between collisional excitation and recombination. This is in contrast to observations of  A1795 and A2597 \citep{Odea2004} where the  Ly$\alpha$ : H$\alpha$ ratio is lower by a factor of $2-3$, pointing to recombination radiation in these cases. Unlike M87, however, the filaments in these clusters are thought to be star-forming, and the presence of young stars in the filaments provides an important additional source of ionizing radiation which is absent in M87. 

If there is any dust in this filament, then it will affect Ly$\alpha$ much more than H$\alpha$, and so the intrinsic Ly$\alpha$ : H$\alpha$ ratio will be pushed upwards, favoring collisional excitation more strongly. 

There is also the possibility of photoionization by EUV lines with $h \nu > 13.6$ eV, which can arise from the boundary layer itself, even in the absence of an external photoionizing field. We discuss this further in sect. 6.3.2. We also note that the intermediate Ly$\alpha$ : H$\alpha$ ratio rules out photoionization by the counter-jet as the dominant mechanism for FUV emission in the filament,  since collisional excitation is clearly important as well.

\subsection{Differential emission measure}
Finally, while we showed that recombinations are somewhat important for Ly$\alpha$ and H$\alpha$, we can still gain insight by assuming in this section that CIE holds for each of the species we consider in this paper. With this assumption, we can estimate the emission measure of the plasma producing each line. We do this in Figure 8, following a similar analysis as in \citet{Anderson2016}. In this figure, we plot the emission measure, defined here as 
\begin{equation}
\text{EM} \equiv \int n_e^2 dV = \frac{L}{\epsilon(T) \times Z}
\end{equation}

 \noindent where $L$ is the line luminosity for a given transition, $\epsilon = \epsilon(T)$ is the volume emissivity of the line-emitting species in CIE, which includes the variation of ionization fraction as a function of temperature, and $Z$ is the chemical abundance of the relevant element relative to Hydrogen. We assume solar abundances using the abundance tables of \citet{Grevesse1998}, and we obtain $\epsilon(T)$ from the CHIANTI database v. 8.0.2 (\citealt{Dere1997}, \citealt{Delzanna2015}). The resulting EM curves are shown in Figure 8 for Ly$\alpha$, \ion{C}{ii}, \ion{C}{iv}, \ion{N}{v}, \ion{He}{ii}, and [\ion{Fe}{xxi}]. In general, it can be expected that the EM of a species is close to the minimum value along its curve in Figure 8. It is natural that lines with a higher atomic number are radiated at higher temperatures (or equivalently, at lower densities due to pressure equilibrium), and occupy a higher part of the filament volume. We also remind the reader that under typical astrophysical conditions, plasma at $T \sim 10^5$ K is highly thermally unstable (e.g., \citealt{Mckee1977b}).

For comparison, we also plot in Figure 8 the emission measure of various phases of the hot ICM, as measured by \citet{Werner2013}, scaled to our distance and to solar abundance. They identify three components in the ICM along this sightline, at 0.5 keV, 1 keV, and 2 keV, and give emission measures for each. It is reassuring that the EM of 1 keV ($10^7$ K) plasma implied by our COS observation of [\ion{Fe}{xxi}] matches the \citet{Werner2013} X-ray measurement. 

\begin{figure*}
\begin{center}
{\includegraphics[width=18.5cm]{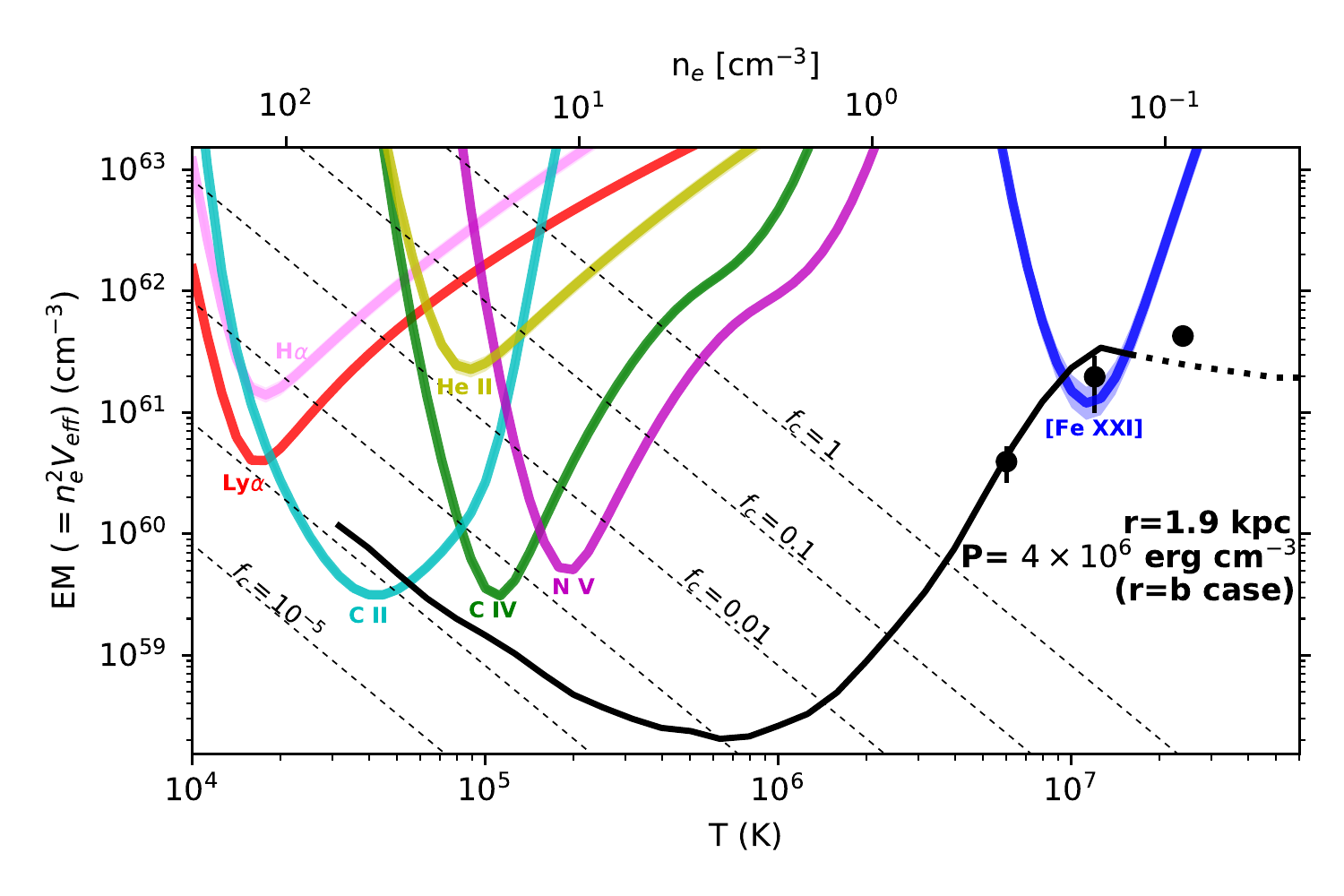}}
\end{center}
\caption{Emission measure curves for H$\alpha$ and each of the FUV lines we measure, assuming the line is produced under CIE and using equation 1. In general it can be expected that the EM of a species is close to the minimum value along its curve, and the $1\sigma$ uncertainties on line flux are propagated into this plot as the shaded regions around each curve (which are usually smaller than the width of the curve used in the plot). For comparison, we also plot with black points the emission measure of various phases of the hot ICM, as measured by Werner et al. (2013), and we note that the EM of $T\sim1\times10^7$ K plasma implied by our measurement of [\ion{Fe}{xxi}] matches the X-ray measurement. Next, we specify a pressure of $4\times10^6$ cm$^{-3}$ K \citep{Churazov2008} which assumes no projection effects, so that distance $r$ from the nucleus of M87 is equal to the observed impact parameter $b$. Our physical model is a cylindrical filament with a radius $r_c = 10$ pc (see sect. 8.1) filled with clumps that occupy a volume filling factor $f_c$ of the filament volume (see equation 2). Assuming the FUV lines are emitted from the filament, we plot lines of constant $f_c$. The FUV-emitting plasma must have a volume filling factor $10^{-5} \lapprox f_c \lapprox 10^{-2}$, with the exception of \ion{He}{ii} which we discuss in sect. 6.3.2. Finally, we overlay a normalized model of the EM for a solar flare \citep{Dere1979}, which offers a rough match to the observed FUV lines, with the exception of \ion{He}{ii} and H$\alpha$, which have contributions from photoionization as we discuss in the text.  }
\end{figure*}

Next, as in \citet{Anderson2016}, we estimate the corresponding electron density of the line-emitting plasma under the additional assumption of pressure equilibrium, specifying a pressure of $P = 1.91P_e = 4\times10^6$ cm$^{-3}$ K \citep{Churazov2008}. This is the pressure measured for the Virgo ICM at a deprojected radius of 1.9 kpc, so we refer to this case as the "r = b" case and note that the true pressures and densities are likely to be lower (see sect. 7.1). The implied densities at $10^4$ K agree to within a factor of two with the densities measured by \citet{Werner2013} for the nearby S filament based on observations of the [\ion{S}{ii}] $\lambda$6716,6731 doublet. 

\subsubsection{Solar flare DEM}

We compare these emission measures to observations of the differential emission measure (DEM) for an M2 X-ray class solar flare on 1973 August 9 \citep{Dere1979}, which are included in the CHIANTI package (\verb"flare_ext.dem"). At the highest temperatures (log $T \ge 7.2$ K) the values have been extended by the CHIANTI team, so we denote this portion of the DEM with a dotted line. solar flares are the most prominent source of [\ion{Fe}{xxi}] emission from the Sun and provide a unique environment with multiphase gas ranging from below $10^4$ K to above $10^7$ K, which is primarily collisionally excited. Under the hypothesis that our filament is also a collisional multiphase plasma, we adopt a solar flare DEM as a crude approximation for what we might expect in this sightline. We multiply the DEM (DEM = d$EM$/d$T$) by temperature and plot the result in Figure 8. We have multiplied the solar flare DEM by an arbitrary normalization factor to shift it in the vertical direction so that its magnitude matches the emission measures appropriate for the M87 filament instead of a solar flare.

\subsubsection{Anomalous \ion{He}{ii}}

The solar flare DEM is a reasonable match for every observed FUV line except \ion{He}{ii} and the optical H$\alpha$ line which we discussed in sect. 6.2. We pointed out in \citet{Anderson2016} that there is no obvious collisional solution for reproducing the observed \ion{He}{ii} : \ion{C}{iv} ratio in this filament spectrum. Moreover, \ion{He}{ii} presents the same difficulty in the solar spectrum. \citet{Jordan1975} first showed this for \ion{He}{ii} $\lambda$304 and for \ion{He}{i} $\lambda$584, which were observed to be 6 and 15 times brighter than the expectations from collisional ionization equilibrium (CIE) models. \ion{He}{ii} $\lambda$304 is the Ly$\alpha$ transition for He$^{+}$, and if Helium recombinations occur, the He$^{+}$ Balmer-$\alpha$ transition (which is \ion{He}{ii} $\lambda$1640) will be boosted significantly over the value from collisional excitation. This has been explored by many authors in the solar context (e.g., \citealt{Kohl1977}, \citealt{Raymond1979}, \citealt{Laming1992}, \citealt{Laming1993}, \citealt{Golding2017}) and there seems to be a consensus that a combination of collisional excitation, recombinations from photoionization, and repeated resonant scattering of \ion{He}{ii} EUV lines is required to explain their unexpectedly high intensity.

We postulate that similar conditions should be present in this filament, which explains the mixture of recombinations and collisional excitation determined in sect. 6.2 from the H$\alpha$ flux. H$\alpha$ and \ion{He}{ii} $\lambda$1640 are probably the lines most affected by these effects, so that recombinations would be less important for the other FUV lines. 

If this is correct, then \ion{He}{ii} $\lambda$304 from this filament should be extremely bright, and its line profile also exhibit similar resonant broadening as seen in Ly$\alpha$. To get a sense of the expected luminosity of \ion{He}{ii} $\lambda$304, we make predictions in CHIANTI using the solar flare DEM described above. For the pressure specified above, CHIANTI predicts a \ion{He}{ii} 304\AA{} flux which is nearly 90\% of the Ly$\alpha$ flux. We use this value in Table 1. However, if this is underpredicted by a factor of 6, as in the solar case, the true \ion{He}{ii} 304\AA{} luminosity could be as high as $2\times10^{39}$ erg s$^{-1}$ just along this sightline. This will have implications for the ionization fraction of the Virgo ICM (which should already be very highly ionized due to its temperature). A large luminosity of \ion{He}{ii} 304\AA{} photons will further reduce the neutral fraction at $T \sim 1\times10^7$ K below the CIE value of $7\times10^{-9}$ (\citealt{Dere1997}, \citealt{Delzanna2015}) since these photons are more effective at ionizing H atoms compared to X-ray photons. 

\subsection{Filling factor of FUV-emitting plasma in the filaments}
In Figures 8 and 13 we have drawn curves of constant filament volume filling factor $f_c$. We define this filling factor as 
\begin{equation}
f_c \equiv \frac{V_{\text{eff}}}{V_{\text{cyl}}} = \frac{  \text{EM} / n_e^2}{\pi r_c^2 l} = \text{EM} \times \frac{(1.91 k T_e / P)^2}{\pi r_c^2 l} 
\end{equation}

It corresponds to the fraction of the volume in the cylindrical filament which is producing the line emission, and does not depend directly on the fraction of the aperture which is filled by the filament. $V_{\text{eff}}$ is the effective volume of plasma which produces a given emission line and is obtained from the emission measure as $V_{\text{eff}} = \text{EM} / n_e^2$. $V_{\text{cyl}}$ is the volume of the cylindrical filament, which we take to be $\pi r_c^2 l$ with $r_c = 10$ pc (see sect. 8.1) and $l = 202$ pc is the diameter of our COS aperture. If the filament is inclined relative to the aperture at an angle $\theta$ then this will be an underestimate by a factor of sin($\theta$), and therefore the filling factor will be overestimated by a factor of sin($\theta$). 

In Figure 8, Ly$\alpha$, \ion{C}{ii}, \ion{C}{iv}, and \ion{N}{v} all have filling factors $0.01 > f_c  > 10^{-5}$. Since \ion{He}{ii} is so much weaker than \ion{C}{iv} in CIE, \ion{He}{ii} photons produced following photoionizations (see sect. 6.3.2) likely comprise a higher fraction of the total for this line than the other permitted lines, so that our purely collisional assumptions imply an anomalously high $f_c$ of 0.2. H$\alpha$ is also inconsistent with Ly$\alpha$, which proves that photoionization is also important for this line; see sect. 6.2. Ly$\alpha$ is also a resonant line, so for this line we should interpret $f_c$ in connection with the scattering volume, which may larger than the effective volume we use for other lines at similar temperatures. In Figure 13, which places the filament at a greater distance from the nucleus with a correspondingly lower pressure, the inferred filling factors for these lines are about nine times higher ($0.1 > f_c > 10^{-4}$). As we will discuss in sect. 6.5, these two cases roughly bracket the plausible values for $f_c$.

In either case, it is clear that the filling factor of the FUV-emitting plasma is very low; even within the filament, the FUV-emitting material only comprises a small fraction of the total volume. We hypothesize this plasma occurs in very thin conductive boundary layers between the hotter $T \sim10^7$ K plasma and a cooler phase which we discuss in sect. 8. Since the permitted FUV lines are such powerful coolants, maintaining constant energy flow through the boundary layer requires the intermediate-temperature plasma to occupy a very small volume (e.g., \citealt{Basko1973}for the case of illumination by X-rays of the companion in a close X-ray binary system, also see \citealt{McKee1977}), so this very low $f_c$ may indeed be physical. It is well-known that a similar rapid change of electron temperature occurs in the solar chromosphere between $T_e \sim 2\times10^3$ K and $T_e \sim 2\times10^6$ K (e.g., \citealt{Mariska1992} and references therein). Moreover, as \ion{N}{v} has the highest ionization potential among these permitted lines, it likely arises from the outer layers of the boundary, which may be the most turbulent and unstable, explaining the extra broadening observed in this line. 

[\ion{Fe}{xxi}], on the other hand, has an implied filling factor $f_c > 1$. This is evidence that this line, and the $T\sim10^7$ K plasma which produces it, are not physically associated with the filament, so that the above analysis does not apply for this line. Instead, as we describe in the next section, this plasma is more likely to be approximately volume-filling and associated with a cool phase of the M87 ISM (or alternatively the cool core of the Virgo ICM). Thus, we can use the observed velocity dispersion of this line to measure the turbulence in the Virgo ICM, which we do in the next section. The 0.5 keV and 2 keV components identified by \citet{Werner2013} and shown in Figure 8 also have $f_c > 1$, and can be interpreted in a similar way, though unfortunately we do not have any FUV lines in our observation which can be used to measure their kinematics.

Figure 9 shows a sketch of the implied geometry. Since our model is somewhat more complicated than a uniform-density cylinder, it is worth emphasizing the robustness of the major features of our model. The major assumptions are collisional excitation of the FUV lines, pressure equilibrium with the hot ICM, and an approximately cylindrical geometry for the filament which is based on H$\alpha$ imaging and other sources (sect. 8.1). For collisional excitation, we showed in sect. 6.2 that collisional excitation is important even for recombination lines like H$\alpha$, and it can be expected to be more important for the other FUV lines (also see sect. 6.3.2). For pressure equilibrium, underpressurizing the filament by a significant degree would destroy the filament within a few crossing times. The \citet{Werner2013} optical density diagnostics also imply that these filaments are approximately in pressure equilibrium, and \citet{Heckman1989} have showed similar results for other filamentary systems as well. 

Thus we conclude that the FUV emission from the permitted lines is genuinely localized within sub-pc scales, as implied by Figures 8 and 13. With these small emitting regions in mind, it is worth pointing out that the FUV lines still represent significantly more cooling than the X-ray emission. \citet{Werner2013} measure an average surface brightness in the 0.7-0.9 keV band of $1.2\times10^{-16}$ erg s$^{-1}$ cm$^{-2}$ arcsec$^{-2}$ from these filaments, corresponding to a flux of $5.9\times10^{-16}$ erg s$^{-1}$ cm$^{-2}$ from a field of view matching the COS aperture. If we assume the emission is dominated by the 2 keV component we can use an APEC model \citep{Smith2001} with solar abundance and assume a Galactic Hydrogen column of $1.9\times10^{20}$ cm$^{-2}$ \citep{Kalberla2005} to convert to the 0.5-7.0 keV band. This yields an unabsorbed X-ray flux of $7.8 \times10^{-15}$ erg s$^{-1}$ cm$^{-2}$, which corresponds to about $2/3$ of the luminosity in the Ly$\alpha$ line alone. Thus, while the effective path length of the 2 keV emission is close to 10 kpc, the sub-pc FUV-emitting regions are still more important for cooling, which attests to the strength and importance of these permitted FUV transitions in the much smaller field of view of COS.

\begin{figure*}
\begin{center}
{\includegraphics[width=17cm]{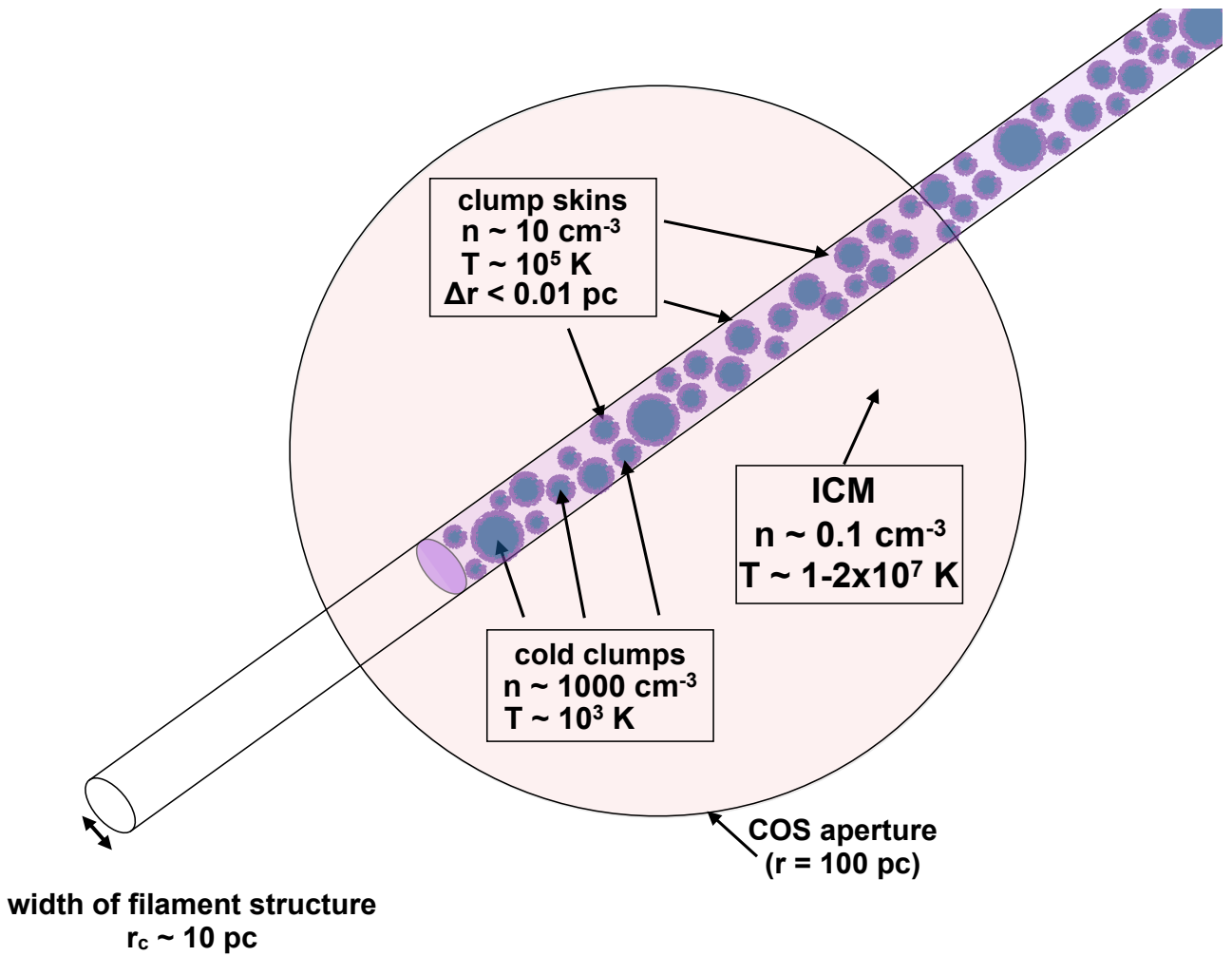}}
\end{center}
\caption{ Schematic of filament geometry. The large circle shows the HST-COS aperture, which is a circle with a radius of 1.25" (101 pc) at the distance of M87. The ambient ICM temperature is $T \approx 2 $ keV, but there is also a significant $T\approx1$ keV component within our field of view, which we argue is projected in front of the filament and is associated with the other side of the eastern radio lobe. In our model, the filament is a cylindrical structure with a radius $r_c \sim 10$ pc, containing cold ($T \sim 10^3$ K) clumps with a low volume filling fraction $f_c$. The clumps are surrounded by thin ''skins'' or ''shells'' which are multiphase and contain the intermediate-temperature plasma that produces the bright FUV emission lines studied in this paper.}
\end{figure*}

\subsection{Inferred physical conditions of FUV-emitting plasma in the filaments}

The small scales also imply short timescales in the filament and the boundary layer. The exact values for these timescales depend on the ambient pressure of the hot ICM (assuming the filament is in pressure equilibrium with the hot gas), and as we will show in sect. 7.1 this pressure is somewhat uncertain due to the uncertain three-dimensional position of the filament. We know that the filament lies at least $r=1.9$ kpc from the nucleus of M87, as this is the observed impact parameter, and we adopt a fiducial distance $r=7.3$ kpc as explained in sect. 7.1. Eq (3) in that section gives a parameterization of the pressure as a function of radius, which is valid up to about $r=10$ kpc. Using this equation, here we compute a number of derived parameters for the intermediate-temperature gas in the filament. 

First, the electron density at a given temperature is just $n_e = P / 1.91 T_e$. Using \ion{C}{iv} as an example, at its peak emissivity we expect $T_e = 1.1\times10^5$ K, so for our fiducial distance we have an electron density of 4 cm$^{-3}$ for the \ion{C}{iv}-emitting plasma. The adiabatic sound speed is given by $c = \sqrt{\gamma P/ \rho}$, so for the the \ion{C}{iv}-emitting plasma, the adiabatic sound speed is 51 km s$^{-1}$ (this parameter is pressure-independent). The characteristic path length is derived in \citet{Anderson2016}, and for the \ion{C}{iv}-emitting plasma at our fiducial distance, it is 0.006 pc. Combining this with the adiabatic sound speed we infer a dynamical time of 114 yr.

The cooling time can be estimated using the density and the cooling functions tabulated by \citet{Sutherland1993}. For the \ion{C}{iv}-emitting plasma at the fiducial distance, the cooling time is 185 yr, assuming solar metallicity. Finally, the mass in this $T \sim 10^5$ K phase within our COS field of view, as traced by the \ion{C}{iv} emission line, is $30 M_{\odot}$. 

Using equation (3), it is possible to infer these parameters for any other choice of pressure and temperature as well. For a different choice of pressure $P$ as compared to our fiducial choice $P_0$, most of these quantities are decreased by a factor of $(P/P_0)^2$, except the cooling time which is just decreased by a factor of $P/P_0$.

\section{Analysis of FUV results, part 2: [\ion{Fe}{xxi}] and the $T \sim 10^7$ K plasma}

\subsection{Blueshifted [\ion{Fe}{xxi}]}

In this subsection we address the measured blueshift for [\ion{Fe}{xxi}] relative to M87 and relative to the redshifted filaments. To visualize this more clearly, in Figure 10 we plot the Gaussian models we have fit to each FUV line in sects. 3 and 4. These Gaussians do not necessarily look like the line profiles in Figures 3-7 since most of these lines are multiplets, and here we are only plotting the underlying Gaussian model used for each component of the multiplet. It is clear that \ion{C}{ii}, \ion{N}{v}, and Ly$\alpha$ are approximately consistent with one another, while [\ion{Fe}{xxi}] has a clearly different bulk velocity and velocity dispersion. The COS G130M relative wavelength calibration is accurate to 15 km s$^{-1}$, and the \ion{C}{ii} velocity measured from the same segment agrees with similar precision with our independent Herschel measurements (sect. 3.2), so an instrumental error is unlikely to explain this difference. The flat-top on the model uncertainty region for [\ion{C}{ii}] in the N filament arises due to the large uncertainty in its bulk velocity. The wavelength of [\ion{Fe}{xxi}] is also well-known; we use a rest-frame wavelength of 1354.1 for the line, and this value seems accurate to about 1 km s$^{-1}$ based on solar measurements \citep{Young2015}.

\begin{figure*}
\begin{center}
{\includegraphics[width=17cm]{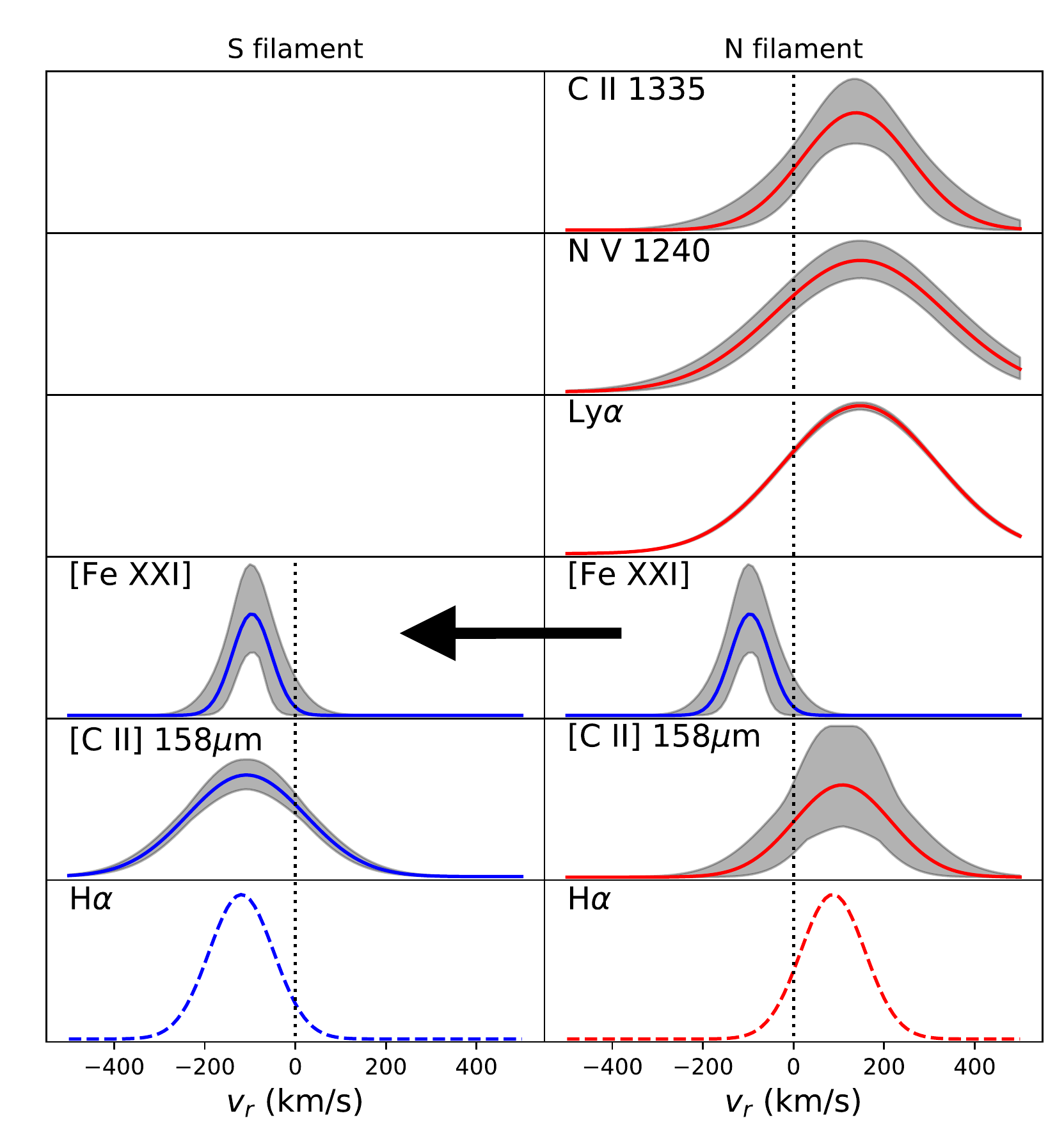}}
\end{center}
\caption{Model Gaussian profiles used in fitting each component for emission lines or multiplets toward the N and S filaments in M87, with velocity $v_r$ relative to M87. Lines fit in this paper are shown with solid lines (red for redshifted lines and blue for blueshifted lines), and data for H$\alpha$ come from \citet{Ford1979}. The shaded region denotes $1\sigma$ uncertainty in the model profile. All of the lines in the N filament have approximately the same profiles, except for [\ion{Fe}{xxi}], which approximately matches the line profiles observed for the nearby S filament. We hypothesize that the two filaments trace opposite sides of the radio lobe, and in our sightline the approaching side of the filament has not cooled much below $10^7$ K while the receding side has cooled down to $10^3$ K or below. We note that [\ion{Fe}{xxi}] is only observed in the N filament, but we show it in both columns to allow for easier comparison with the other lines.}  
\end{figure*}

The only other system where bulk velocities have been measured for both filaments and the ICM is the Perseus cluster \citep{Hitomi2016}, and in that case the X-ray calorimeter measurements of a velocity gradient with radius in the ICM are consistent with the observed velocity gradient of an H$\alpha$- and  CO-emitting filament in the same region (\citealt{Hitomi2016}, \citealt{Salome2011}). 

In both Virgo and Perseus, the filaments are observed near the interface between the ICM and radio lobes from the SMBH (\citealt{Hines1989}, \citealt{McNamara1996}). However, the geometries and orientation angles of these two filament systems are quite different from one another, and unlike M87, the filaments in Perseus host massive reservoirs of CO molecular gas and active star formation \citep{Salome2006}. 

In M87, our target filament is located near the edge of a radio bubble inflated by the counter-jet (\citealt{Hines1989}, \citealt{Sparks1992}), and if the counter-jet has a similar orientation angle as the jet, then it is misaligned from pointing directly away from Earth by just $10^{\circ}-25^{\circ}$ (\citealt{Biretta1999}, \citealt{Acciari2009}, \citealt{Wang2009}). The stellar velocity dispersion at the projected radius of the filament is $300$ km s$^{-1}$ and the profile is fairly flat with radius \citep{Gebhardt2009}, so it is unlikely the filament is in a stable circular orbit around M87.

We hypothesize, following \citet{Hines1989}, that the filament is associated with the radio lobe and is outflowing, which explains the observed redshift. Figure 11 shows another image of the region, along with our annotations. We use the H$\alpha$+[\ion{N}{ii}] image from Figure 1, with a mask applied to remove the noise in regions away from the filaments, as well as the F814W continuum image plotted using a logarithmic scale in order to show the extent of the stellar component. We also overplot 1.69 Ghz radio contours downloaded from the VLA image archive (project BC079), showing the extent of the inner radio lobes and emphasizing the well-known result that the H$\alpha$ filaments avoid the radio lobes and are often found at the edges of the lobes (e.g., \citealt{Hines1989}, \citealt{Sparks1993}, \citealt{Young2002}).

\begin{figure}
\begin{center}
{\includegraphics[width=8.5cm]{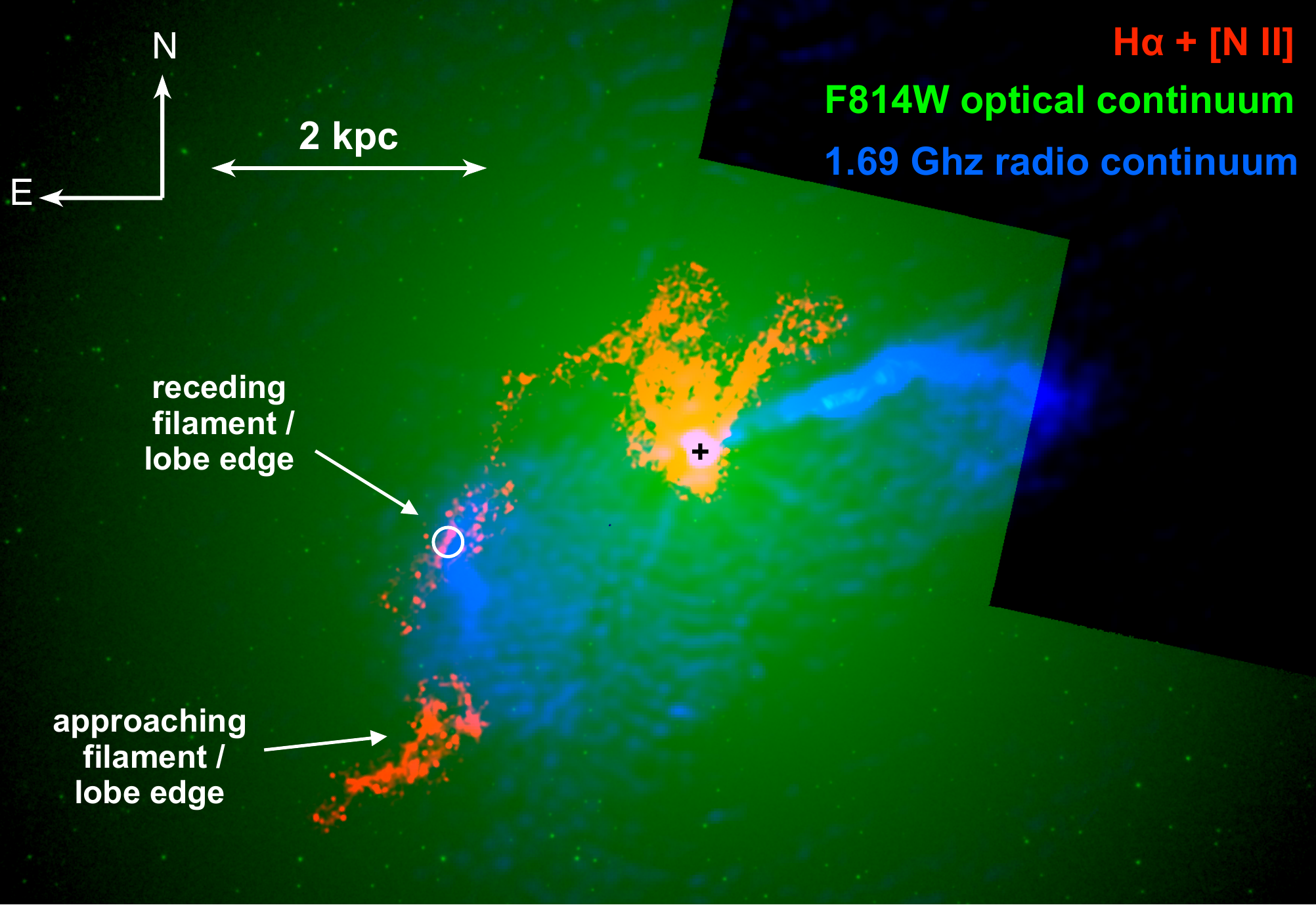}}
\end{center}
\caption{Multiband image of the center of M87, constructed from archival HST-WFPC2 (red and green) and VLA (blue) data as described in sect. 7.1. Note that the superposition of red and green can appear yellow in the central nebula if this figure is printed. The presence of the filaments at the outer projected edge of the E radio lobe is obvious, and based on H$\alpha$ and [\ion{C}{ii}]$\lambda$158$\mu$m measurements, plus our own FUV measurements, the northern filament is receding relative to M87 at $v_r \approx 140$ km s$^{-1}$, while H$\alpha$ and [\ion{C}{ii}]$\lambda$158$\mu$m show that the southern filament is moving toward the Earth relative to the motion of M87 by about 100 km s$^{-1}$. We hypothesize that these filaments represent the two faces of the expanding radio bubble. A natural explanation for the blueshift of [\ion{Fe}{xxi}] in our sightline (indicated by the white circle) is that the filamentary emission arises from the receding edge of the lobe while the [\ion{Fe}{xxi}] emission arises from the approaching edge of the lobe, which is at a different temperature $T \sim 10^7$ K.}  
\end{figure}

\subsection{What is the true distance between the northern filament and the M87 nucleus?}

The impact parameter $b$ between the N filament and the nucleus of M87 is 1.9 kpc, so this sets a lower limit on the true distance $r$ between the filament and the nucleus. This is the case we consider in Figure 8, which we will call the "$r=b$" case.  

For the geometry we discussed in sect. 7.1, where we instead place the filament along the propagation axis of the counter-jet, we can estimate the true distance between the filament and M87. Figure 12 shows a schematic of this proposed model, which we will call the fiducial case. The northern and southern filaments are located on the edges of the expanding radio bubble, but due to its expansion the southern filament is blueshifted relative to Earth while the northern filament is redshifted.  In the simplest case where the northern filament is located at the intersection of the counter-jet and the radio bubble (i.e., $\theta_{\text{fil}} = \theta_{\text{jet}}$), we can estimate its distance $r$ from the nucleus to be 1.9 kpc / sin $15^{\circ} \approx 7.3$ kpc.

\begin{figure}
\begin{center}
{\includegraphics[width=8.5cm]{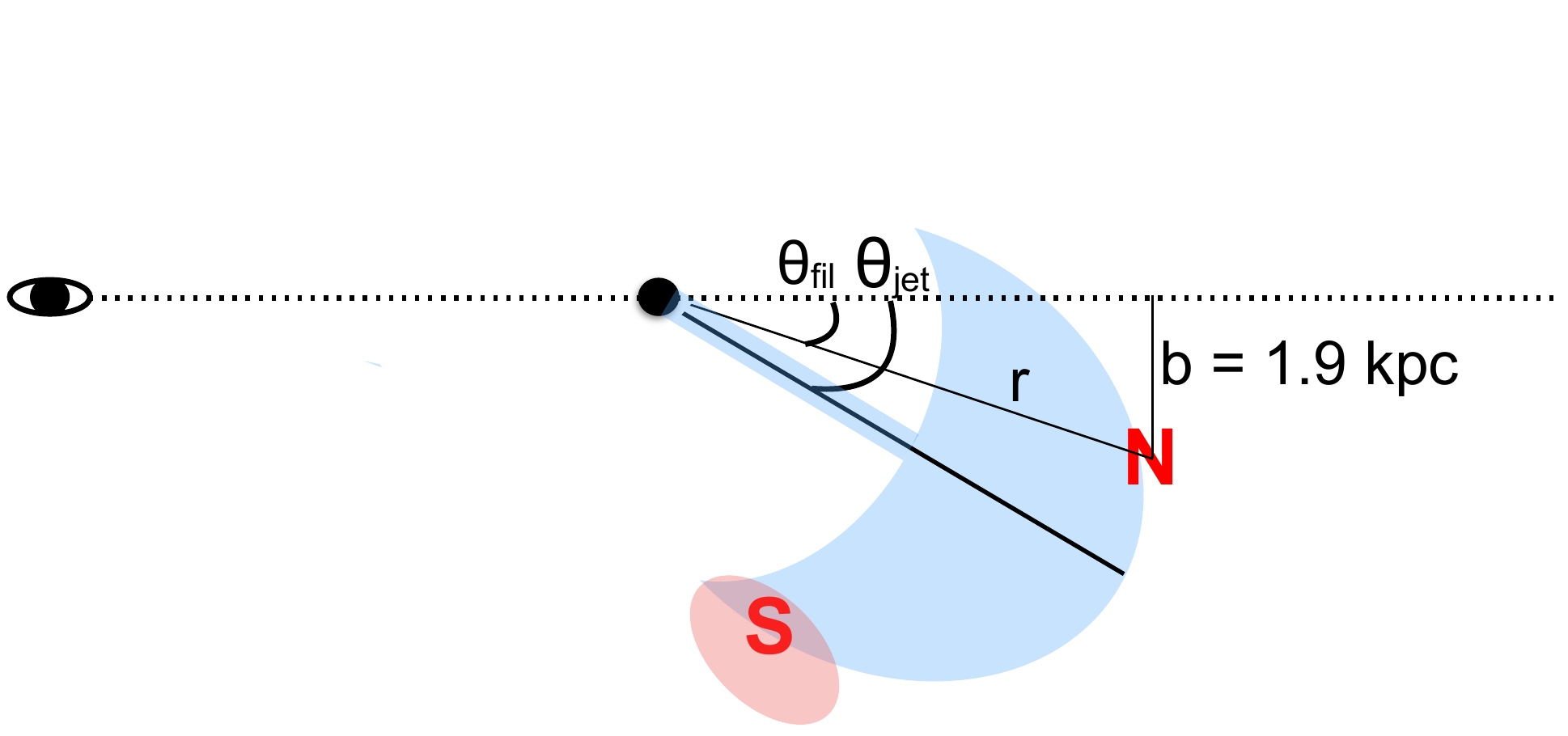}}
\end{center}
\caption{Schematic of our "fiducial" model for the locations of the northern and southern filaments to the east of the M87 nucleus. The two filaments are indicated with red letters, and we place them on the edge of the radio bubble (see Figure 11). We also place a red oval near the southern filament which contains the 1 keV plasma detected in our sightline. Due to the expansion of the radio bubble, the northern filament appears redshifted relative to Earth while the southern filament appears blueshifted. While the impact parameter $b$ between the northern filament and the M87 nucleus (indicated with the black dot) is just 1.9 kpc, in our fiducial model the true distance $r$ between the filament and the nucleus is much higher: $r = b / $sin $\theta_{\text{fil}}$. In our fiducial model we take $\theta_{\text{fil}} = \theta_{\text{jet}}$ = 15$^{\circ}$, so that $r \approx 7.3$ kpc. The pressure in the filament depends on the radius $r$, as we discuss in sect. 7.3 and in equation 3.} 
\end{figure}

We discuss the implications of these two cases on the pressure, filling factor, and velocity of the filament next. First, we parameterize the pressure as a function of radius using Figure 5 of \citet{Churazov2008}. We obtain a reasonable fit for 1 kpc $\le r \le$ 10 kpc using the third-order polynomial 

\begin{equation}
P(r) \approx -4667r^3 + (1.495\times10^5)r^2 - (1.573\times10^6)r + 6.159\times10^6
\end{equation}
\noindent for $r$ in kpc and $P$ in erg cm$^{-3}$. Thus, the the deprojected pressure in the hot phase decreases by roughly a factor of 4.5 between $r=1.9$ kpc and $r=7.3$ kpc. In Figure 13 we therefore recompute the densities and volume filling factors $f_c$ from Figure 8, now assuming a pressure of $8\times10^5$ erg cm$^{-3}$. This corresponds to the geometry considered in this section, with the impact parameter at $b=1.9$ kpc but the true distance from the nucleus at $r=7.3$ kpc.  This geometry is plausible for the FUV lines corresponding to the filament. 

However, as we showed in \citet{Anderson2016}, the effective path length goes as $P^{-2}$, so decreasing the pressure by a factor of 4.5 increases the effective path length by a factor of 20. For the [\ion{Fe}{xxi}]-emitting plasma, the effective path length increases from about 0.5 kpc (Figure 8) to 10 kpc (Figure 13). We think the former value is much more likely, given the impact parameter of 1.9 kpc and the preference of 1 keV plasma to lie in and around the filaments. This leads us to the idea that the $T\sim10^7$ K plasma lies closer to the nucleus, while the filament is projected behind this plasma and lies at a larger radius from the nucleus of M87.

The 0.5 keV component of the hot phase has an effective path length about 20 times smaller than the 1 keV component. It is still larger than the estimated size of the filament (see sect. 8.1), but in the $r=b$ case the difference is only a factor of two or so. It is therefore possible to interpret the 0.5 keV component as a high-temperature extension of the filaments, and the spatial association of the 0.5 keV components with the filaments in \citet{Werner2013} supports this hypothesis. The 1 keV component, however, has a much higher emission measure and extends spatially far beyond the filaments, so it is possible to imagine 1 keV plasma existing on both sides of the radio bubble, with the approaching face dominating the signal, as we propose in our fiducial model.

\begin{figure*}
\begin{center}
{\includegraphics[width=18.5cm]{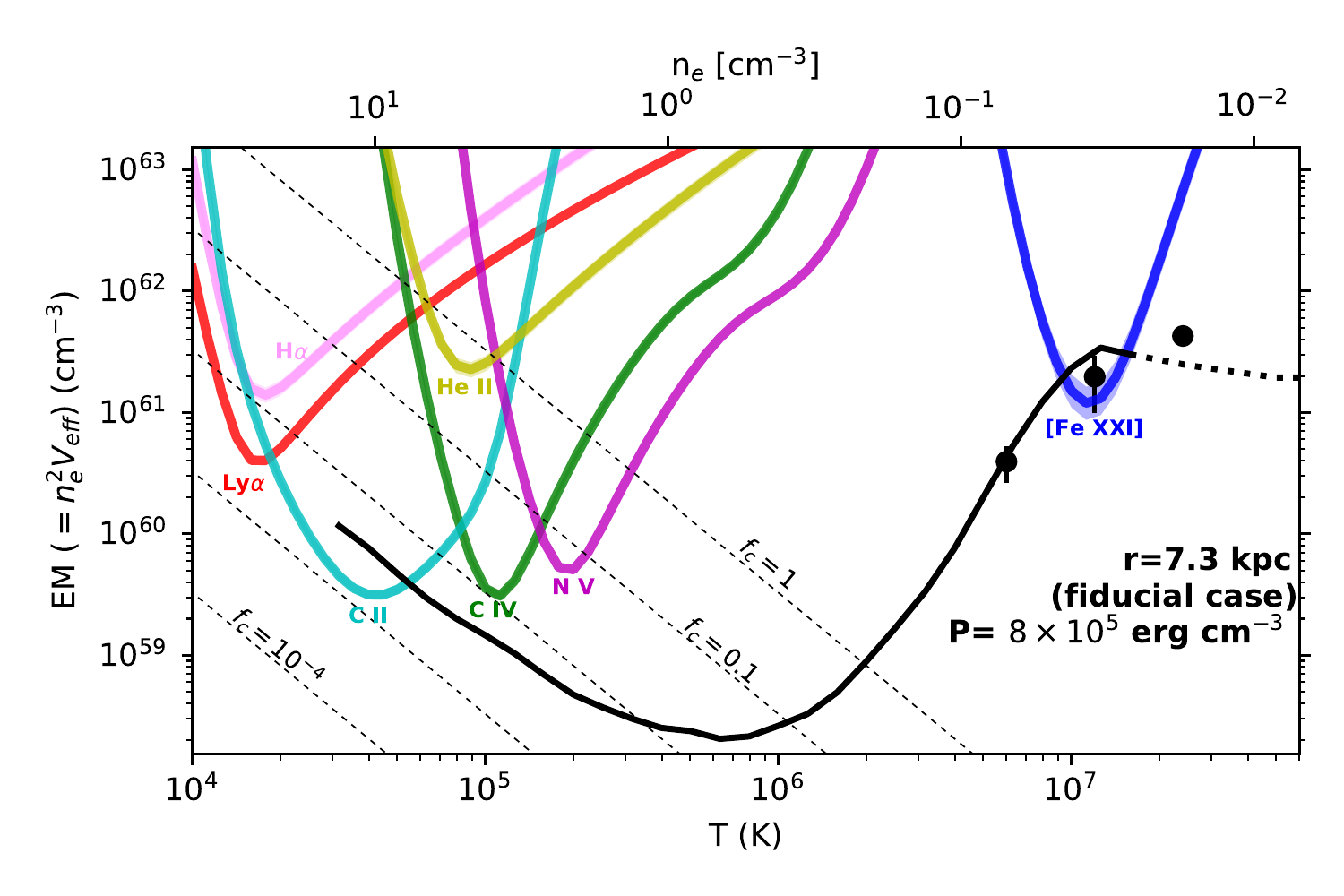}}
\end{center}
\caption{Emission measure curves for H$\alpha$ and each of the FUV lines we measure, assuming the line is produced under CIE and the plasma has solar abundances. In this plot, however, we assume a pressure several times lower than the pressure used for that Figure 8. The motivation for this change is the geometry discussed in sect. 7.1, where the filament is behind the E radio lobe and the counter-jet, so its true distance from the nucleus of M87 is $r=7.3$ kpc while the impact parameter is still $b=1.9$ kpc. We argue geometry likely applies to all the lines in this plot except for [\ion{Fe}{xxi}], so we refer to this as the fiducial case for most of the subsequent discussion. Compared to the $r=b$ case, the corresponding densities are therefore reduced by a factor of 4.5, and the corresponding filament volume filling factors $f_c$ are increased by a factor of 20 (see sects. 6.5 and 7.2). Thus the FUV-emitting plasma must have a volume filling factor $10^{-3} \lapprox f_c \lapprox 10^{-1}$.}
\end{figure*}

In sect. 6.5, we present a number of derived quantities (density, effective path length, cooling time, etc.) for intermediate-temperature plasma as traced by the FUV permitted lines, as a function of $r$.

In terms of the velocity of the filament, the radial component $v_r \approx 140$ km s$^{-1}$, so our fiducial geometry would imply a total speed relative to M87 of 140 km s$^{-1}$ / cos$15^{\circ} \approx 145$ km s$^{-1}$. In the unrealistic case of constant speed (deceleration is more likely) the travel time for the outflow is 49 Myr.

 On the other hand, since the filament is located at the edge of the radio lobe which may be expanding buoyantly, this expansion may introduce a strong tangential component to the velocity of the filament, so that its total speed would be significantly higher. This agrees with the predictions of \citet{Churazov2001}, who estimate a buoyant velocity of about half the local ambient sound speed in simulations of M87. Since $c_s \approx 730$ ($510$) km s$^{-1}$ for an ambient temperature of 2 (1) keV assuming $\mu = 0.6$ and $\gamma = 5/3$, their expected buoyant velocity is several times larger than the observed $v_r$ of our filament. For comparison, based on X-ray and radio data \citet{Forman2017} estimate that this radio lobe and a shock front at larger radii were both produced in an outburst about 12 Myr ago, which is reasonably consistent with our estimate if the deceleration is not too strong or if the true velocity is somewhat higher. 
 
We can further support this idea by noticing the connection between the bulk velocity of the nearby southern filament (see Figures 1 and 10) and the observed [\ion{Fe}{xxi}] blueshift. Our COS spectrum does not cover this filament, but we measure the velocity for [\ion{C}{ii}] $\lambda$158$\mu$m from the archival Herschel-PACS data, finding $v_r = -108\pm5$ km s$^{-1}$ and $\sigma_r = 130 \pm 6$ km s$^{-1}$. Kinematics for H$\alpha$ are also measured for both filaments by \citet{Ford1979} and \citet{Sparks1993}, and we adopt the former values for this figure ($v_r = 119$ km s$^{-1}$ and $v_r = -87$ km s$^{-1}$). \citet{Ford1979} measure $\sigma_r = 71$ km s$^{-1}$ for the S filament as well but do not report $\sigma_r$ for the N filament, so we use the same value for both filaments. 

It is notable that both lines in the S filament have very similar values of $v_r$ and $\sigma_r$ as the values we measure for [\ion{Fe}{xxi}]. One wonders if the southern filament might represent the approaching side of the radio lobe as it expands tangentially to the east, while the northern filament might represent the receding side of the lobe. If this is true, the blueshifted [\ion{Fe}{xxi}] might also be physically associated with the approaching side of the lobe, which in our sightline might not have cooled as far as the receding side, so instead of forming a $T \sim 10^3$ K filament it forms a reservoir of $10^7$ K plasma to emit [\ion{Fe}{xxi}].  This is one potential explanation for the different velocities for these different phases, and it is worth re-emphasizing the complicated and distributed geometry of the 1 keV plasma in this field (see Figure 2 of \citealt{Werner2013}).

A different possibility relates to magnetic fields. If the filament is produced from a cooling instability in the ICM, as the density increases the magnetic field strength could correspondingly increase, from a few $\mu$G potentially to a few $mG$. Magnetized multiphase plasma can certainly exhibit multi-directional flows associated with different phases; an example of exactly this behavior is seen in solar flares  (e.g., \citealt{Innes2001}, \citealt{Young2013}). While the conditions in the M87 filament are obviously very different than in a solar flare, the role of magnetic fields in galaxy cluster filaments is already appreciated and is the subject of ongoing study (e.g., \citealt{Brueggen2001}, \citealt{Ferland2009}, \citealt{Sharma2010}). 

Finally, we note that the S filament is likely projected along the radial direction as well. Optical [\ion{S}{ii}] measurements of the S filament by \citet{Werner2013} show that it has an electron density of approximately 30 cm$^{-3}$, which at $T \approx 1.6\times10^4$ K corresponds to a total pressure of $9\times10^5$ erg cm$^{-3}$. The impact parameter for the S filament relative to the M87 nucleus is around 3-4 kpc, but the pressure from eq. (3) is still at least twice this high at a distance of 4 kpc. A distance of 6.8 kpc for the S filament can reproduce the observed pressure, which is fairly similar to our fiducial estimate for the N filament as well.

\subsection{ICM turbulence measured by [\ion{Fe}{xxi}]}
In Figure 14 we present the probability distribution function for the velocity dispersion in [\ion{Fe}{xxi}], marginalized over the integrated flux and the radial velocity, based on our analysis in sect. 4. As described previously, our procedure is to adopt the median of the pdf as the central value for the parameter and then adopt the 68\% credible interval as the 1$\sigma$ uncertainties. This procedure yields $\sigma_r = 69^{+79}_{-27}$ km s$^{-1}$, which is indicated in the plot as well. However, the pdf is skewed and non-Gaussian, such that the maximum likelihood velocity dispersion is 47 km s$^{-1}$, which barely lies within our 68\% credible interval. 

 As discussed in sect. 2.2, the thermal velocity of Iron ions at the temperature corresponding to the peak emissivity of [\ion{Fe}{xxi}] (log $T = 7.05$) is 58 km s$^{-1}$. We subtract the radial component of this velocity in quadrature from the total velocity dispersion in order to estimate the turbulent velocity along the line of sight: $v_{\text{LOS}} =  \sqrt{\sigma_r^2 - (58 \text{ km s}^{-1})^2/3} \approx 60^{+84}_{-35}$ km s$^{-1}$. The maximum likelihood value is $v_{\text{LOS}} = 33$ km s$^{-1}$, but the central 68\% credible interval extends to $v_{\text{LOS}} = 144$ km s$^{-1}$ and the 95\% upper limit on the turbulence is 227 km s$^{-1}$. The tail toward large values of $\sigma_r$ is due to our uniform prior on $\sigma_r$; deeper observations with a higher SN would allow us to constrain $\sigma_r$ more precisely. As discussed in sect. 2.2 and in Appendix A, there may also be artificial broadening from the [\ion{Fe}{xxi}] emission being extended, which if the emission is uniformly aperture-filling can be as large as 76 km s$^{-1}$. We do not include this term explicitly in the above analysis since we do not have any direct observational constraints on the extent of [\ion{Fe}{xxi}] along the dispersion axis, but we note that it can fully explain the remaining velocity dispersion, so that the resulting turbulent velocities can be interpreted as upper limits. Including both thermal broadening and 76 km s$^{-1}$ for fully extended emission, the 95\% upper limit on the turbulence becomes $v_{\text{LOS}} < 214$ km s$^{-1}$.

Next we estimate the dynamical implications of this turbulence. We write the total pressure due to turbulence as $P_{\text{turb}} = \rho v_{\text{LOS}}^2 = \mu m_p n v_{\text{LOS}}^2$, where $n$ is the total (ion + electron) density, and we take $\mu = 0.6$. The thermal pressure is  $P_{\text{th}} = n k T$.  Combining these equations yields (see also \citealt{Werner2009}):

\begin{equation}
\frac{P_{\text{turb}}}{ P_{\text{th}}} = \gamma  \left(\frac{v_{\text{LOS}}}{c_s}\right)^2
\end{equation}

\noindent Assuming log $T = 7.05$ and $\gamma = 5/3$, the adiabatic sound speed is 507 km s$^{-1}$. The resulting turbulent pressure we infer is $2^{+11}_{-1}$\% of the thermal pressure. As pointed out by \citet{Inogamov2003}, the large mass of Fe ions keeps their thermal velocity low, which is what allows us to measure turbulent velocities even at these moderate values.

\begin{figure}
\begin{center}
{\includegraphics[width=8.5cm]{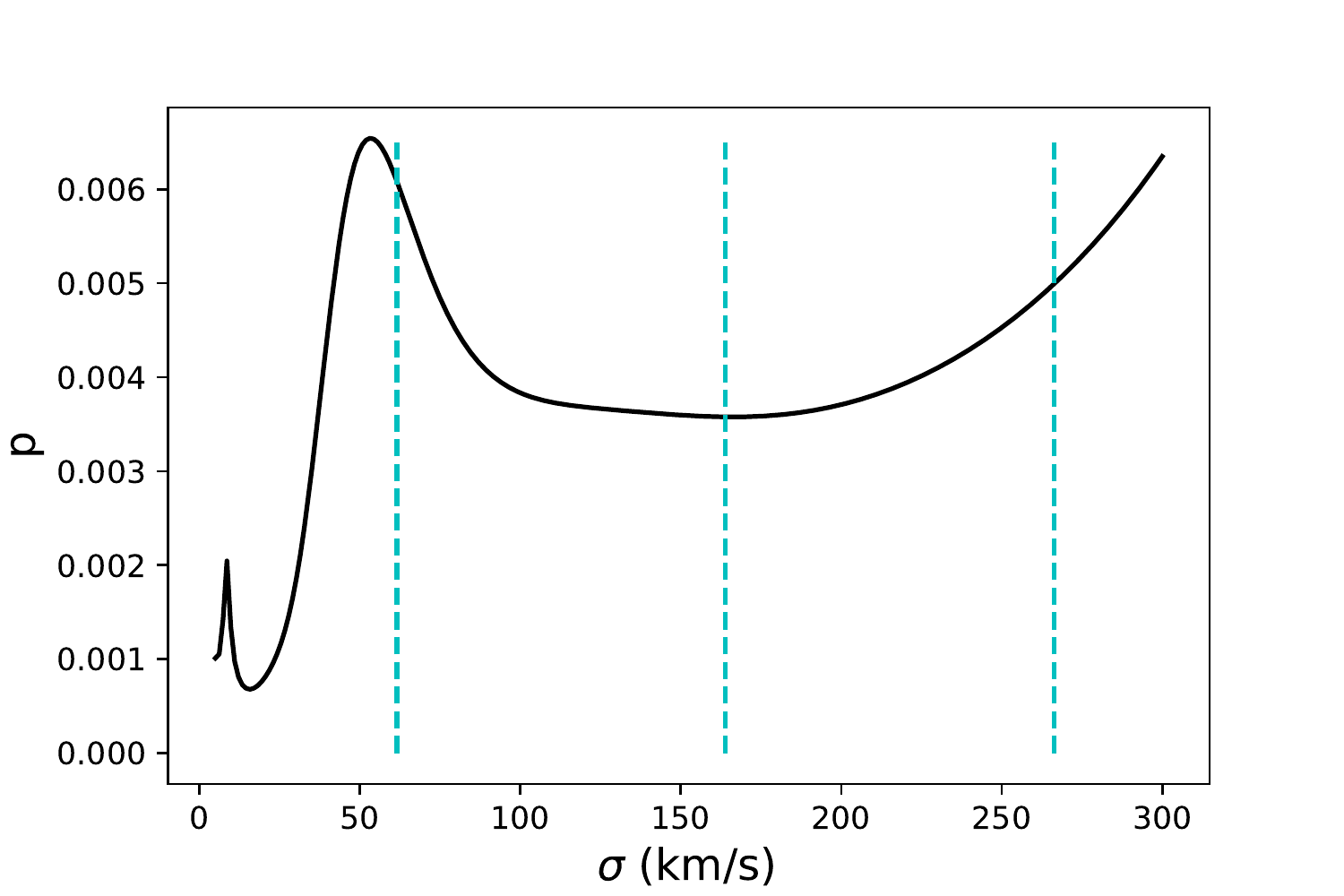}}
\end{center}
\caption{Probability distribution function for $\sigma_r$, the velocity distribution along the line of sight for [\ion{Fe}{xxi}], marginalized over $v_r$ and $F_0$.  The three dashed lines indicate the median value and the central 68\% credible interval, ($\sigma_r = 69^{+79}_{-27}$ km s$^{-1}$)  which just barely includes the maximum likelihood value of  47 km s$^{-1}$. Subtracting the thermal velocity we obtain an estimate of the turbulent velocity along the line of sight for the 1 keV phase of the ICM: $v_{\text{LOS}} =  60^{+84}_{-35}$ km s$^{-1}$.  The 95\% upper limit on $v_{\text{LOS}}$ is $227$ km s$^{-1}$ after subtracting the thermal velocity, or $214$ km s$^{-1}$ if we assume the emission is also aperture-filling and artificially broadened by this effect as well.} The long tail to the right is due to our uniform prior on $\sigma_r$; deeper observations with a higher SN are needed in order to allow us to constrain the posterior more tightly.
\end{figure}

 \subsection{Upper limit on Ly$\alpha$ absorption from M87 - constraints on photoionization in the hot halo}
Since we interpret the filament as lying behind the $T\sim10^7$ K gas, we also checked for evidence of absorption in the filament spectrum from this hot gas in M87. We first explored adding a component to our Ly$\alpha$ modeling to account for the small neutral Hydrogen column in the hot M87 ISM, for which the 2 keV component has the largest column density.  We fix this absorption component to have a velocity relative to M87 of $v_0 = -100$ km s$^{-1}$ and we fix its velocity dispersion to $\sigma_0 = 730$ km s$^{-1}$, which is roughly the expected thermal broadening for H atoms at 2 keV, although the final result is not extremely sensitive to either of these choices. We find a $2\sigma$ upper limit $N_H < 1\times10^{13}$ cm$^{-2}$. For purely collisional excitation, the neutral fraction expected in $T \approx 2\times10^7$ K plasma is about $1\times10^{-8}$ (\citealt{Dere1997}, \citealt{Delzanna2015}), so the electron column in the hot phase should be  $N_e < 1\times10^{21}$ cm$^{-2}$. 

By applying equation 1 from \citet{Anderson2016} to the emission measure of 2 keV plasma measured by \citet{Werner2013}, we find an effective electron column of $3\times10^{21}$ cm$^{-2}$ for the 2 keV plasma within our aperture, assuming a constant pressure of $4\times10^6$ erg cm$^{-3}$ as in Figure 8. This a factor of three higher than our 2$\sigma$ upper limit. The inferred effective electron column scales inversely with the pressure, so this discrepancy can be mitigated somewhat by assuming a lower average pressure for the column of 2 keV plasma. However, this analysis shows there is probably some need for photoionizations to reduce the neutral H fraction in the M87 ISM below its CIE value (see sect. 6.2).

\section{The cold phase of the filament as traced by [\ion{C}{ii}] $\lambda$158$\mu$\texorpdfstring{\MakeLowercase{m}}{m} }

In this section we discuss the three-dimensional structure of the filament system in more detail. Recall from sect. 6.4 that we associate the bulk of the FUV emission from very thin ($\Delta r \ll 1$ pc) ''skins'' which form a conductive interface between the ambient $T \sim 10^7$ K medium and cooler clumps which comprise the core of the filament. Now we turn to the emission directly associated with the cores. In sect. 3.2 we presented observations of the highly efficient transitions \ion{C}{ii} $\lambda$1335 and [\ion{C}{ii}] $\lambda$158$\mu$m which demonstrated that energy losses in both these lines are comparable, within the coarse spatial resolution offered by Herschel-PACS. This means the filament region radiates several thousand times more infrared photons than FUV photons. We attempt to understand that result here.

Figure 15 shows the expected ratio of these two lines in CIE and in the low-density limit. We have used the effective collision strengths tabulated in \citet{Liang2012} in order to estimate this ratio. For additional context, we plot in the inset the emissivities of these two lines in the low-density limit, as derived from CHIANTI. The FUV multiplet peaks in emissivity at around $T = 4\times10^4$ K, dropping dramatically toward higher or lower temperatures, and at the peak of its emissivity the ratio of the FUV flux to the FIR line flux is about 150. The observed ratio (see sect. 3.2) is about 0.2, with uncertainty of a factor of a few due to the difference in aperture sizes (see sect. 2.1). Thus the overall discrepancy between the CIE expectation and the observation is around a factor of 750  - much larger than can be explained by the different aperture sizes or the uncertainty in COS flux calibration. 

\begin{figure}
\begin{center}
{\includegraphics[width=8.5cm]{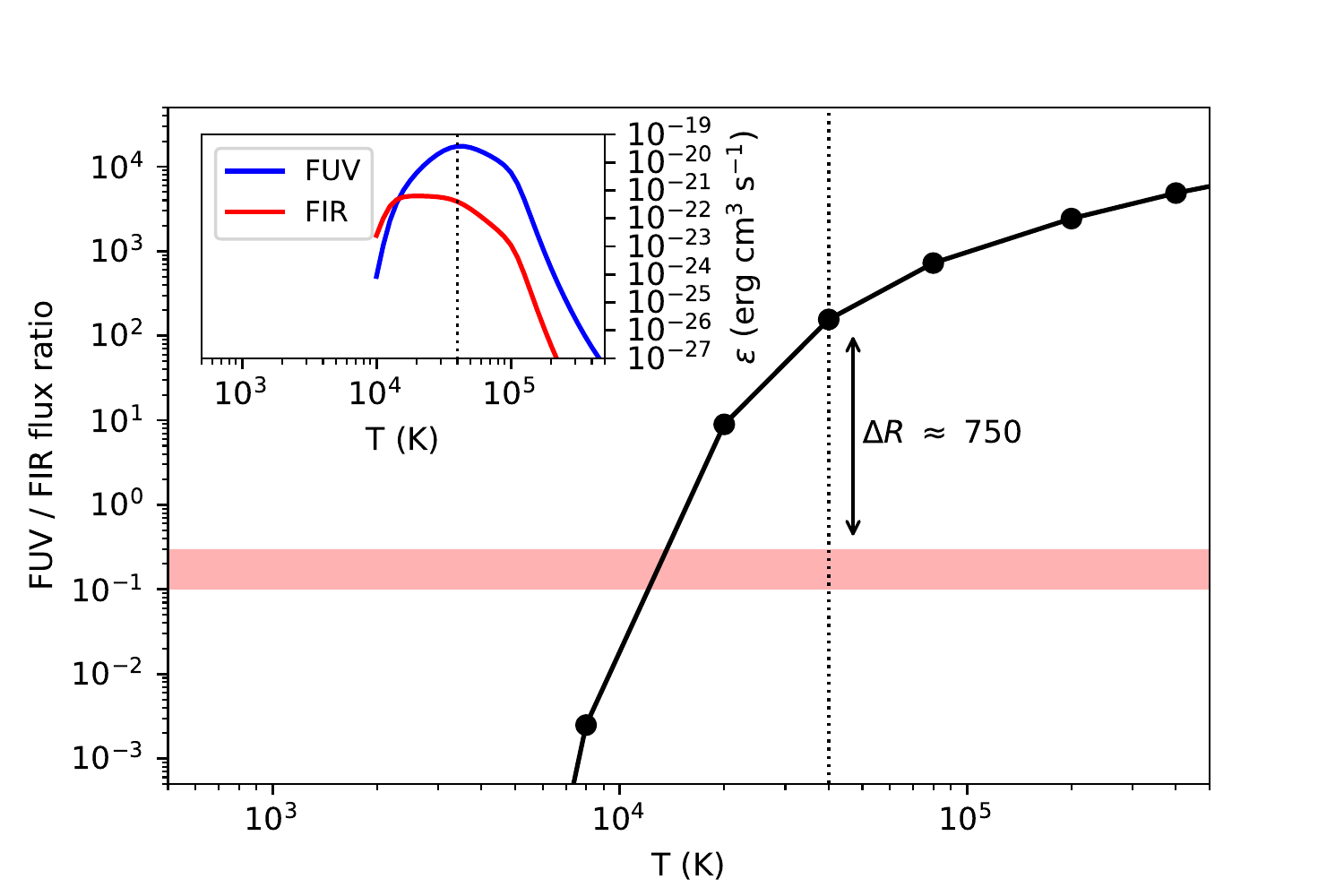}}
\end{center}
\caption{The main plot shows the expected flux ratio between the  \ion{C}{ii} $\lambda$1335 triplet and the  [\ion{C}{ii}] $\lambda$158$\mu$m forbidden line as a function of temperature, in the low-density limit ($n \lapprox 10-100$ cm$^{-3}$). The inset shows the emissivities of the FUV triplet and the FIR line as a function of temperature, again in the low-density limit. The FUV emissivity is strongly peaked around $T \sim 4\times10^4$ K (indicated with the dotted black line in both plots), and at this temperature the expected flux ratio is about 150.  The observed ratio, denoted with the red shaded region, is about 0.2, so the discrepancy between the CIE expectation and the observation is a factor of 750. We therefore invoke a two-phase model: the \ion{C}{ii} $\lambda$1335 emission can be associated with a $T\sim4\times10^4$ K phase of the thin "skin" surrounding the filament core, while the [\ion{C}{ii}] $\lambda$158$\mu$m emission can be associated with clumps at $T\sim10^3$ K which comprise the core of the filament, which we discuss in sect. 8.}
\end{figure}

This discrepancy can be understood in the case of strongly dust-obscured star-forming galaxies, where matter is transparent for IR radiation but fully opaque for the FUV. However, as we have shown there is no evidence for this degree of absorption by dust within the filament. Importantly, the ratios of the three \ion{C}{ii} FUV lines are consistent with their expected values in the optically thin case, without any resonant scattering effects. 

Another possibility we consider is that the \ion{C}{ii} $\lambda$1335 and [\ion{C}{ii}] $\lambda$158$\mu$m trace different locations along the filament. The Herschel-PACS image from \citet{Werner2013} uses 6"$\times$6" extraction regions, from which we estimated the [\ion{C}{ii}] $\lambda$158$\mu$m flux in sect. 3.2 by multiplying the flux per spaxel by the ratio of the length of the COS aperture to the length of the spaxel, which assumes the [\ion{C}{ii}] $\lambda$158$\mu$m emission is associated with the filament. If we instead multiply the flux within the spaxel by the ratio of areas (which would be appropriate if the emission were spatially uniform), the [\ion{C}{ii}] $\lambda$158$\mu$m flux inferred within our aperture would be lower by about 35\%. This does not help significantly, and we do not consider uniform, aperture-filling [\ion{C}{ii}] $\lambda$158$\mu$m further here. However, this analysis does introduce the question of how to explain the [\ion{C}{ii}] $\lambda$158$\mu$m flux in this field at all, which we address in the rest of this section.

\subsection{Size of filament}

We begin by summarizing observational constraints on the size of the filament. From visual inspection of the F658N images and the images in \citet{Sparks2009}, the filaments appear to be marginally resolved with a half-width of 0.4" or so, but this number is rather uncertain. We adopt a fiducial radius $r_c$ for the filament $r_c \sim 10$ pc. This corresponds to a radius of 0.12", which is about the right size to be broadened by the WFPC2 psf into the observed filament. This radius is uncertain by about a factor of two, so that $r_c$ could plausibly be anywhere from about 5 pc to about 20 pc.

We have further evidence for $r_c$ around this order of magnitude from the values of $\sigma_r$ measured for \ion{C}{iv} and \ion{He}{ii} in the low-resolution G140L archival data (sect. 3.6.2). We find $\sigma_r = 465\pm50$ km s$^{-1}$ and $\sigma_r = 375^{+65}_{-50}$ km s$^{-1}$ respectively for these lines, which is much larger than the $\sigma_r \approx 130$ km s$^{-1}$ we infer for the other lines from the filament. In \citet{Anderson2016}, we attribute this large velocity dispersion to the extended nature of the emission; for the lower-resolution G140L spectrograph, aperture-filling emission can have a measured velocity dispersion as high as 2000 km s$^{-1}$. If the emission fills 10\%-20\% of the aperture, then the spatial broadening would be about the right order of magnitude to produce the observed $\sigma_r$ for the G140L lines.

\subsection{Is the filament really cylindrical?}
In this subsection, we briefly justify our assumption that the filament is, at least to first order, cylindrical in shape, extending more than a kpc in length but having a width of just a few tens of pc. This physical picture is supported by the H$\alpha$+[\ion{N}{ii}] imaging in Figure 1, in which the filament appears cylindrical. We further justify it in Figure 16, which shows the average HST-WFPC2 surface brightness profile in the F658N filter (which includes the H$\alpha$+[\ion{N}{ii}] feature) measured across the northern filament from NE to SW. We average the surface brightness profile in a box of length 1" which extends along the filament, centered in the location of our COS aperture. 

The filament is clearly visible above the red stellar continuum from M87 (the continuum is visible in the F658N filter as well as the line emission, and is also dominant in the F851W filter which we plot for comparison), and it seems to have a width of about half an arcsecond, as described in the previous subsection. The edges of the filament are also relatively well-defined, which suggests the filament truly is intrinsically thin. This argues against the possibility that the filament is actually a sheet or pancake "draped" over the interface between the radio lobe and the Virgo ICM, since the curvature of the radio lobe would tend to smooth out the edges of the sheet as seen in projection. 

\begin{figure}
\begin{center}
{\includegraphics[width=8.5cm]{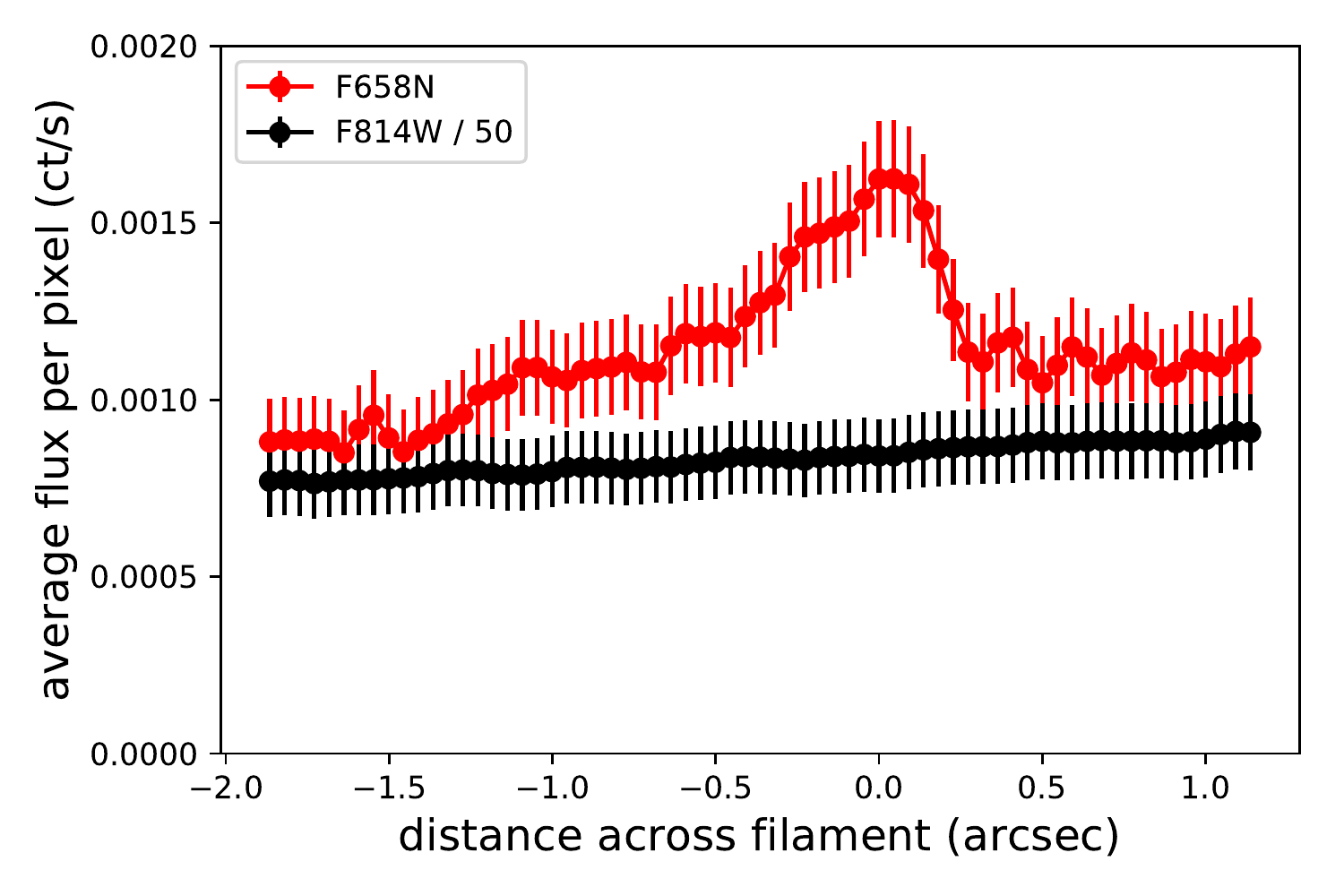}}
\end{center}
\caption{Average HST-WFPC2 surface brightness profiles in the F851W filter (which measures the red stellar continuum) and the F658N filter (which includes the H$\alpha$+[\ion{N}{ii}] feature as well as some continuum) measured across the northern filament from NE to SW. We average the surface brightness profiles in a box of length 1" which extends along the filament, centered in the location of our COS aperture. The filament is clearly visible above the red stellar continuum from M87. The relatively well-defined edges of the filament show that it has a roughly cylindrical geometry, as suggested by the imaging in Figure 1, and it would require an unlikely geometrical coincidence for the filament to be a sheet or pancake seen in projection.  }
\end{figure}

\subsection{Production and excitation of C$^+$ ions in filament}

One obvious solution which could explain the [\ion{C}{ii}] $\lambda$158$\mu$m flux is to assume the existence of a rather cold ($T_e \lapprox 8000$ K) phase in the core of the filament where the FUV \ion{C}{ii} $\lambda$1335 emission cannot be excited by collisions due to the rather high energy of the transition ($\Delta E = 9.3$ eV). 

For such a cold phase, in CIE the fraction of C in the ionized C$^{+}$ phase is near zero, but we see in our spectrum an FUV continuum from the stars in M87 which is capable of ionizing C atoms above the CIE expectation. We estimate this flux by integrating the extrapolated dereddened continuum which we fit in sect. 3.1 from 912\AA{} (the Lyman limit) to 1101\AA{} (the ionization edge for C$^{+}$), inferring a C-ionizing flux of $5\times10^{-15}$ erg s$^{-1}$ within our COS aperture. To convert this value into an ionizing flux at the surface of the filament, we assume a uniform spatial distribution of stars within M87, so that the incident flux on the filament is higher than our observed flux by the ratio of $4\pi$ sterad to the angular size of the COS aperture on the sky ($4.9$ square arcsec). This gives a flux incident on the filaments $F \sim 5\times10^{-4}$ erg s$^{-1}$ cm$^{-2}$. If the counter-jet is illuminating the filament, the ionizing flux may be even higher, but as we will show next, the fraction of C in the C$^{+}$ state is already nearly 100\%, and the Ly$\alpha$:H$\alpha$ ratio (sect. 6.2) also shows that photoionizations do not dominate the excitation of the filament. 

We next assume C ionizations are the dominant source of free electrons in the cold filament core, which is consistent with expectations from chemical networks (e.g., \citealt{Sternberg1995}) for the C$^{+}$-dominated portion of photodissociation regions. (If the FUV flux is so high that Hydrogen becomes significantly ionized, then the free electron density goes up by many orders of magnitude, but also the recombination rate rises accordingly, which is not observed.) We can then find a steady-state solution for the C ionization balance in the cold core of the filament. The recombination rate is given by $\alpha_C n_e n_{C+} \approx \alpha_C n_{C+}^2$, where $\alpha_C \approx 1\times10^{-12}$ cm$^3$ s$^{-1}$ is the recombination coefficient for C$^+$ ions at a temperature of a few thousand K \citep{Nahar1997}. The photoionization rate is $(F/h\nu) \sigma_{C} n_{C}$ where $F$ is the flux from the previous paragraph, we take $h\nu = 2\times10^{-11}$ erg,  and estimate $\sigma_C \approx 2\times10^{-17}$ cm$^{-2}$ \citep{Nahar1997}. Finally, the total amount of Carbon is fixed; for a given temperature, we can estimate the total Hydrogen density from pressure equilibrium arguments, and then assuming solar abundance we know the total Carbon density. We will use $T=1000$ K as a reference value. 

We have a range of possible pressures for the filament, and here we consider two choices which approximately bracket the possible cases (see sect. 6.5 and 7.2); in one case the distance from the nucleus is equal to the projected distance of 1.9 kpc (as in Figure 8) and in the other the filament is located behind the observed radio lobe in the direction of the counter-jet, in which case the distance is 7.3 kpc and the pressure is several times lower (as in Figure 13). We consider the latter case as the more probable geometry for the filament but we provide results under both cases for completeness. For the latter case (which we call the "fiducial case" in this section) we have $n_H \approx 500($1000 K$/T)$ cm$^{-3}$, and for the former case (which we call the "$r=b$" case in this section) we have $n_H \approx 2000($1000 K$/T)$ cm$^{-3}$. Then at solar abundance the total Carbon density in these two respective cases is $n_{C} + n_{C+} \approx 0.2($1000 K$/T)$ cm$^{-3}$ or $0.7($1000 K$/T)$ cm$^{-3}$. Solving this system of equations, we find respective ionization fractions for the C in the filament of 99.96\% or 99.86\% at $T=1000$ K. Even at a lower temperature of $T=100$ K, the ionization fraction is still 99.7\% or 98.6\% for the fiducial or $r=b$ cases respectively. 

This yields an optical depth for C-ionizing photons at $T=1000$ K of 

\begin{equation}
  \tau \equiv \sigma n_{C} r_0 \approx 0.003 \left(\frac{r_0}{1 \text{ pc}}\right) \tag{4a}
\end{equation}
\noindent for the fiducial case, or 
\begin{equation}
  \tau \approx 0.06 \left(\frac{r_0}{1 \text{ pc}}\right) \tag{4b}
\end{equation}

\noindent for the $r=b$ case. As long as the cold clumps are less than either $r_0 \sim 360$ pc (fiducial) or $r_0 \sim 17$ pc ($r=b$) in size, Carbon in the filament will nearly all be ionized to the C$^+$ state. Additionally, absorption of soft X-ray photons leading to the production of secondary electrons and an electron-photon cascade could further increase the C$^{+}$ fraction and allow the Carbon in larger clumps to remain ionized as well. We also have an upper limit on the size of clumps from the observed width of the filament ($r_c \sim 10$ pc).

Collisional excitation of these C$^{+}$ ions will produce the observed [\ion{C}{ii}] $\lambda$158$\mu$m emission. The most important relevant species for collisional excitations are free electrons, neutral H atoms, or molecular Hydrogen (see, e.g., \citealt{Draine2011}). Since these clumps are small enough to be optically thin to C-ionizing photons, they are unlikely to be able to sustain much molecular Hydrogen. The solar abundance of Carbon relative to Hydrogen (and therefore also the ratio of electrons to neutral H atoms) is $3.3\times10^{-4}$ \citep{Grevesse1998}, while the effective cross sections for collisions with free electrons are about $10^3$ times larger than the effective cross sections for collisions with atomic Hydrogen \citep{Hayes1984}. Thus we expect that collisions with neutral H atoms should be the most important. Our picture is therefore a cylindrical filament of radius $r_c$ composed of number of small, cold clumps with a relatively low volume filling factor $f_c$. 

We also point out that saturation may be important since [\ion{C}{ii}] $\lambda$158$\mu$m will be near the critical density. At temperatures near $10^3$ K, the critical density for collisions with H atoms goes approximately as $2300 T_3^{-0.15}$ cm$^{-3}$ \citep{Draine2011}, which is close to the expected Hydrogen density in pressure equilibrium ($n_H \approx 2000$ cm$^{-3}$ or $n_H \approx 700$ cm$^{-3}$ for the two respective pressures). The critical density for collisions with electrons is $17 T_3^{0.5}$ cm$^{-3}$ \citep{Draine2011}, and as long as the cold core is mostly neutral $(n_e / n_H \lapprox 0.0085$, which is guaranteed if the electrons are produced from ionizations of Carbon at the solar abundance) [\ion{C}{ii}] $\lambda$158$\mu$m will also be sub-critical for electron collisions. We note that [\ion{C}{ii}] $\lambda$158$\mu$m will become saturated at temperatures significantly lower than $10^3$ K, as the neutral H density will quickly rise above the critical density.

\subsection{Mass and filling factor of clumps in filament}

The total [\ion{C}{ii}] $\lambda$158$\mu$m flux within the $9.4$"$\times9.4$" region which covers the COS sightline is $2.4\pm0.6\times10^{-15}$ erg s$^{-1}$ cm$^{-2}$, corresponding to a [\ion{C}{ii}] $\lambda$158$\mu$m luminosity of $8\times10^{37}$ erg s$^{-1}$. The filament passes diagonally through the spaxel, subtending a distance $l \approx 1$ kpc.  We use the [\ion{C}{ii}] $\lambda$158$\mu$m cooling functions for collisions with atomic Hydrogen from \citet{Tielens1985}, for an assumed temperature of $T = 1000$ K. For an optically thin cylinder we have 
\begin{equation}
L_{CII} = \Lambda_{CII} n_H n_{C+} (\pi r_c^2 l f_c) \tag{5}
\end{equation}

As we found above that nearly all of the C atoms are ionized, we solve for $f_c$ for each case.  For the fiducial case we obtain
\begin{equation}
f_c = 0.010 (r_c / 10\text{ pc})^{-2} \tag{6a}
\end{equation}
\noindent and for the $r=b$ case we find 
\begin{equation}
f_c = 0.001 (r_c / 10\text{ pc})^{-2}\tag{6b}
\end{equation}

In Figure 17 we extend this analysis to other temperatures and compare the emission measure of [\ion{C}{ii}] $\lambda$158$\mu$m-emitting gas with the emission measure of the plasma which emits the \ion{C}{ii} $\lambda$1335 FUV triplet. We have again estimated the emissivity of the FIR line for C$^{+}$ excited by collisions with neutral H based on \citet{Tielens1985}, and then applied equation 1 to compute the corresponding emission measure. The emission measure curve for \ion{C}{ii} $\lambda$1335 is the same as in Figures 8 and 13.  We also plot curves of constant $f_C$ for each line, assuming a filament radius $r_c = 10$, using equation 5.

\begin{figure}
\begin{center}
{\includegraphics[width=8.5cm]{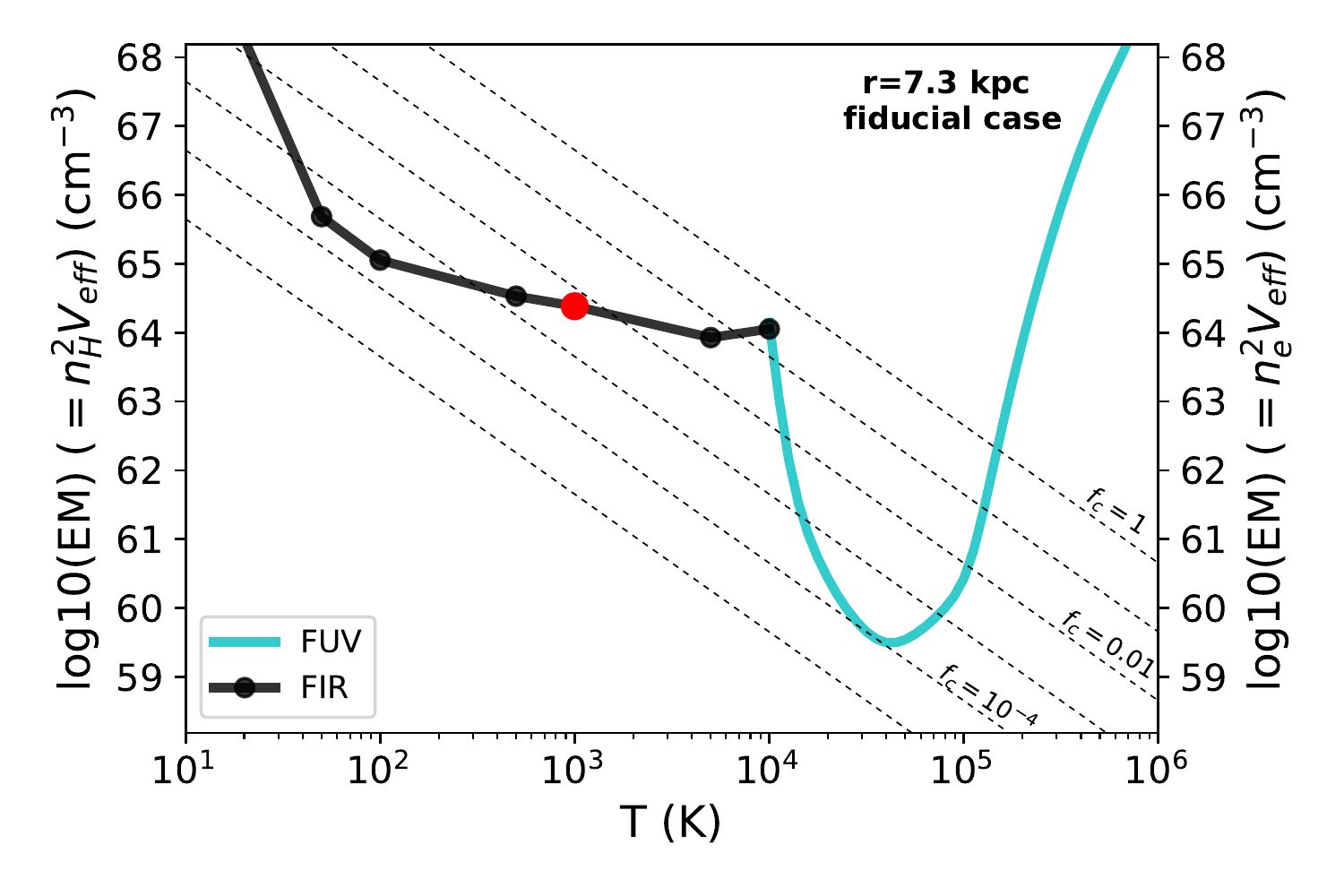}}
{\includegraphics[width=8.5cm]{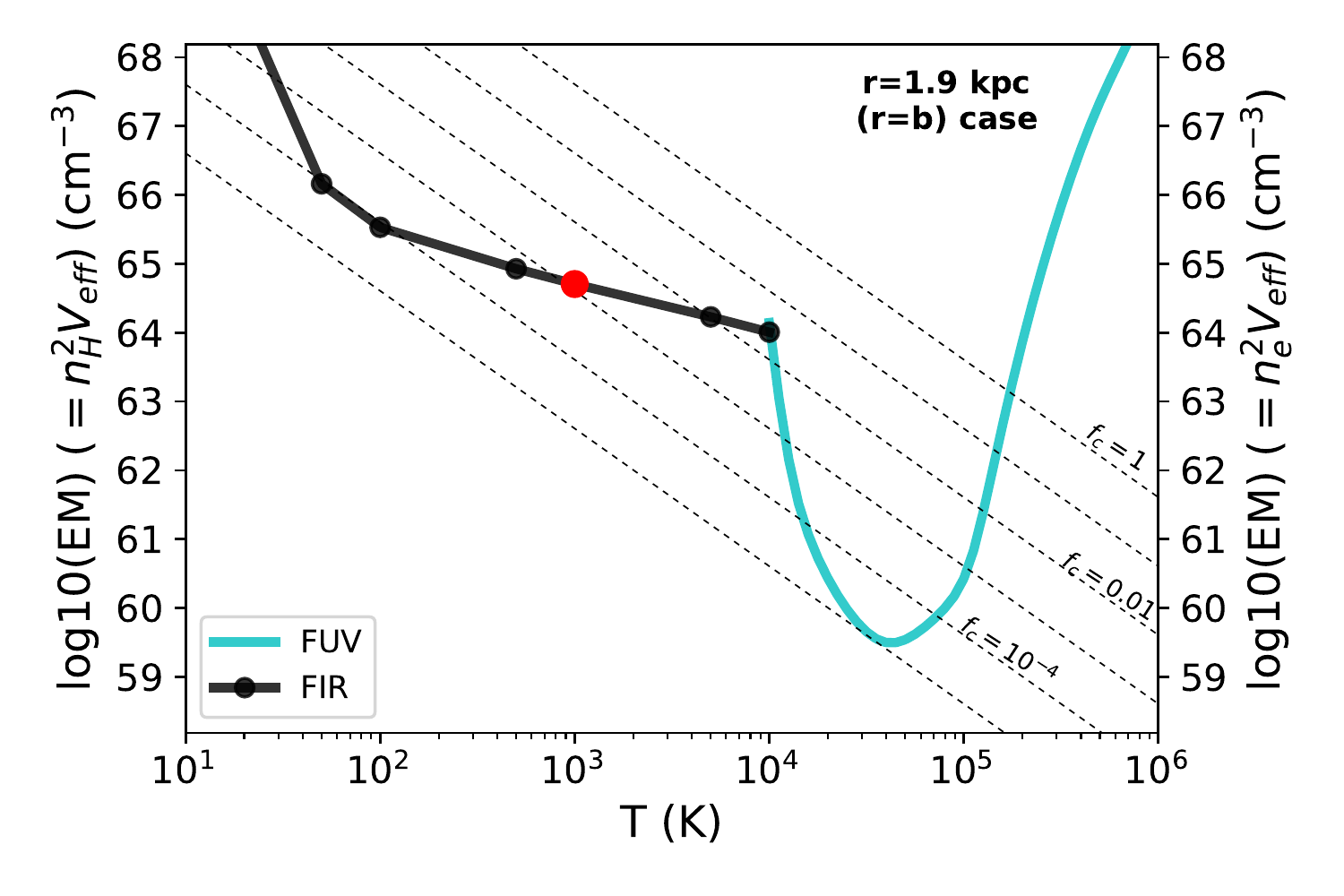}}
\end{center}
\caption{Estimated emission measures of gas emitting in the [\ion{C}{ii}] $\lambda$158$\mu$m (FIR) line and in the \ion{C}{ii}$\lambda$1335 (FUV) triplet in our sightline. The latter curve is identical to the curve for \ion{C}{ii} in Figs. 8 and 13, and the FIR curve is derived as described in sect. 8.4, assuming nearly all the C in the filament is in the C$^{+}$ state, and that collisions with atomic Hydrogen excite the  [\ion{C}{ii}] $\lambda$158$\mu$m transition. The red point shows our fiducial choice of temperature for the cold clumps, $T = 1000$ K, although we expect any value from $100 \lapprox T \lapprox 5000$ K is possible. We have also drawn lines of constant filament volume filling factor $f_c$, as in Figures 8 and 13, using equation 5. We note that the FIR-emitting gas must have a higher $f_c$ than the FUV-emitting gas assuming the latter is at its peak emissivity at $T \approx 4\times10^4$ K. This result holds for both choices of pressure, indicated by the upper and lower panels. This supports our model of cold FIR-emitting clumps in the filament surrounded by thin conductive boundary layers which emit the FUV lines.  }
\end{figure}

There is no clear best temperature for [\ion{C}{ii}] $\lambda$158$\mu$m, and probably any value from $100 \lapprox T \lapprox 5000$ K is possible. For both choices of pressure, any value of $T$ in this range still results in the FIR line-emitting phase having a larger filling factor than the FUV line-emitting phase, supporting our general picture of cold clumps which emit [\ion{C}{ii}] $\lambda$158$\mu$m, surrounded by thin conductive boundary layers which produce  \ion{C}{ii} $\lambda$1335 and the other permitted FUV lines.

With estimates of the density, size, and volume filling fraction, we can compute the mass of the cold phase within our COS aperture. For the fiducial case, with T=1000 K we have $M_c \approx 8000 M_{\odot}$. The filling factor of clumps scales approximately linearly with temperature and their density scales inversely with temperature, so the inferred mass does not depend strongly on the assumed temperature. There are small variations (factor of $\sim 3$ over the $100 \lapprox T \lapprox 5000$ K temperature range) due to the temperature dependence of the emissivity of C$^{+}$ ions, however. For the $b=r$ case, with T=1000 K we have $M_c \approx 4000 M_{\odot}$, and again this value can vary by about a factor of three across the temperature range. The final plausible range of masses is therefore $1000 M_{\odot} \lapprox M_c \lapprox 2\times10^4 M_{\odot}$.

If we assume equipartition between atomic and molecular gas phases and scale to the larger and brighter southern filament, this is still a bit smaller than the estimate for the molecular gas mass in the southern filament ($0.4-2 \times10^6 M_{\odot}$) obtained by \citet{Werner2013} using somewhat different considerations. However, we still infer a fairly large amount of cold gas, compared with the $\sim10-100 M_{\odot}$ of intermediate-temperature gas within our aperture, and we discuss some implications of this result in the next subsection.

The average neutral Hydrogen column density from the cold clumps can depend on the temperature, size of the clumps, and filling factor in detail, but if we assume a constant covering fraction within the cylinder we can make a simple estimate. For the fiducial case with $T=1000$ K, $r_c= 10$ pc, and $f_c = 0.01$, we have $N_H \sim 3\times10^{20}$ cm$^{-2}$, and again there are variations with temperature on the order of a factor of three. For the $r=b$ case with $T=1000$ K, $r_c= 10$ pc, and $f_c = 0.001$, we have $N_H \sim 1\times10^{20}$ cm$^{-2}$ with the same factor-of-three variations. Thus the plausible range of column densities is $3 \times 10^{19}$ cm$^{-2}\lapprox N_H \lapprox 9 \times10^{20}$ cm$^{-2}$.

\subsection{Consistency with other observations}

First, we check our results against existing observational limits. The most stringent comes from \citet{Sparks1993}, who measure visual extinction in the nucleus of M87 and place a limit $A_V < 0.02$ at the location of this filament. For dust with a Galactic dust-to-gas ratio of $1.9\times10^{21}$ cm$^{-2}$ / $A_V$ \citep{Bohlin1978}, this upper limit implies $N_H \lapprox 4\times10^{19}$ cm$^{-2}$, which is within the range of values we estimated in the previous subsection. (We note that these are all measurements of intrinsic absorption within M87; the Galactic column is around $1.4\times10^{20}$ cm$^{-2}$ in this direction.)

An alternative estimate of the extinction is possible for the nearby southern filament, where the H$\alpha$ : H$\beta$ ratio implies $E(B-V) = 0.28$ (\citealt{Ford1979}, \citealt{Werner2013}) and therefore  the total $N_H  \approx 2\times10^{21}$ cm$^{-2}$. Neutral columns $N_H > 10^{21}$ cm$^{-2}$ are also measured in filaments in the cores of other nearby massive galaxy clusters, such as Centaurus \citep{Mittal2011} and Perseus \citep{Mittal2012}. Our estimate of $N_H$ for the northern filament is below these larger columns inferred for other filaments, and closer to the smaller columns implied by the \citet{Sparks1993} upper limit.

Second, \citet{diseregoalighieri2013} detect dust in the center of M87, with a total dust mass of $2\times10^5 M_{\odot}$ within the central arcminute. For dust : gas mass ratios of $10^3 - 10^4$, as are observed in the Perseus filaments \citep{Mittal2012}, this implies a mass of $10^8 - 10^9 M_{\odot}$ in neutral gas in the center of M87. If the average filament has a column $N_H \sim 3\times10^{20}$ cm$^{-2}$ and we use the result from \citet{Sparks1993} that the filaments subtend an area of 1250 arcsec$^2$, then the total cold gas mass implied is $3\times10^7 M_{\odot}$ for the center of M87. There are large enough uncertainties in both of these estimates that an unusual dust : gas ratio in M87 is not necessarily implied, although it is possible.

Third, depending on the exact geometry, it is possible for the neutral Hydrogen in the filaments to absorb soft X-rays ($\tau \sim 0.1$ for 1 keV photons, depending on gas-phase metallicity). However, the location of our filament (as well as the nearby southern filament) shows enhanced soft X-ray surface brightness relative to its surroundings (\citealt{Young2002}, \citealt{Sparks2004}, \citealt{Forman2007}, \citealt{Million2010}). This can actually be explained in the context of our model (sect. 7.1), however, since we place the filament behind the $T\sim1$ keV plasma along our line of sight; it therefore does not contribute to the absorption of the soft X-ray emission.
 
Finally, a key difference between the M87 filaments and the filaments in Perseus and Centaurus is the lack of star formation in M87. This is probably due to the lower mass of atomic gas in the M87 filaments as compared to these other systems.

While the exact dimensions and mass of the cold phase of the filament are therefore somewhat uncertain, the large ratio of  [\ion{C}{ii}] $\lambda$158$\mu$m to \ion{C}{ii} $\lambda$1335 demonstrates conclusively the existence of a cold phase with $T \lapprox $ several thousand K, and it is clear that this phase contains the majority of the material associated with the filaments. More work is needed in order to fully understand its mass, dimensions, and phase structure. In particular, ALMA observations would be especially useful, as ALMA has the sensitivity and angular resolution to detect molecular gas in these filaments, if it exists. Detection or non-detection of molecular gas at sub-arcsecond resolution would allow us to constrain the size, mass, and temperature of the clumps or strands which comprise the cold phase of the filament.

\subsection{The velocity dispersion in the filament is explained by the random motion of clumps}

Figure 18 shows two different ways to produce velocity dispersion which we consider in this paper. The left panel is a standard turbulent cascade in a single-phase plasma. This is how we understand the observed $\sigma_r$ for [\ion{Fe}{xxi}], which traces turbulence in the $T \sim 10^7$ K volume-filling intracluster medium in front of the radio lobe in our sightline (see sect. 6.4). The right panel is velocity dispersion arising from random motions  of clumps confined within a cylindrical filament. In this case, the measured $\sigma_r$ does not relate to pressure support within the clumps, but rather to their characteristic velocity as they move around in the filament. We also remind the reader that broadening can also come from thermal motions and from instrumental effects if the emission is extended, as we discuss in sect. 2.2 and 7.3, but that these effects are not likely to be very important except possibly in the case of [\ion{Fe}{xxi}].

\begin{figure}
\begin{center}
{\includegraphics[width=8.5cm]{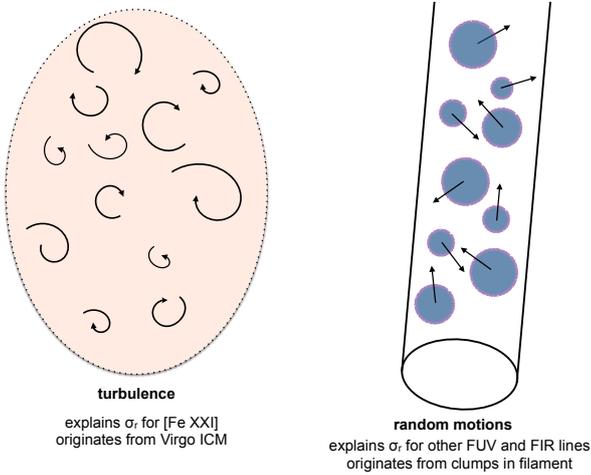}}
\end{center}
\caption{A diagram showing the two types of velocity dispersion we consider in this paper. The left panel is our explanation for the observed velocity dispersion in [\ion{Fe}{xxi}] -- a turbulent cascade in the hot, volume-filling intracluster medium (see sect. 7.3). The right panel is our explanation for the observed velocity dispersion in the other lines we discuss in this paper. These other lines all have about the same value of $\sigma_r \approx 130$ km s$^{-1}$ even though they are emitted from gas at characteristic temperatures ranging from below $10^3$ K up to several times $10^5$ K, we interpret the velocity dispersion we measure for these lines as the result of random motions from clumps which are magnetically confined inside the filament. We also note that Ly$\alpha$ is broadened additionally by the large optical depth that Ly$\alpha$ photons experience as they escape from the boundary layers around these clumps. }
\end{figure}

This latter interpretation of velocity dispersion is our explanation for the observed velocity dispersion $\sigma_r \approx 130$ km s$^{-1}$ which we measure for \ion{C}{ii} $\lambda$1335, \ion{N}{v}, and Ly$\alpha$, and which is also seen in [\ion{C}{ii}] $\lambda$158$\mu$m and H$\alpha$ from this region as well. A powerful piece of evidence for this interpretation is the fact that we measure approximately the same $\sigma_r$ in the FUV lines which trace the boundary layer and in [\ion{C}{ii}] $\lambda$158$\mu$m which traces the cold core of the filament. We therefore explain the observed velocity dispersion in these lines as the random motions of an ensemble of cold clouds in the filament. These random motions naturally produce a Gaussian line profile, somewhat analogously to the broad-line region of an AGN. The non-Gaussian profile observed for \ion{N}{v} might be related to insufficient \ion{N}{v}-emitting clumps causing the underlying clump velocity distribution to be undersampled for this line. Similar clumps are also seen in filaments in the nearby dwarf galaxy NGC 5253 \citep{Consiglio2017}, and we discussed the need for clumps with a low volume filling factor $f_c$ in sect. 8.1.

 While the measured $\sigma_r$ is close to the value measured by \citet{Hitomi2016} for the ICM in Perseus, these motions are supersonic for the cooler gas in the filament. At $T = 10^3$ K, the adiabatic sound speed is just 3 km s$^{-1}$, and so the Mach number of these motions is nearly 40. This makes it difficult to interpret $\sigma_r$ for the cold phase as a turbulent cascade. 

We have examined the 2D HST-COS spectrum in order to search for any clumpiness in the cross-dispersion direction, but the observed spectrum is indistinguishable from the line spread function in the cross-dispersion direction, which sets an upper limit of around 40 pc on the distance between bright clumps within the field of view. This upper limit is larger than the $\sim 5$ pc size of the clumps in filaments in NGC 5253, and also larger than the observed radius of the filament. It is likely that the filament is instead composed of many hundreds or thousands of clumps, would naturally produce a Gaussian velocity distribution while appearing smooth at HST-COS resolution.

We can get a lower limit on the characteristic clump size from simple mean free path arguments. The mean free path $l$ for clumps in the filament can be related to the size of a clump $r_0$ by $l \sim (n r_c^2)$, where $n$ is the volume density of clumps (clumps per unit volume). From here we can derive 
\begin{equation}
l \sim \frac{r_0}{f_c} \tag{7}
\end{equation}
\noindent For $f_c \sim 0.01$ and $l > r_c \sim 10$ pc (since we want clumps not to collide frequently, in order for the system to survive), the clumps must have $r_0 > 0.1$ pc. For the fiducial case, our range of plausible values for $f_c$ spans $10^{-3} \lapprox f_c \lapprox 0.1$, which corresponds to a range of lower limits on $r_0$ between 0.01 pc and 1 pc.

There must also be some mechanism to confine the clumps within the filament, which can be either pressure confinement (from the hot ICM), magnetic confinement, or some combination of the two. For pressure confinement, the ram pressure force that a clump of radius $r_0$ experiences is approximately $\pi r_0^2 \rho_{ICM} \sigma^2$, so the work done while crossing the filament is $\pi r_0^2 \rho_{ICM} \sigma^2 r_c$, which can be equated to the kinetic energy of a clump $(2/3) \pi r_0^3 \rho_{c} \sigma^2$. Thus we have 
\begin{equation}
\rho_{ICM} / \rho_{c} \gapprox r_0/r_c \tag{8}
\end{equation}
 as the condition for ram pressure to be able to confine a clump. This condition gives 
 \begin{equation}
 r_0 \lapprox 0.0005 (T / 1000\text{ K}) \text{ pc} \tag{9}
 \end{equation}
 \noindent Thus ram pressure from the Virgo ICM is unlikely to be sufficient to confine the clumps within the filament, and assistance from magnetic fields is probably necessary.

\subsection{Mass transfer rate?}

We argued in sect. 7.1 that the filament system has traveled from the nucleus for a few $10^7$ yr and in sect. 6.4 that the FUV emission comes from plasma with cooling times less than $10^3$ yr. Such a situation can only arise if the filaments are a dynamical system, continually changing in mass and possibly in temperature. Thus we expect there to be times when the system is globally cooling (gaining mass from the ICM) and times when the filaments are globally heating (losing mass to the ICM). 

Considering the latter possibility first, we also remind the reader of the slight asymmetry in the Ly$\alpha$ profile (sect. 3.4) which is also a sign of a  general outflow with a velocity $ v_{r} \sim 10$ km s$^{-1}$. In order to estimate the properties of this putative outflow, we check whether heat conduction from the ICM into the filament is saturated (i.e., whether the mean free path of hot electrons is longer than the temperature scale height). This calculation requires an estimate of the clump size within the filament. We showed in sect. 6.4 that this size is bounded by the range $0.001$ pc $< r_0 < 10$ pc. For now we assume a clump size $r_0 = 1$ pc, which yields a saturation parameter $\sigma_0$ as defined by \citet{Cowie1977} of $\sigma_0 \approx 4 \times ($1 pc$/r_0)$ (taking $T = 10^7$ K and $n_e = 0.1$ cm$^{-3}$ in the ICM), so this clump size corresponds to moderately saturated conduction. Smaller clump sizes will yield more highly saturated conduction. The saturated heat flux (eq. 7 of \citealt{Cowie1977}) is about 0.004 erg s$^{-1}$ cm$^{-2}$ for the $T \sim 10^5$ K intermediate-temperature shell in the fiducial case, and about a factor of three higher in the $r=b$ case.

Taking eqs. (21) and (22) from \citet{Cowie1977}, which apply in the unsaturated case, we can estimate the mass loss rate $\dot{m}$ and the evaporation rate $\tau$ for the filament, obtaining $\dot{m} \approx 0.0001 (r_0/$1 pc$) M_{\odot}$ yr$^{-1}$ for both cases,  $\tau \approx 0.7 (r_0/$1 pc$)^2 (1000$ K$/T)^{2.5} $ Myr for the fiducial case, and $\tau \approx 2.0 (r_0/$1 pc$)^2 (1000$ K$/T)^{2.5} $ Myr for the $r=b$ case. Using eqs. (63) and (64) instead, which apply in the case of highly saturated conduction, we find $ \dot{m} \approx 7 (r_0/$1 pc$)^2 \times10^{-5} M_{\odot}$ yr$^{-1}$ and for both cases,  $\tau \approx 1 (r_0/$1 pc$) $ Myr for the fiducial case, and  $\tau \approx 1 (r_0/$1 pc$) $ Myr for the $r=b$ case. These latter relations have an additional dependence on $r_0$ as well, through a factor of $F(\sigma_0)$ which is of order unity; we compute this for $r_0=1$ in the above estimates. These sets of values are close to one another and bracket the moderately saturated case which we infer for $r_0 = 1$ pc.

The inferred evaporation time is much shorter than the travel time for the filaments which we estimated in sect. 7.1. However, if there is a significant turbulent magnetic field around the clumps, this will strongly decrease the mean free path of electrons, which will correspondingly reduce the thermal conductivity. This will decrease the mass-loss rate and increase the lifetime of the filament system. 

However, the correspondence in X-ray images between the soft X-ray flux and the H$\alpha$ filaments (e.g., \citealt{Young2002}, \citealt{Sparks2004}, \citealt{Forman2007}, \citealt{Million2010}) also raises the possibility that the ICM may be cooling onto the filaments rather than evaporating them. The 0.7-0.9 keV band shows a significant flux excess over the other nearby bands, and in this band the flux from within our COS aperture is $5.9\times10^{-16}$ erg s$^{-1}$ cm$^{-2}$ (see sect. 6.4). If this flux were entirely due to a cooling flow, we can estimate the cooling rate using the \verb"cflow" model in Xspec (v. 12.9.1). We set the metallicity to solar, the low-energy cutoff to 0.1 keV (the exact value is not important for this calculation), the high-energy cutoff to 3 keV. For the redshift of M87, we find that this flux corresponds to a cooling rate of $3\times10^{-4} M_{\odot}$ yr$^{-1}$. This is within a factor of three of the evaporation rate computed above, and while this cooling rate is likely an upper limit it still seems somewhat possible for cooling to balance heating in this filament. It therefore remains unclear whether the filament is actually growing or evaporating. 


\subsection{Collisions of clumps?}

It is also worth considering the role of collisions between clumps in the filament. The crossing time for clumps within the filament is less than $10^5$ yr, and there will be shocks when clumps collide since the velocity of random motions $\sigma_r \sim 130$ km s$^{-1}$ is so much higher than the sound speed within the clumps (see sect. 8.6). For a velocity dispersion $\sigma_r \approx 130$ km s$^{-1}$, the thermalization temperature is of order $T\sim10^6$ K, with the exact value depending on the effective equation of state. In this temperature regime plasma is highly unstable \citep{Mckee1977b} and can be expected to cool, so these shocks will produce intermediate-temperature plasma which radiates in the permitted FUV lines we observe. The cooling time for the shock-heated plasma is also on the order of $10^5$ yr \citep{Sutherland1993}, comparable to the crossing time across the filament.  It is possible to imagine a sort of equilibrium arising where the average rate of clump destruction through collisions is comparable to the production rate of clumps through cooling of shock-heated million-degree plasma.

This would provide an alternative source of intermediate-temperature $T \sim 10^5$ K plasma, which could supplement the conductively-powered boundary layers around the cold clumps. This picture merits further investigation. For clumps of size $r_0 \sim 1$ pc, the mean free path is about $10^2$ filament widths, so the typical time between collisions is about $10^7$ years, which is an order of magnitude larger than the evaporation time estimated in the previous section. For smaller values of $r_0$, both timescales vary linearly with $r_0$, so thermal conduction will continue to dominate. For larger values of $r_0$, the conduction becomes less saturated and $\tau \propto r_0^2$, so if clumps are sufficiently large then collisions may become more important than conduction. On the other hand, for sufficiently large clumps (i.e., $r_0 = 3$ pc, $r_c = 10$ pc, and $f_c = 0.001$), we expect less than one clump within the COS aperture, and this would not produce a Gaussian distribution of velocities as we observe for most lines, nor would it produce the area-filling H$\alpha$ emission seen in Figure 1. Thus our initial impression is that collisions of clumps are not the primary source of intermediate temperature plasma in this filament. 

There may also be a radiative shock associated with the filament, since it occurs at the interface between an expanding radio lobe and the ICM. Recall also that the presence of filaments at such interfaces is a general phenomenon, and the filaments in M87 and other nearby clusters are usually observed to be perpendicular to these interface regions. \citet{Dopita1996} considered this possibility in the general case and showed that photoionization in front of the shock can be important as well, and their model can reproduce the observed optical and FUV lines from the nucleus of M87 \citep{Dopita1997}. Inclusion of radiative shocks also complicates the picture we have developed. At this level of detail, hydrodynamical simulations are required, which is outside the scope of this paper. We remind the reader, however, that we have showed that most of the lines we observe can be adequately explained collisionally.

\section{Conclusions}

M87 is an ideal laboratory for studying the problem of coupling feedback from accreting SMBHs to the surrounding gas in the context of AGN feedback. In this paper we study an FUV spectrum which we obtained with the G130M grating on HST-COS, which covers an H$\alpha$- and \ion{C}{iv}-emitting filament projected 1.9 kpc to the E of the nucleus of M87. This is a well-studied region full of multiphase gas at the edge of a radio lobe which is thought to be inflated by the SMBH in the nucleus.

We detect a faint FUV continuum we attribute to the so-called ''FUV excess", a fairly common phenomenon in early-type galaxies with old stellar populations which is thought to be connected to emission from evolved stars in these galaxies. 

We also detect a number of lines in emission from this region, most of which we attribute to the multiphase plasma in the filament. In our HST-COS G130M spectrum, these lines include \ion{C}{ii} $\lambda$1335, Ly$\alpha$, and \ion{N}{v} $\lambda$1238, and we supplement them with archival COS G140L measurements of \ion{C}{iv} $\lambda$1549 and \ion{He}{ii} $\lambda$1640 (\citealt{Sparks2012}, \citealt{Anderson2016}). The former three lines all show approximately the same radial velocity $v_r \approx 140$ km s$^{-1}$. This radial velocity is also seen in H$\alpha$ \citep{Sparks1993} and [\ion{C}{ii}] $\lambda$158$\mu$m measurements of the same filament. For \ion{C}{ii} $\lambda$1335 we measure a velocity dispersion $\sigma_r = 128^{+23}_{-17}$ km s$^{-1}$, which is also consistent with the $\sigma_r = 107\pm20$ km s$^{-1}$ we measured in [\ion{C}{ii}] $\lambda$158$\mu$m. Ly$\alpha$ appears broader ($\sigma_r = 171\pm2$ km s$^{-1}$) which we attribute to the effects of resonant scattering due to the large optical depth for Ly$\alpha$ photons. \ion{N}{v} also appears broader than \ion{C}{ii} ($\sigma_r = 189^{+12}_{-11}$ km s$^{-1}$) but the lines are also noticeably non-Gaussian; we hypothesize \ion{N}{v} is formed in a narrow boundary region between the cooler filament and the warmer ICM which may be especially unstable. 

The overall agreement in kinematics for these lines is remarkable; the characteristic temperatures of these lines span four orders of magnitude, so that the observed $\sigma_r$ is subsonic for the hotter gas and highly supersonic for the coolest gas in this sightline. We infer that the velocity dispersion is the product of random motions of clumps in the filament. The large velocity dispersion implies that cold clouds (and the warm gas surrounding the clouds) are moving with large relative velocities relative to one another, or possibly that the radial velocity of the gas in the filament fluctuates very strongly over scales of less than 100 pc. In the first case, collisions of clouds are inevitable, although their frequency depends on the cloud size. Under certain conditions, these collisions might be a very important reason for the heating of warm gas.

In addition to the overall bulk velocity  $v_r \approx 140$ km s$^{-1}$ and velocity dispersion $\sigma_r \approx 130$ km s$^{-1}$ in the filament, we also find evidence of a $v_r \approx 10$ km s$^{-1}$ outflow in the filament based on the asymmetry of the (Galactic-absorption-corrected) Ly$\alpha$ profile, which implies evaporation in the filament. However, the spatial coincidence between a soft X-ray excess in M87 and the H$\alpha$ filaments could be evidence of cooling from the hot ICM onto the filaments. At the moment it is still unclear whether the filaments are in fact growing or evaporating. 

We also detect, at $4-5 \sigma$ depending on the binning, [\ion{Fe}{xxi}] $\lambda$1354 from M87 in our sightline. This transition occurs in $10^7$ K plasma, and we therefore do not associate this line with the filament, but rather with gas which associated with the 1 keV phase of the ambient ICM. Our [\ion{Fe}{xxi}] is slightly blueshifted ($v_r = -92^{+34}_{-22}$ km s$^{-1}$) which is different from the observed $v_r$ for the filament emission. However, this is the same velocity as observed for the nearby southern filament. We propose that the southern filament traces the approaching side of the radio lobe, while the northern filament traces the receding side; our sightline naturally passes through both sides of the lobe and we speculate that the permitted FUV lines are coming from the latter while the [\ion{Fe}{xxi}] is emitted by 1 keV plasma associated with the former. We also measure the velocity dispersion in [\ion{Fe}{xxi}], which is only the second direct measurement to date of turbulence in the ICM of a galaxy cluster, and find a turbulent velocity along the line of sight $v_{\text{LOS}} =  60^{+84}_{-35}$ km s$^{-1}$. It will be very interesting to obtain HST-COS spectroscopy of the southern filament as well in order to check the kinematics of [\ion{Fe}{xxi}]. If our picture is correct and the 1 keV plasma lies on the approaching side of the radio lobe, then our model makes a clear prediction that $v_r$ and $\sigma_r$ should be very similar for [\ion{Fe}{xxi}] in both sightlines.

We also estimate the Ly$\alpha$ : H$\alpha$ ratio along our sightline and find intermediate values (17-23), which points to a mixture of collisional excitation and recombination involved in producing these lines. Under the assumption of CIE we also estimate the emission measures of plasma emitting each of the FUV lines, and find generally good agreement between our observations (including [\ion{Fe}{xxi}]) and the differential emission measure from a solar flare, which is a multiphase collisional plasma. This supports our conclusion that these lines are generally collisionally powered, but we note that \ion{He}{ii} is a significant outlier and probably requires a boost from recombinations, as does H$\alpha$. 

These observations underscore the importance of studying the FUV lines from filaments in clusters. The bulk of the emission is radiated in the FUV (and the EUV, although there are currently no instruments available to observe extragalactic EUV radiation), and line ratios of FUV lines are very powerful for diagnosing physical conditions in the filamentary plasma. The total cooling rate in the FUV is likely over $10^{41}$ erg s$^{-1}$, even though the volume filing fraction of these filaments is negligible, underscoring their importance in the energy balance of the ICM in Virgo. 

Additionally, [\ion{Fe}{xxi}] is a uniquely powerful FUV line which can be exploited to measure the kinematics of the 1 keV component of the ICM. These observations are otherwise unobtainable at present, and they are very complementary to Hitomi and to the upcoming microcalorimeter studies of the hotter components of the ICM through kinematics of Fe K$\alpha$. Our observations along this sightline are currently photon-limited, and not systematics-limited, so deeper observations of 1 keV plasma using HST-COS promise to measure the kinematics of this plasma with higher precision in the future. 
 
Finally, we also consider the cold ($T \lapprox $ a few thousand K) phase of the filament, which is probed by archival Herschel-PACS data showing strong  [\ion{C}{ii}] $\lambda$158$\mu$m emission from the filament region. The high  [\ion{C}{ii}] $\lambda$158$\mu$m : \ion{C}{ii} $\lambda$ 1335 ratio is very clear evidence that a cold phase must exist in the filament, and the inferred mass in this phase is in range $1\times10^3 - 2\times10^4 M_{\odot}$ within our COS aperture. This is orders of magnitude larger than the tens of solar masses in intermediate temperature ($T \sim 10^4 - 10^6$ K) gas at the interface between the filament and the Virgo ICM,  but still probably not massive enough to form stars. The filament is likely composed of a bundle of strands or clumps with a low volume filling factor. Observations of the cold phase with ALMA would be very helpful for understanding the mysterious cores of these filaments.
\begin{appendix}

\section{Cross-dispersion profile of FUV emission}

In this Appendix we examine the cross-dispersion profiles of the FUV emission in our COS observations. To do this, we add together the individual \verb"flt" files from each observation to produce a summed 2D spectrum, and then measure profiles in the cross-dispersion direction. Figure A.1 shows the results. In general the SN is rather low, so we can only perform this exercise for the brightest lines (Ly$\alpha$ and \ion{N}{v}). 

For comparison, we also plot the COS line spread function in the cross-dispersion direction at 1221\AA{}, which roughly corresponds to the redshifted wavelength of Ly$\alpha$ from M87. We downloaded this profile from the COS website\footnote{\url{http://www.stsci.edu/hst/cos/performance/spectral_resolution/}} for COS G130M at lifetime position 3 with the \verb"CENWAVE=1318" setting. To bracket the opposite case, where extended emission uniformly fills the aperture, we also measure the cross-dispersion profile for the Ly$\alpha$ airglow line in our observations. We fit Gaussians to each profile, estimating a standard deviation of $3.3\pm0.1$ pixels for the line spread function in the cross-dispersion direction and a standard deviation of $8.5\pm0.2$ pixels for Ly$\alpha$ airglow. 

For Ly$\alpha$, we can also fit a Gaussian to the profile, and obtain a standard deviation of $8.0\pm0.2$ pixels. For \ion{N}{v} the SN is not very high but we obtain a standard deviation of $9.8^{+2.7}_{-1.8}$ pixels .This analysis is difficult due to vignetting and other effects, and the observed profiles are probably not simply linear combinations of the cross-dispersion line spread function and the uniform aperture-filling profile, but it seems that the emission comes close to spanning the entire cross-dispersion axis. Our expectation (see sect. 2.2) is that the filament extends roughly along the cross-dispersion axis, so it is not surprising for the emission to fill the aperture along this axis.

\begin{figure}
\begin{center}
{\includegraphics[width=8.5cm]{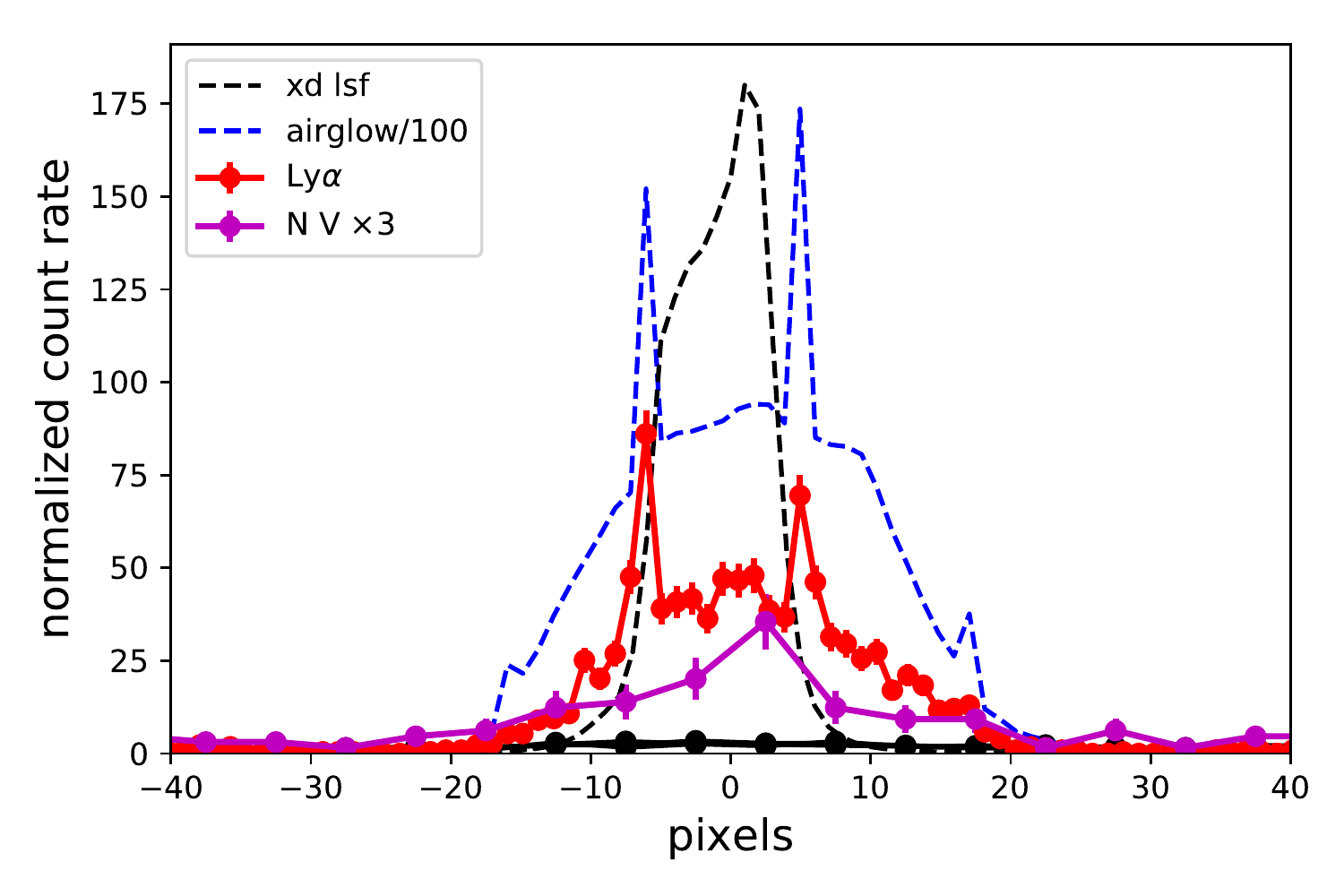}}
\end{center}
\caption{Cross-dispersion profiles for the two brightest lines in our FUV spectrum, along with the COS cross-dispersion line spread function and the cross-dispersion profile for Ly$\alpha$ airglow for comparison. These latter two profiles bracket the range of spatial extension in the cross-dispersion direction, with the lsf corresponding to zero extent and the airglow corresponding to uniformly aperture-filling emission. The filament approximately extends along the cross-dispersion axis in our observations. We infer standard deviations of $3.3\pm0.1$ pixels, $8.5\pm0.2$ pixels,  $8.0\pm0.2$ pixels, and $9.8^{+2.7}_{-1.8}$ pixels for the four lines respectively, suggesting that Ly$\alpha$ and \ion{N}{v} approximately extend along the entire cross-dispersion axis. This does not imply that they are similarly extended along the dispersion axis, however, and we expect that they are in fact not very extended in the dispersion direction.  }
\end{figure}

\section{Effects of COS line spread function}

Since this paper employs a careful kinematic analysis of emission lines with COS, it is reasonable to check that the COS line spread function does not affect our results. We would not expect it to be a major issue, since the COS resel size is about 15 km s$^{-1}$, but the line spread function does have non-Gaussian wings which could theoretically be important to model for bright lines like Ly$\alpha$. To test this, we downloaded the COS line spread function from the same website as in Appendix A, again selecting the line spread function centered at 1221\AA{} to correspond to redshifted Ly$\alpha$ from M87. We convolved our best-fit model for Ly$\alpha$ from M87 (see Figure 5) with this line spread function, and plot the result in Figure B.1. 

The two models are nearly identical. The model convolved with the lsf is formally a poorer fit, but the $\Delta \chi^2$ is just 0.8, so the difference is not very significant. The systematic errors in fitting the complicated Ly$\alpha$ profile with our simple model clearly dominate over uncertainties caused by the line spread function. We therefore conclude that the line spread function is not a significant source of error in our model fitting.

\begin{figure}
\begin{center}
{\includegraphics[width=8.5cm]{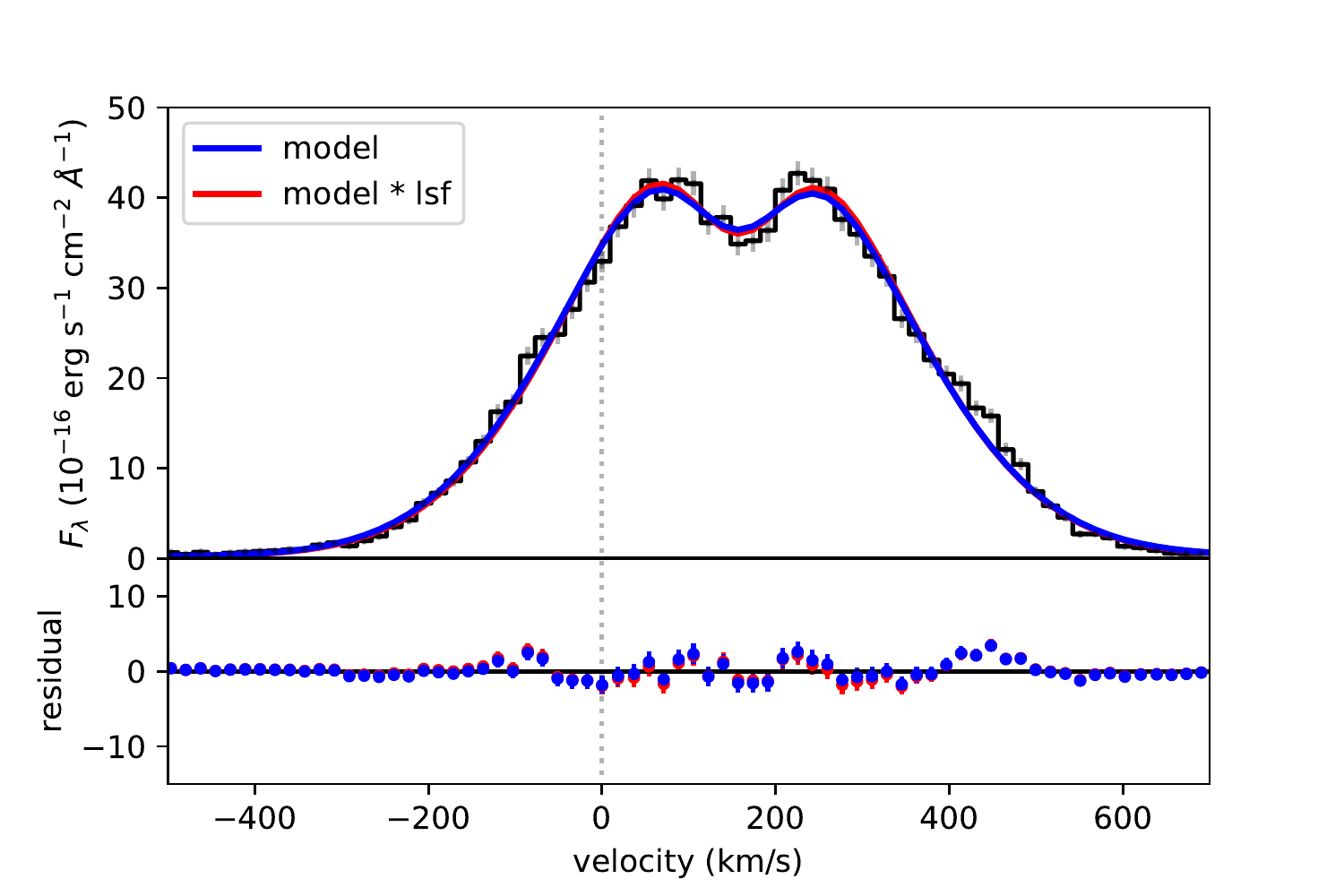}}
\end{center}
\caption{Portion of the spectrum around Ly$\alpha$, along with our best-fit model from sect. 3.4 (blue) and this best-fit model convolved with the COS line spread function (red). The lower panel shows the residuals between both models and the data. The two models are nearly identical, and the convolution of the line spread function with the model increased the $\chi^2$ by only 0.8 when including a systematic error term as described in sect. 3.4. We therefore conclude that the line spread function is not a significant source of error in our model fitting, and can safely be neglected in our analysis.   }
\end{figure}

\end{appendix}

\begin{acknowledgements}
 Based on observations made with the NASA-ESA Hubble Space Telescope, obtained at the Space Telescope Science Institute, which is operated by the Association of Universities for Research in Astronomy, Inc., under NASA contract NAS 5-26555. These observations are associated with program \#14623. This work has been partially supported by the Russian Science Foundation through grant 14-22-00271. The authors would like to thank the referee for thoughtful and helpful comments, as well as Patricia Ar\'{e}valo, Eugene Churazov, and Sara Heap for helpful discussions during the preparation of this paper, and Elaine Snyder at the HST-COS Helpdesk for assistance with understanding the COS calibrations.
\end{acknowledgements}

\bibliographystyle{aa.bst} 
\bibliography{paper.bib} 

\begin{thebibliography}{120}
\expandafter\ifx\csname natexlab\endcsname\relax\def\natexlab#1{#1}\fi

\bibitem[{{Acciari} {et~al.}(2009){Acciari}, {Aliu}, {Arlen}, {Bautista},
  {Beilicke}, {Benbow}, {Bradbury}, {Buckley}, {Bugaev}, {Butt}, \&
  et~al.}]{Acciari2009}
{Acciari}, V.~A., {Aliu}, E., {Arlen}, T., {et~al.} 2009, Science, 325, 444

\bibitem[{{Anderson} \& {Sunyaev}(2016)}]{Anderson2016}
{Anderson}, M.~E. \& {Sunyaev}, R. 2016, \mnras, 459, 2806

\bibitem[{{Arp}(1967)}]{Arp1967}
{Arp}, H.~C. 1967, \aplett, 1, 1

\bibitem[{{Basko} \& {Sunyaev}(1973)}]{Basko1973}
{Basko}, M.~M. \& {Sunyaev}, R.~A. 1973, \apss, 23, 117

\bibitem[{{Bertola} {et~al.}(1980){Bertola}, {Capaccioli}, {Holm}, \&
  {Oke}}]{Bertola1980}
{Bertola}, F., {Capaccioli}, M., {Holm}, A.~V., \& {Oke}, J.~B. 1980, \apjl,
  237, L65

\bibitem[{{Binette} {et~al.}(1993){Binette}, {Fosbury}, \&
  {Parker}}]{Binette1993}
{Binette}, L., {Fosbury}, R.~A., \& {Parker}, D. 1993, \pasp, 105, 1150

\bibitem[{{Biretta} {et~al.}(1999){Biretta}, {Sparks}, \&
  {Macchetto}}]{Biretta1999}
{Biretta}, J.~A., {Sparks}, W.~B., \& {Macchetto}, F. 1999, \apj, 520, 621

\bibitem[{{Blakeslee} {et~al.}(2009){Blakeslee}, {Jord{\'a}n}, {Mei},
  {C{\^o}t{\'e}}, {Ferrarese}, {Infante}, {Peng}, {Tonry}, \&
  {West}}]{Blakeslee2009}
{Blakeslee}, J.~P., {Jord{\'a}n}, A., {Mei}, S., {et~al.} 2009, \apj, 694, 556

\bibitem[{{Boehringer} \& {Fabian}(1989)}]{Boehringer1989}
{Boehringer}, H. \& {Fabian}, A.~C. 1989, \mnras, 237, 1147

\bibitem[{{Bohlin} {et~al.}(1978){Bohlin}, {Savage}, \& {Drake}}]{Bohlin1978}
{Bohlin}, R.~C., {Savage}, B.~D., \& {Drake}, J.~F. 1978, \apj, 224, 132

\bibitem[{{Boissier} {et~al.}(2018){Boissier}, {Cucciati}, {Boselli}, {Mei}, \&
  {Ferrarese}}]{Boissier2018}
{Boissier}, S., {Cucciati}, O., {Boselli}, A., {Mei}, S., \& {Ferrarese}, L.
  2018, \aap, 611, A42

\bibitem[{{Brown}(1999)}]{Brown1999}
{Brown}, T.~M. 1999, in Astronomical Society of the Pacific Conference Series,
  Vol. 192, Spectrophotometric Dating of Stars and Galaxies, ed. I.~{Hubeny},
  S.~{Heap}, \& R.~{Cornett}, 315

\bibitem[{{Br{\"u}ggen} \& {Kaiser}(2001)}]{Brueggen2001}
{Br{\"u}ggen}, M. \& {Kaiser}, C.~R. 2001, \mnras, 325, 676

\bibitem[{{Cardelli} {et~al.}(1989){Cardelli}, {Clayton}, \&
  {Mathis}}]{Cardelli1989}
{Cardelli}, J.~A., {Clayton}, G.~C., \& {Mathis}, J.~S. 1989, \apj, 345, 245

\bibitem[{{Cavagnolo} {et~al.}(2008){Cavagnolo}, {Donahue}, {Voit}, \&
  {Sun}}]{Cavagnolo2008}
{Cavagnolo}, K.~W., {Donahue}, M., {Voit}, G.~M., \& {Sun}, M. 2008, \apjl,
  683, L107

\bibitem[{{Churazov} {et~al.}(2001){Churazov}, {Br{\"u}ggen}, {Kaiser},
  {B{\"o}hringer}, \& {Forman}}]{Churazov2001}
{Churazov}, E., {Br{\"u}ggen}, M., {Kaiser}, C.~R., {B{\"o}hringer}, H., \&
  {Forman}, W. 2001, \apj, 554, 261

\bibitem[{{Churazov} {et~al.}(2008){Churazov}, {Forman}, {Vikhlinin},
  {Tremaine}, {Gerhard}, \& {Jones}}]{Churazov2008}
{Churazov}, E., {Forman}, W., {Vikhlinin}, A., {et~al.} 2008, \mnras, 388, 1062

\bibitem[{{Churazov} {et~al.}(2002){Churazov}, {Sunyaev}, {Forman}, \&
  {B{\"o}hringer}}]{Churazov2002}
{Churazov}, E., {Sunyaev}, R., {Forman}, W., \& {B{\"o}hringer}, H. 2002,
  \mnras, 332, 729

\bibitem[{{Code}(1969)}]{Code1969}
{Code}, A.~D. 1969, \pasp, 81, 475

\bibitem[{{Consiglio} {et~al.}(2017){Consiglio}, {Turner}, {Beck}, {Meier},
  {Silich}, \& {Zhao}}]{Consiglio2017}
{Consiglio}, S.~M., {Turner}, J.~L., {Beck}, S., {et~al.} 2017, ArXiv e-prints
  [\eprint[arXiv]{1710.10282}]

\bibitem[{{Cowie} \& {McKee}(1977)}]{Cowie1977}
{Cowie}, L.~L. \& {McKee}, C.~F. 1977, \apj, 211, 135

\bibitem[{{Danforth} {et~al.}(2016){Danforth}, {Stocke}, {France}, {Begelman},
  \& {Perlman}}]{Danforth2016}
{Danforth}, C.~W., {Stocke}, J.~T., {France}, K., {Begelman}, M.~C., \&
  {Perlman}, E. 2016, \apj, 832, 76

\bibitem[{{Del Zanna} {et~al.}(2015){Del Zanna}, {Dere}, {Young}, {Landi}, \&
  {Mason}}]{Delzanna2015}
{Del Zanna}, G., {Dere}, K.~P., {Young}, P.~R., {Landi}, E., \& {Mason}, H.~E.
  2015, \aap, 582, A56

\bibitem[{{Dere} \& {Cook}(1979)}]{Dere1979}
{Dere}, K.~P. \& {Cook}, J.~W. 1979, \apj, 229, 772

\bibitem[{{Dere} {et~al.}(1997){Dere}, {Landi}, {Mason}, {Monsignori Fossi}, \&
  {Young}}]{Dere1997}
{Dere}, K.~P., {Landi}, E., {Mason}, H.~E., {Monsignori Fossi}, B.~C., \&
  {Young}, P.~R. 1997, \aaps, 125, 149

\bibitem[{{di Serego Alighieri} {et~al.}(2013){di Serego Alighieri}, {Bianchi},
  {Pappalardo}, {Zibetti}, {Auld}, {Baes}, {Bendo}, {Corbelli}, {Davies},
  {Davis}, {De Looze}, {Fritz}, {Gavazzi}, {Giovanardi}, {Grossi}, {Hunt},
  {Magrini}, {Pierini}, \& {Xilouris}}]{diseregoalighieri2013}
{di Serego Alighieri}, S., {Bianchi}, S., {Pappalardo}, C., {et~al.} 2013,
  \aap, 552, A8

\bibitem[{{Dijkstra} {et~al.}(2006){Dijkstra}, {Haiman}, \&
  {Spaans}}]{Dijkstra2006}
{Dijkstra}, M., {Haiman}, Z., \& {Spaans}, M. 2006, \apj, 649, 14

\bibitem[{{Dopita} {et~al.}(1997){Dopita}, {Koratkar}, {Allen}, {Tsvetanov},
  {Ford}, {Bicknell}, \& {Sutherland}}]{Dopita1997}
{Dopita}, M.~A., {Koratkar}, A.~P., {Allen}, M.~G., {et~al.} 1997, \apj, 490,
  202

\bibitem[{{Dopita} \& {Sutherland}(1996)}]{Dopita1996}
{Dopita}, M.~A. \& {Sutherland}, R.~S. 1996, \apjs, 102, 161

\bibitem[{{Doschek} {et~al.}(1976){Doschek}, {Vanhoosier}, {Bartoe}, \&
  {Feldman}}]{Doschek1976}
{Doschek}, G.~A., {Vanhoosier}, M.~E., {Bartoe}, J.-D.~F., \& {Feldman}, U.
  1976, \apjs, 31, 417

\bibitem[{{Draine}(2011)}]{Draine2011}
{Draine}, B.~T. 2011, {Physics of the Interstellar and Intergalactic Medium}

\bibitem[{{Emsellem} {et~al.}(2011){Emsellem}, {Cappellari}, {Krajnovi{\'c}},
  {Alatalo}, {Blitz}, {Bois}, {Bournaud}, {Bureau}, {Davies}, {Davis}, {de
  Zeeuw}, {Khochfar}, {Kuntschner}, {Lablanche}, {McDermid}, {Morganti},
  {Naab}, {Oosterloo}, {Sarzi}, {Scott}, {Serra}, {van de Ven}, {Weijmans}, \&
  {Young}}]{Emsellem2011}
{Emsellem}, E., {Cappellari}, M., {Krajnovi{\'c}}, D., {et~al.} 2011, \mnras,
  414, 888

\bibitem[{{Eracleous} {et~al.}(2010){Eracleous}, {Hwang}, \&
  {Flohic}}]{Eracleous2010}
{Eracleous}, M., {Hwang}, J.~A., \& {Flohic}, H.~M.~L.~G. 2010, \apj, 711, 796

\bibitem[{{Fabian} {et~al.}(2011){Fabian}, {Sanders}, {Williams}, {Lazarian},
  {Ferland}, \& {Johnstone}}]{Fabian2011}
{Fabian}, A.~C., {Sanders}, J.~S., {Williams}, R.~J.~R., {et~al.} 2011, \mnras,
  417, 172

\bibitem[{{Ferland} {et~al.}(2009){Ferland}, {Fabian}, {Hatch}, {Johnstone},
  {Porter}, {van Hoof}, \& {Williams}}]{Ferland2009}
{Ferland}, G.~J., {Fabian}, A.~C., {Hatch}, N.~A., {et~al.} 2009, \mnras, 392,
  1475

\bibitem[{{Ford} \& {Butcher}(1979)}]{Ford1979}
{Ford}, H.~C. \& {Butcher}, H. 1979, \apjs, 41, 147

\bibitem[{{Ford} {et~al.}(1994){Ford}, {Harms}, {Tsvetanov}, {Hartig},
  {Dressel}, {Kriss}, {Bohlin}, {Davidsen}, {Margon}, \& {Kochhar}}]{Ford1994}
{Ford}, H.~C., {Harms}, R.~J., {Tsvetanov}, Z.~I., {et~al.} 1994, \apjl, 435,
  L27

\bibitem[{{Forman} {et~al.}(2017){Forman}, {Churazov}, {Jones}, {Heinz},
  {Kraft}, \& {Vikhlinin}}]{Forman2017}
{Forman}, W., {Churazov}, E., {Jones}, C., {et~al.} 2017, \apj, 844, 122

\bibitem[{{Forman} {et~al.}(2007){Forman}, {Jones}, {Churazov}, {Markevitch},
  {Nulsen}, {Vikhlinin}, {Begelman}, {B{\"o}hringer}, {Eilek}, {Heinz},
  {Kraft}, {Owen}, \& {Pahre}}]{Forman2007}
{Forman}, W., {Jones}, C., {Churazov}, E., {et~al.} 2007, \apj, 665, 1057

\bibitem[{{Forman} {et~al.}(2005){Forman}, {Nulsen}, {Heinz}, {Owen}, {Eilek},
  {Vikhlinin}, {Markevitch}, {Kraft}, {Churazov}, \& {Jones}}]{Forman2005}
{Forman}, W., {Nulsen}, P., {Heinz}, S., {et~al.} 2005, \apj, 635, 894

\bibitem[{{Gebhardt} \& {Thomas}(2009)}]{Gebhardt2009}
{Gebhardt}, K. \& {Thomas}, J. 2009, \apj, 700, 1690

\bibitem[{{Gehrels}(1986)}]{Gehrels1986}
{Gehrels}, N. 1986, \apj, 303, 336

\bibitem[{{Golding} {et~al.}(2017){Golding}, {Leenaarts}, \&
  {Carlsson}}]{Golding2017}
{Golding}, T.~P., {Leenaarts}, J., \& {Carlsson}, M. 2017, \aap, 597, A102

\bibitem[{{Grevesse} \& {Sauval}(1998)}]{Grevesse1998}
{Grevesse}, N. \& {Sauval}, A.~J. 1998, \ssr, 85, 161

\bibitem[{{Harms} {et~al.}(1994){Harms}, {Ford}, {Tsvetanov}, {Hartig},
  {Dressel}, {Kriss}, {Bohlin}, {Davidsen}, {Margon}, \& {Kochhar}}]{Harms1994}
{Harms}, R.~J., {Ford}, H.~C., {Tsvetanov}, Z.~I., {et~al.} 1994, \apjl, 435,
  L35

\bibitem[{{Harrington}(1973)}]{Harrington1973}
{Harrington}, J.~P. 1973, \mnras, 162, 43

\bibitem[{{Harris} {et~al.}(1997){Harris}, {Biretta}, \& {Junor}}]{Harris1997}
{Harris}, D.~E., {Biretta}, J.~A., \& {Junor}, W. 1997, \mnras, 284, L21

\bibitem[{{Hayes} \& {Nussbaumer}(1984)}]{Hayes1984}
{Hayes}, M.~A. \& {Nussbaumer}, H. 1984, \aap, 134, 193

\bibitem[{{Heckman} {et~al.}(1989){Heckman}, {Baum}, {van Breugel}, \&
  {McCarthy}}]{Heckman1989}
{Heckman}, T.~M., {Baum}, S.~A., {van Breugel}, W.~J.~M., \& {McCarthy}, P.
  1989, \apj, 338, 48

\bibitem[{{Heiles} \& {Troland}(2003)}]{Heiles2003}
{Heiles}, C. \& {Troland}, T.~H. 2003, \apj, 586, 1067

\bibitem[{{Hines} {et~al.}(1989){Hines}, {Eilek}, \& {Owen}}]{Hines1989}
{Hines}, D.~C., {Eilek}, J.~A., \& {Owen}, F.~N. 1989, \apj, 347, 713

\bibitem[{{Hitomi Collaboration} {et~al.}(2016){Hitomi Collaboration},
  {Aharonian}, {Akamatsu}, {Akimoto}, {Allen}, {Anabuki}, {Angelini}, {Arnaud},
  {Audard}, {Awaki}, {Axelsson}, {Bamba}, {Bautz}, {Blandford}, {Brenneman},
  {Brown}, {Bulbul}, {Cackett}, {Chernyakova}, {Chiao}, {Coppi}, {Costantini},
  {de Plaa}, {den Herder}, {Done}, {Dotani}, {Ebisawa}, {Eckart}, {Enoto},
  {Ezoe}, {Fabian}, {Ferrigno}, {Foster}, {Fujimoto}, {Fukazawa}, {Furuzawa},
  {Galeazzi}, {Gallo}, {Gandhi}, {Giustini}, {Goldwurm}, {Gu}, {Guainazzi},
  {Haba}, {Hagino}, {Hamaguchi}, {Harrus}, {Hatsukade}, {Hayashi}, {Hayashi},
  {Hayashida}, {Hiraga}, {Hornschemeier}, {Hoshino}, {Hughes}, {Iizuka},
  {Inoue}, {Inoue}, {Ishibashi}, {Ishida}, {Ishikawa}, {Ishisaki}, {Itoh},
  {Iyomoto}, {Kaastra}, {Kallman}, {Kamae}, {Kara}, {Kataoka}, {Katsuda},
  {Katsuta}, {Kawaharada}, {Kawai}, {Kelley}, {Khangulyan}, {Kilbourne},
  {King}, {Kitaguchi}, {Kitamoto}, {Kitayama}, {Kohmura}, {Kokubun}, {Koyama},
  {Koyama}, {Kretschmar}, {Krimm}, {Kubota}, {Kunieda}, {Laurent}, {Lebrun},
  {Lee}, {Leutenegger}, {Limousin}, {Loewenstein}, {Long}, {Lumb}, {Madejski},
  {Maeda}, {Maier}, {Makishima}, {Markevitch}, {Matsumoto}, {Matsushita},
  {McCammon}, {McNamara}, {Mehdipour}, {Miller}, {Miller}, {Mineshige},
  {Mitsuda}, {Mitsuishi}, {Miyazawa}, {Mizuno}, {Mori}, {Mori}, {Moseley},
  {Mukai}, {Murakami}, {Murakami}, {Mushotzky}, {Nagino}, {Nakagawa},
  {Nakajima}, {Nakamori}, {Nakano}, {Nakashima}, {Nakazawa}, {Nobukawa},
  {Noda}, {Nomachi}, {O'Dell}, {Odaka}, {Ohashi}, {Ohno}, {Okajima}, {Ota},
  {Ozaki}, {Paerels}, {Paltani}, {Parmar}, {Petre}, {Pinto}, {Pohl}, {Porter},
  {Pottschmidt}, {Ramsey}, {Reynolds}, {Russell}, {Safi-Harb}, {Saito},
  {Sakai}, {Sameshima}, {Sato}, {Sato}, {Sato}, {Sawada}, {Schartel},
  {Serlemitsos}, {Seta}, {Shidatsu}, {Simionescu}, {Smith}, {Soong}, {Stawarz},
  {Sugawara}, {Sugita}, {Szymkowiak}, {Tajima}, {Takahashi}, {Takahashi},
  {Takeda}, {Takei}, {Tamagawa}, {Tamura}, {Tamura}, {Tanaka}, {Tanaka},
  {Tanaka}, {Tashiro}, {Tawara}, {Terada}, {Terashima}, {Tombesi}, {Tomida},
  {Tsuboi}, {Tsujimoto}, {Tsunemi}, {Tsuru}, {Uchida}, {Uchiyama}, {Uchiyama},
  {Ueda}, {Ueda}, {Ueno}, {Uno}, {Urry}, {Ursino}, {de Vries}, {Watanabe},
  {Werner}, {Wik}, {Wilkins}, {Williams}, {Yamada}, {Yamaguchi}, {Yamaoka},
  {Yamasaki}, {Yamauchi}, {Yamauchi}, {Yaqoob}, {Yatsu}, {Yonetoku}, {Yoshida},
  {Yuasa}, {Zhuravleva}, \& {Zoghbi}}]{Hitomi2016}
{Hitomi Collaboration}, {Aharonian}, F., {Akamatsu}, H., {et~al.} 2016, \nat,
  535, 117

\bibitem[{{Ho}(2008)}]{Ho2008}
{Ho}, L.~C. 2008, \araa, 46, 475

\bibitem[{{Innes} {et~al.}(2001){Innes}, {Curdt}, {Schwenn}, {Solanki},
  {Stenborg}, \& {McKenzie}}]{Innes2001}
{Innes}, D.~E., {Curdt}, W., {Schwenn}, R., {et~al.} 2001, \apjl, 549, L249

\bibitem[{{Inogamov} \& {Sunyaev}(2003)}]{Inogamov2003}
{Inogamov}, N.~A. \& {Sunyaev}, R.~A. 2003, Astronomy Letters, 29, 791

\bibitem[{{Inogamov} \& {Sunyaev}(2010)}]{Inogamov2010}
{Inogamov}, N.~A. \& {Sunyaev}, R.~A. 2010, Astronomy Letters, 36, 835

\bibitem[{{Jordan}(1975)}]{Jordan1975}
{Jordan}, C. 1975, \mnras, 170, 429

\bibitem[{{Kalberla} {et~al.}(2005){Kalberla}, {Burton}, {Hartmann}, {Arnal},
  {Bajaja}, {Morras}, \& {P{\"o}ppel}}]{Kalberla2005}
{Kalberla}, P.~M.~W., {Burton}, W.~B., {Hartmann}, D., {et~al.} 2005, \aap,
  440, 775

\bibitem[{{Kohl}(1977)}]{Kohl1977}
{Kohl}, J.~L. 1977, \apj, 211, 958

\bibitem[{{Kormendy} {et~al.}(2009){Kormendy}, {Fisher}, {Cornell}, \&
  {Bender}}]{Kormendy2009}
{Kormendy}, J., {Fisher}, D.~B., {Cornell}, M.~E., \& {Bender}, R. 2009, \apjs,
  182, 216

\bibitem[{{Krajnovi{\'c}} {et~al.}(2011){Krajnovi{\'c}}, {Emsellem},
  {Cappellari}, {Alatalo}, {Blitz}, {Bois}, {Bournaud}, {Bureau}, {Davies},
  {Davis}, {de Zeeuw}, {Khochfar}, {Kuntschner}, {Lablanche}, {McDermid},
  {Morganti}, {Naab}, {Oosterloo}, {Sarzi}, {Scott}, {Serra}, {Weijmans}, \&
  {Young}}]{Krajnovic2011}
{Krajnovi{\'c}}, D., {Emsellem}, E., {Cappellari}, M., {et~al.} 2011, \mnras,
  414, 2923

\bibitem[{{Kwan}(1984)}]{Kwan1984}
{Kwan}, J. 1984, \apj, 283, 70

\bibitem[{{Kwan} \& {Krolik}(1981)}]{Kwan1981}
{Kwan}, J. \& {Krolik}, J.~H. 1981, \apj, 250, 478

\bibitem[{{Laming} \& {Feldman}(1992)}]{Laming1992}
{Laming}, J.~M. \& {Feldman}, U. 1992, \apj, 386, 364

\bibitem[{{Laming} \& {Feldman}(1993)}]{Laming1993}
{Laming}, J.~M. \& {Feldman}, U. 1993, \apj, 403, 434

\bibitem[{{Liang} {et~al.}(2012){Liang}, {Badnell}, \& {Zhao}}]{Liang2012}
{Liang}, G.~Y., {Badnell}, N.~R., \& {Zhao}, G. 2012, \aap, 547, A87

\bibitem[{{Macchetto} {et~al.}(1997){Macchetto}, {Marconi}, {Axon}, {Capetti},
  {Sparks}, \& {Crane}}]{Macchetto1997}
{Macchetto}, F., {Marconi}, A., {Axon}, D.~J., {et~al.} 1997, \apj, 489, 579

\bibitem[{{Makarov} {et~al.}(2014){Makarov}, {Prugniel}, {Terekhova},
  {Courtois}, \& {Vauglin}}]{Makarov2014}
{Makarov}, D., {Prugniel}, P., {Terekhova}, N., {Courtois}, H., \& {Vauglin},
  I. 2014, \aap, 570, A13

\bibitem[{{Mariska}(1992)}]{Mariska1992}
{Mariska}, J.~T. 1992, {The Solar Transition Region}, 290

\bibitem[{{McDonald} {et~al.}(2012){McDonald}, {Veilleux}, \&
  {Rupke}}]{McDonald2012}
{McDonald}, M., {Veilleux}, S., \& {Rupke}, D.~S.~N. 2012, \apj, 746, 153

\bibitem[{{McKee} \& {Cowie}(1977)}]{McKee1977}
{McKee}, C.~F. \& {Cowie}, L.~L. 1977, \apj, 215, 213

\bibitem[{{McKee} \& {Ostriker}(1977)}]{Mckee1977b}
{McKee}, C.~F. \& {Ostriker}, J.~P. 1977, \apj, 218, 148

\bibitem[{{McNamara} {et~al.}(1996){McNamara}, {O'Connell}, \&
  {Sarazin}}]{McNamara1996}
{McNamara}, B.~R., {O'Connell}, R.~W., \& {Sarazin}, C.~L. 1996, \aj, 112, 91

\bibitem[{{Million} {et~al.}(2010){Million}, {Werner}, {Simionescu}, {Allen},
  {Nulsen}, {Fabian}, {B{\"o}hringer}, \& {Sanders}}]{Million2010}
{Million}, E.~T., {Werner}, N., {Simionescu}, A., {et~al.} 2010, \mnras, 407,
  2046

\bibitem[{{Mittal} {et~al.}(2011){Mittal}, {O'Dea}, {Ferland}, {Oonk}, {Edge},
  {Canning}, {Russell}, {Baum}, {B{\"o}hringer}, {Combes}, {Donahue}, {Fabian},
  {Hatch}, {Hoffer}, {Johnstone}, {McNamara}, {Salom{\'e}}, \&
  {Tremblay}}]{Mittal2011}
{Mittal}, R., {O'Dea}, C.~P., {Ferland}, G., {et~al.} 2011, \mnras, 418, 2386

\bibitem[{{Mittal} {et~al.}(2012){Mittal}, {Oonk}, {Ferland}, {Edge}, {O'Dea},
  {Baum}, {Whelan}, {Johnstone}, {Combes}, {Salom{\'e}}, {Fabian}, {Tremblay},
  {Donahue}, \& {Russell}}]{Mittal2012}
{Mittal}, R., {Oonk}, J.~B.~R., {Ferland}, G.~J., {et~al.} 2012, \mnras, 426,
  2957

\bibitem[{{Moore}(1970)}]{Moore1970}
{Moore}, C.~E. 1970, {Selected tables of atomic spectra}

\bibitem[{{Nahar} \& {Pradhan}(1997)}]{Nahar1997}
{Nahar}, S.~N. \& {Pradhan}, A.~K. 1997, \apjs, 111, 339

\bibitem[{{Netzer}(1987)}]{Netzer1987}
{Netzer}, H. 1987, \mnras, 225, 55

\bibitem[{{Neufeld}(1990)}]{Neufeld1990}
{Neufeld}, D.~A. 1990, \apj, 350, 216

\bibitem[{{Nipoti} \& {Binney}(2004)}]{Nipoti2004}
{Nipoti}, C. \& {Binney}, J. 2004, \mnras, 349, 1509

\bibitem[{{O'Connell}(1999)}]{Oconnell1999}
{O'Connell}, R.~W. 1999, \araa, 37, 603

\bibitem[{{O'Dea} {et~al.}(2004){O'Dea}, {Baum}, {Mack}, {Koekemoer}, \&
  {Laor}}]{Odea2004}
{O'Dea}, C.~P., {Baum}, S.~A., {Mack}, J., {Koekemoer}, A.~M., \& {Laor}, A.
  2004, \apj, 612, 131

\bibitem[{{Ohl} {et~al.}(1998){Ohl}, {O'Connell}, {Bohlin}, {Collins},
  {Dorman}, {Fanelli}, {Neff}, {Roberts}, {Smith}, \& {Stecher}}]{Ohl1998}
{Ohl}, R.~G., {O'Connell}, R.~W., {Bohlin}, R.~C., {et~al.} 1998, \apjl, 505,
  L11

\bibitem[{{Osterbrock} \& {Ferland}(2006)}]{Osterbrock2006}
{Osterbrock}, D.~E. \& {Ferland}, G.~J. 2006, {Astrophysics of gaseous nebulae
  and active galactic nuclei}

\bibitem[{{Peeples} {et~al.}(2017){Peeples}, {Tumlinson}, {Fox}, {Aloisi},
  {Fleming}, {Jedrzejewski}, {Oliveira}, {Ayres}, {Danforth}, {Keeney}, \&
  {Jenkins}}]{Peeples2017}
{Peeples}, M., {Tumlinson}, J., {Fox}, A., {et~al.} 2017, {The Hubble
  Spectroscopic Legacy Archive}, Tech. rep.

\bibitem[{{Perlman} {et~al.}(2011){Perlman}, {Adams}, {Cara}, {Bourque},
  {Harris}, {Madrid}, {Simons}, {Clausen-Brown}, {Cheung}, {Stawarz},
  {Georganopoulos}, {Sparks}, \& {Biretta}}]{Perlman2011}
{Perlman}, E.~S., {Adams}, S.~C., {Cara}, M., {et~al.} 2011, \apj, 743, 119

\bibitem[{{Poglitsch} {et~al.}(2010){Poglitsch}, {Waelkens}, {Geis},
  {Feuchtgruber}, {Vandenbussche}, {Rodriguez}, {Krause}, {Renotte}, {van
  Hoof}, {Saraceno}, {Cepa}, {Kerschbaum}, {Agn{\`e}se}, {Ali}, {Altieri},
  {Andreani}, {Augueres}, {Balog}, {Barl}, {Bauer}, {Belbachir}, {Benedettini},
  {Billot}, {Boulade}, {Bischof}, {Blommaert}, {Callut}, {Cara}, {Cerulli},
  {Cesarsky}, {Contursi}, {Creten}, {De Meester}, {Doublier}, {Doumayrou},
  {Duband}, {Exter}, {Genzel}, {Gillis}, {Gr{\"o}zinger}, {Henning},
  {Herreros}, {Huygen}, {Inguscio}, {Jakob}, {Jamar}, {Jean}, {de Jong},
  {Katterloher}, {Kiss}, {Klaas}, {Lemke}, {Lutz}, {Madden}, {Marquet},
  {Martignac}, {Mazy}, {Merken}, {Montfort}, {Morbidelli}, {M{\"u}ller},
  {Nielbock}, {Okumura}, {Orfei}, {Ottensamer}, {Pezzuto}, {Popesso},
  {Putzeys}, {Regibo}, {Reveret}, {Royer}, {Sauvage}, {Schreiber}, {Stegmaier},
  {Schmitt}, {Schubert}, {Sturm}, {Thiel}, {Tofani}, {Vavrek}, {Wetzstein},
  {Wieprecht}, \& {Wiezorrek}}]{Poglitsch2010}
{Poglitsch}, A., {Waelkens}, C., {Geis}, N., {et~al.} 2010, \aap, 518, L2

\bibitem[{{Raymond} {et~al.}(1979){Raymond}, {Noyes}, \& {Stopa}}]{Raymond1979}
{Raymond}, J.~C., {Noyes}, R.~W., \& {Stopa}, M.~P. 1979, \solphys, 61, 271

\bibitem[{{Roussel-Dupre}(1982)}]{Roussel1982}
{Roussel-Dupre}, D. 1982, \apj, 256, 284

\bibitem[{{Russell} {et~al.}(2015){Russell}, {Fabian}, {McNamara}, \&
  {Broderick}}]{Russell2015}
{Russell}, H.~R., {Fabian}, A.~C., {McNamara}, B.~R., \& {Broderick}, A.~E.
  2015, \mnras, 451, 588

\bibitem[{{Sabra} {et~al.}(2003){Sabra}, {Shields}, {Ho}, {Barth}, \&
  {Filippenko}}]{Sabra2003}
{Sabra}, B.~M., {Shields}, J.~C., {Ho}, L.~C., {Barth}, A.~J., \& {Filippenko},
  A.~V. 2003, \apj, 584, 164

\bibitem[{{Salom{\'e}} \& {Combes}(2008)}]{Salome2008}
{Salom{\'e}}, P. \& {Combes}, F. 2008, \aap, 489, 101

\bibitem[{{Salom{\'e}} {et~al.}(2006){Salom{\'e}}, {Combes}, {Edge},
  {Crawford}, {Erlund}, {Fabian}, {Hatch}, {Johnstone}, {Sanders}, \&
  {Wilman}}]{Salome2006}
{Salom{\'e}}, P., {Combes}, F., {Edge}, A.~C., {et~al.} 2006, \aap, 454, 437

\bibitem[{{Salom{\'e}} {et~al.}(2011){Salom{\'e}}, {Combes}, {Revaz}, {Downes},
  {Edge}, \& {Fabian}}]{Salome2011}
{Salom{\'e}}, P., {Combes}, F., {Revaz}, Y., {et~al.} 2011, \aap, 531, A85

\bibitem[{{Sargent} {et~al.}(1978){Sargent}, {Young}, {Lynds}, {Boksenberg},
  {Shortridge}, \& {Hartwick}}]{Sargent1978}
{Sargent}, W.~L.~W., {Young}, P.~J., {Lynds}, C.~R., {et~al.} 1978, \apj, 221,
  731

\bibitem[{{Shankar} {et~al.}(2009){Shankar}, {Weinberg}, \&
  {Miralda-Escud{\'e}}}]{Shankar2009}
{Shankar}, F., {Weinberg}, D.~H., \& {Miralda-Escud{\'e}}, J. 2009, \apj, 690,
  20

\bibitem[{{Sharma} {et~al.}(2010){Sharma}, {Parrish}, \&
  {Quataert}}]{Sharma2010}
{Sharma}, P., {Parrish}, I.~J., \& {Quataert}, E. 2010, \apj, 720, 652

\bibitem[{{Simionescu} {et~al.}(2008){Simionescu}, {Werner}, {Finoguenov},
  {B{\"o}hringer}, \& {Br{\"u}ggen}}]{Simionescu2008}
{Simionescu}, A., {Werner}, N., {Finoguenov}, A., {B{\"o}hringer}, H., \&
  {Br{\"u}ggen}, M. 2008, \aap, 482, 97

\bibitem[{{Smith} {et~al.}(2001){Smith}, {Brickhouse}, {Liedahl}, \&
  {Raymond}}]{Smith2001}
{Smith}, R.~K., {Brickhouse}, N.~S., {Liedahl}, D.~A., \& {Raymond}, J.~C.
  2001, \apjl, 556, L91

\bibitem[{{Sparks} {et~al.}(2004){Sparks}, {Donahue}, {Jord{\'a}n},
  {Ferrarese}, \& {C{\^o}t{\'e}}}]{Sparks2004}
{Sparks}, W.~B., {Donahue}, M., {Jord{\'a}n}, A., {Ferrarese}, L., \&
  {C{\^o}t{\'e}}, P. 2004, \apj, 607, 294

\bibitem[{{Sparks} {et~al.}(1993){Sparks}, {Ford}, \& {Kinney}}]{Sparks1993}
{Sparks}, W.~B., {Ford}, H.~C., \& {Kinney}, A.~L. 1993, \apj, 413, 531

\bibitem[{{Sparks} {et~al.}(1992){Sparks}, {Fraix-Burnet}, {Macchetto}, \&
  {Owen}}]{Sparks1992}
{Sparks}, W.~B., {Fraix-Burnet}, D., {Macchetto}, F., \& {Owen}, F.~N. 1992,
  \nat, 355, 804

\bibitem[{{Sparks} {et~al.}(2012){Sparks}, {Pringle}, {Carswell}, {Donahue},
  {Martin}, {Voit}, {Cracraft}, {Manset}, \& {Hough}}]{Sparks2012}
{Sparks}, W.~B., {Pringle}, J.~E., {Carswell}, R.~F., {et~al.} 2012, \apjl,
  750, L5

\bibitem[{{Sparks} {et~al.}(2009){Sparks}, {Pringle}, {Donahue}, {Carswell},
  {Voit}, {Cracraft}, \& {Martin}}]{Sparks2009}
{Sparks}, W.~B., {Pringle}, J.~E., {Donahue}, M., {et~al.} 2009, \apjl, 704,
  L20

\bibitem[{{Sternberg} \& {Dalgarno}(1995)}]{Sternberg1995}
{Sternberg}, A. \& {Dalgarno}, A. 1995, \apjs, 99, 565

\bibitem[{{STScI Development Team}(2013)}]{Lim2015}
{STScI Development Team}. 2013, {pysynphot: Synthetic photometry software
  package}, Astrophysics Source Code Library

\bibitem[{{Sutherland} \& {Dopita}(1993)}]{Sutherland1993}
{Sutherland}, R.~S. \& {Dopita}, M.~A. 1993, \apjs, 88, 253

\bibitem[{{Tielens} \& {Hollenbach}(1985)}]{Tielens1985}
{Tielens}, A.~G.~G.~M. \& {Hollenbach}, D. 1985, \apj, 291, 722

\bibitem[{{Tucker} \& {David}(1997)}]{Tucker1997}
{Tucker}, W. \& {David}, L.~P. 1997, \apj, 484, 602

\bibitem[{{Tumlinson} {et~al.}(2013){Tumlinson}, {Thom}, {Werk}, {Prochaska},
  {Tripp}, {Katz}, {Dav{\'e}}, {Oppenheimer}, {Meiring}, {Ford}, {O'Meara},
  {Peeples}, {Sembach}, \& {Weinberg}}]{Tumlinson2013}
{Tumlinson}, J., {Thom}, C., {Werk}, J.~K., {et~al.} 2013, \apj, 777, 59

\bibitem[{{Walsh} {et~al.}(2013){Walsh}, {Barth}, {Ho}, \& {Sarzi}}]{Walsh2013}
{Walsh}, J.~L., {Barth}, A.~J., {Ho}, L.~C., \& {Sarzi}, M. 2013, \apj, 770, 86

\bibitem[{{Wang} \& {Zhou}(2009)}]{Wang2009}
{Wang}, C.-C. \& {Zhou}, H.-Y. 2009, \mnras, 395, 301

\bibitem[{{Werner} {et~al.}(2013){Werner}, {Oonk}, {Canning}, {Allen},
  {Simionescu}, {Kos}, {van Weeren}, {Edge}, {Fabian}, {von der Linden},
  {Nulsen}, {Reynolds}, \& {Ruszkowski}}]{Werner2013}
{Werner}, N., {Oonk}, J.~B.~R., {Canning}, R.~E.~A., {et~al.} 2013, \apj, 767,
  153

\bibitem[{{Werner} {et~al.}(2009){Werner}, {Zhuravleva}, {Churazov},
  {Simionescu}, {Allen}, {Forman}, {Jones}, \& {Kaastra}}]{Werner2009}
{Werner}, N., {Zhuravleva}, I., {Churazov}, E., {et~al.} 2009, \mnras, 398, 23

\bibitem[{{Yi} {et~al.}(2011){Yi}, {Lee}, {Sheen}, {Jeong}, {Suh}, \&
  {Oh}}]{Yi2011}
{Yi}, S.~K., {Lee}, J., {Sheen}, Y.-K., {et~al.} 2011, \apjs, 195, 22

\bibitem[{{Young} {et~al.}(2002){Young}, {Wilson}, \& {Mundell}}]{Young2002}
{Young}, A.~J., {Wilson}, A.~S., \& {Mundell}, C.~G. 2002, \apj, 579, 560

\bibitem[{{Young} {et~al.}(1978){Young}, {Westphal}, {Kristian}, {Wilson}, \&
  {Landauer}}]{Young1978}
{Young}, P.~J., {Westphal}, J.~A., {Kristian}, J., {Wilson}, C.~P., \&
  {Landauer}, F.~P. 1978, \apj, 221, 721

\bibitem[{{Young} {et~al.}(2013){Young}, {Doschek}, {Warren}, \&
  {Hara}}]{Young2013}
{Young}, P.~R., {Doschek}, G.~A., {Warren}, H.~P., \& {Hara}, H. 2013, \apj,
  766, 127

\bibitem[{{Young} {et~al.}(2015){Young}, {Tian}, \& {Jaeggli}}]{Young2015}
{Young}, P.~R., {Tian}, H., \& {Jaeggli}, S. 2015, \apj, 799, 218

\end{thebibliography}
\end{document}